\documentclass[iop]{emulateapj}
\usepackage{apjfonts}

\usepackage{natbib}
\usepackage{graphics}
\usepackage{multirow}
\usepackage{amsbsy}
\usepackage{mathrsfs}
\usepackage{footmisc}
\usepackage{hyperref}
\hypersetup{
  colorlinks = true,
  urlcolor = black,
  linkcolor=black,
  citecolor=black
}

\def\bicep{{\sc Bicep}}

\def\dasi{{\sc Dasi}}

\def\bicepone{{\sc Bicep1}}
\def\biceptwo{{\sc Bicep2}}

\def\planck{{\it Planck}}
\def\quiet{{\sc Quiet}}
\def\spider{{\sc Spider}}

\def\QUAD{{\sc QUaD}}

\def\wmap{WMAP}

\def\keck{{\it Keck Array}}
\def\polarbear{{\sc Polarbear}}
\def\polar{{\sc Polar}}

\def\gcp{GCP}    

\def\muK{~\mu{\rm K}}

\def\scanset{{scan set}}
\def\elnod{{el nod}}
\def\elnods{{el nods}}

\def\eg{{\em e.g.}}
\def\ie{{\em i.e.}}

\def\aap{Astr. \& Astroph.}

\def\apjl{Astrophys.\ J.\ Lett.}
\def\apjs{Astrophys.\ J.\ Suppl.}

\begin{document}

\title{\bicep2 II: Experiment and Three-Year Data Set}

\author{\textsc{Bicep2} Collaboration - P.~A.~R.~Ade\altaffilmark{1}}
\author{R.~W.~Aikin\altaffilmark{2}}
\author{M.~Amiri\altaffilmark{3}}
\author{D.~Barkats\altaffilmark{4}}
\author{S.~J.~Benton\altaffilmark{5}}
\author{C.~A.~Bischoff\altaffilmark{6}}
\author{J.~J.~Bock\altaffilmark{2,7}}
\author{J.~A.~Brevik\altaffilmark{2}}
\author{I.~Buder\altaffilmark{6}}
\author{E.~Bullock\altaffilmark{8}}
\author{G.~Davis\altaffilmark{3}}
\author{P.~K.~Day\altaffilmark{7}}
\author{C.~D.~Dowell\altaffilmark{7}}
\author{L.~Duband\altaffilmark{9}}
\author{J.~P.~Filippini\altaffilmark{2}}
\author{S.~Fliescher\altaffilmark{10}}
\author{S.~R.~Golwala\altaffilmark{2}}
\author{M.~Halpern\altaffilmark{3}}
\author{M.~Hasselfield\altaffilmark{3}}
\author{S.~R.~Hildebrandt\altaffilmark{2,7}}
\author{G.~C.~Hilton\altaffilmark{11}}
\author{K.~D.~Irwin\altaffilmark{12,13,11}}
\author{K.~S.~Karkare\altaffilmark{6}}
\author{J.~P.~Kaufman\altaffilmark{14}}
\author{B.~G.~Keating\altaffilmark{14}}
\author{S.~A.~Kernasovskiy\altaffilmark{12}}
\author{J.~M.~Kovac\altaffilmark{6}}
\author{C.~L.~Kuo\altaffilmark{12,13}}
\author{E.~M.~Leitch\altaffilmark{15}}
\author{N.~Llombart\altaffilmark{7}}
\author{M.~Lueker\altaffilmark{2}}
\author{C.~B.~Netterfield\altaffilmark{5}}
\author{H.~T.~Nguyen\altaffilmark{7}}
\author{R.~O'Brient\altaffilmark{7}}
\author{R.~W.~Ogburn~IV\altaffilmark{12,13,XX}}
\author{A.~Orlando\altaffilmark{14}}
\author{C.~Pryke\altaffilmark{10}}
\author{C.~D.~Reintsema\altaffilmark{11}}
\author{S.~Richter\altaffilmark{6}}
\author{R.~Schwarz\altaffilmark{10}}
\author{C.~D.~Sheehy\altaffilmark{10,15}}
\author{Z.~K.~Staniszewski\altaffilmark{2}}
\author{K.~T.~Story\altaffilmark{15}}
\author{R.~V.~Sudiwala\altaffilmark{1}}
\author{G.~P.~Teply\altaffilmark{2}}
\author{J.~E.~Tolan\altaffilmark{12}}
\author{A.~D.~Turner\altaffilmark{7}}
\author{A.~G.~Vieregg\altaffilmark{6,15}}
\author{P.~Wilson\altaffilmark{7}}
\author{C.~L.~Wong\altaffilmark{6}}
\author{K.~W.~Yoon\altaffilmark{12,13}}

\altaffiltext{1}{School of Physics and Astronomy, Cardiff University, Cardiff, CF24 3AA, UK}
\altaffiltext{2}{Department of Physics, California Institute of Technology, Pasadena, CA 91125, USA}
\altaffiltext{3}{Department of Physics and Astronomy, University of British Columbia, Vancouver, BC, Canada}
\altaffiltext{4}{Joint ALMA Observatory, ESO, Santiago, Chile}
\altaffiltext{5}{Department of Physics, University of Toronto, Toronto, ON, Canada}
\altaffiltext{6}{Harvard-Smithsonian Center for Astrophysics, 60 Garden Street MS 42, Cambridge, MA 02138, USA}
\altaffiltext{7}{Jet Propulsion Laboratory, Pasadena, CA 91109, USA}
\altaffiltext{8}{Minnesota Institute for Astrophysics, University of Minnesota, Minneapolis, MN 55455, USA}
\altaffiltext{9}{Universit\'{e} Grenoble Alpes, CEA INAC-SBT, F-38000 Grenoble, France}
\altaffiltext{10}{Department of Physics, University of Minnesota, Minneapolis, MN 55455, USA}
\altaffiltext{11}{National Institute of Standards and Technology, Boulder, CO 80305, USA}
\altaffiltext{12}{Department of Physics, Stanford University, Stanford, CA 94305, USA}
\altaffiltext{13}{Kavli Institute for Particle Astrophysics and Cosmology, SLAC National Accelerator Laboratory, 2575 Sand Hill Rd, Menlo Park, CA 94025, USA}
\altaffiltext{14}{Department of Physics, University of California at San Diego, La Jolla, CA 92093, USA}
\altaffiltext{15}{University of Chicago, Chicago, IL 60637, USA}
\altaffiltext{XX}{Corresponding author: ogburn@stanford.edu}

\begin{abstract}
We report on the design and performance of the \bicep2 instrument and on its
three-year data set.  \bicep2 was designed
to measure the polarization of the cosmic
microwave background (CMB) on angular scales of 1 to 5 degrees ($\ell=40$--$200$),
near the expected peak of the $B$-mode polarization signature of primordial
gravitational waves from cosmic inflation.  Measuring $B$-modes requires dramatic
improvements in sensitivity combined with exquisite control of systematics.
The \bicep2 telescope observed from the South Pole with a 26~cm aperture and cold, on-axis, refractive optics.
\bicep2 also adopted a new
detector design in which beam-defining slot antenna arrays couple to transition-edge
sensor (TES) bolometers, all fabricated on a common substrate.
The antenna-coupled TES detectors supported scalable fabrication and
multiplexed readout that allowed \bicep2 to achieve a high detector count of 500
bolometers at 150~GHz, giving unprecedented sensitivity to $B$-modes at degree angular scales.
After optimization of detector
and readout parameters, \bicep2 achieved an instrument noise-equivalent
temperature of $15.8~\mu\mathrm{K}\sqrt{\mathrm{s}}$.  The full data set reached
Stokes $Q$ and $U$ map depths of 87.2~nK in square-degree pixels
($5.2~\mu\mathrm{K}\cdot\mathrm{arcmin}$) over an effective area of 384
square degrees within a 1000 square degree field.
These are the deepest CMB polarization maps at degree angular scales to date.
The power spectrum analysis presented in a companion
paper has resulted in a significant detection of $B$-mode polarization at degree scales.
\end{abstract}


\section{Introduction}

During the past two decades the $\Lambda$CDM model has become the
standard framework for understanding the large-scale phenomenology of
our universe.  Observations of the cosmic microwave background (CMB)
radiation have played a prominent role in developing this concordance
model.  The temperature anisotropy measured by the
\wmap~\citep{wmap9params} and \planck~\citep{planckXVI} satellites have
allowed very precise determination of key parameters such as the mean
curvature, the dark energy density, and the baryon fraction.

In addition to this temperature signal, the CMB also possesses a small
degree of polarization.  This arises from Thomson scattering of photons from
free electrons at the time of decoupling in the presence of an anisotropic
distribution of photons~\citep{rees68}.
The largest component of polarization is a curl-free ``$E$-mode'' pattern produced
by the same scalar density fluctuations that give rise to the CMB temperature
anisotropy.  The scalar fluctuations are unable to induce a pure-curl
``$B$-mode'' polarization pattern in the CMB, but $B$-modes can be produced
by primordial gravitational waves~\citep{selzal97,kamikosostebbins97}.
These gravitational waves, or tensor fluctuations, are a generic prediction of
inflationary models ~\citep{starobinskii79,rubakov82,fabbri83}.
The relative amplitude, characterized by the tensor-to-scalar ratio
$r$~\citep{camb2000,leach02}, is a probe of the energy scale of the physics
behind inflation.  The presence and amplitude of primordial $B$-mode polarization
is thus a key tool for understanding the inflationary epoch. 

The first detection of CMB polarization was made in 2002 by the \dasi\
experiment~\citep{kovac02}, leading the way to subsequent measurements with
ever-increasing sensitivity.  Precision measurements of $E$-modes have
been made by \QUAD~\citep{pryke09}, \bicep1~\citep{barkats14},
\wmap~\citep{wmap9yr}, \quiet~\citep{quiet12} and others.
Secondary $B$-modes produced by gravitational lensing~\citep{zaldarriaga98}
have recently been detected by the South Pole Telescope~\citep{sptlensing13}
and \polarbear~\citep{polarbear2013a,polarbear2013b,polarbear2014}.
The inflationary $B$-mode signal has been more difficult to detect
because of its small amplitude.
The excellent sensitivity and exquisite control of instrumental systematics
achieved by \bicep2 have allowed it to make the first detection of $B$-mode
power on degree angular scales.  This analysis is reported in a companion
paper, the \emph{Results Paper}~\citep{b2results14}.
In this paper, we will present the design and performance of \bicep2
and the properties of its three-year data set that have enabled this
exciting first detection.

The organization of the current paper is as follows.
In \S\ref{sec:expapproach} we give an overview of the
 experimental approach used by \bicep2 and the other experiments in
 the \bicep/\keck\ series.
The following sections present the design and construction of the experiment:
    the observing site and telescope mount~(\S\ref{sec:obs_mount});
    the telescope optics~(\S\ref{sec:optics});
    the telescope support tube, with radio frequency and magnetic shielding~(\S\ref{sec:telescope});
    the focal plane unit~(\S\ref{sec:FPU});
    the transition-edge sensor bolometers~(\S\ref{sec:det});
    the cryogenic and thermal design~(\S\ref{sec:cryodesign});
    and the data acquisition and control system~(\S\ref{sec:electronics}).
The detectors will also be described in a dedicated \emph{Detector Paper}~\citep[in preparation]{b2dets14}.

The performance of the detectors is described in \S\ref{sec:det_performance}, which reports 
 the achieved noise level and other parameters that set the ultimate sensitivity of the experiment.
In \S\ref{sec:performance} we describe the characterization of instrumental properties that are relevant to systematics.
 We have developed analysis techniques to mitigate many of these effects and we use detailed
 simulations to show that the remaining systematics are at a sufficiently low level for the
 experiment to remain sensitivity limited.  Full details of these techniques and simulations are presented in
 two companion papers: a \emph{Systematics Paper}~\citep[in preparation]{b2syst14} covering the analysis methods and overall results,
 and a \emph{Beams Paper}~\citep[in preparation]{b2beams14} describing the beam measurement campaign and application of the methods
 to beam systematics.
The observing strategy of \bicep2 is presented in~\S\ref{sec:observing}.
The low-level data reduction, and data quality cuts are described in~\S\ref{sec:pipeline}.
In Section~\ref{sec:data} we describe the three-year data set taken in the years
2010--12, reporting the final map depth and projected $B$-mode sensitivity.  

\section{Experimental approach}
\label{sec:expapproach}
Searching for inflationary $B$-modes requires excellent sensitivity
to detect a small signal and excellent control of systematics to avoid
contamination of that small signal by instrumental effects.
\bicep2 is one of a family of experiments,
the \bicep /\keck\ series, which share a similar experimental approach
to meeting these challenges.
We observe from the South Pole, where atmospheric loading is consistently
very low, and use cryogenically cooled optics for very low internal
loading.  Our sub-kelvin bolometer detectors are
photon-noise limited, while the low optical power keeps the photon noise low.
In combination these properties give excellent sensitivity.  
To minimize systematics we use small, on-axis refracting
telescopes that have low instrumental polarization and can be
extensively characterized in the optical far field.  Rotation
about the telescope boresight cancels many classes of systematic
effects and allows us to form jackknife maps that verify the reliability of
the data.  Our CMB observations are made within a field in the
``Southern Hole'', where Galactic foregrounds are expected to be very low.

The pathfinder for this strategy was \bicep1~\citep{keating03}, which observed
from 2006--2008 with neutron transmutation doped (NTD) germanium
thermistor bolometers at $100$, $150$, and $220~\mathrm{GHz}$.
Its full three-year data set yields the best direct limits to
date on inflationary $B$-modes: $r<0.65$ at 95\%
confidence level~\citep{chiang10, kaufman2013, barkats14}.
\bicep2 leverages the successful design and observing strategy
of \bicep1~\citep{takahashi10},
including many common calibration and analysis techniques that were proven
for \bicep1 to yield noise-limited sensitivity and systematic
contamination at a level below $r=0.1$.

\bicep2 has maintained the simplicity and systematics control of
\bicep1 while continuing to gain in sensitivity.  This was accomplished by
increasing the number of photon-noise-limited, polarization-sensitive bolometers
from 98 to 500 detectors (49 to 250 pairs)---each with lower detector noise and higher optical
efficiency.  We adopted a new detector 
technology: antenna-coupled transition-edge sensor (TES) arrays fabricated at
the Jet Propulsion Laboratory (JPL)~\citep{kuo08}.
These arrays have several key advantages facilitating
high channel counts.  First, the discrete feed horns, filters, absorbers, and
NTD detectors used in \bicep1 were replaced with photolithographically
fabricated planar devices that share a single, monolithic silicon wafer with
the detectors themselves.  This architecture yielded densely packed detector
arrays that can be fabricated rapidly and with high uniformity.  Second, the
detector readout used multiplexing SQUID amplifiers to
reduce the number of wires and therefore the heat load on the focal
plane.  We have continued to apply the \bicep1 methods to achieve low
systematics and remaining noise-limited, and in addition we have developed
new analysis techniques to purify the data of instrumental signals.
These methods and the application to the \bicep2 beams and instrument
will be described in the Systematics Paper and Beams Paper.

As the first experiment to deploy the Caltech-JPL antenna-coupled
TES detectors, \bicep2 has opened a path to larger arrays that will continue to
increase in sensitivity and cover multiple frequencies for possible
foreground removal.  The ongoing development of the arrays is described in
the Detector Paper.
The \keck\ \citep[2010--present;][]{sheehy10,ogburn12} has built on this
design by placing five \bicep2-style receivers in a single mount.
All \keck\ receivers through the 2013 observing season have observed at 150~GHz, the
same frequency used by \bicep2.  Beginning with the 2014 season, the \keck\
also includes receivers sensitive to a second frequency, 100~GHz.
\bicep3 (to deploy in 2014) will extend the basic design to a larger 100~GHz focal
plane, with five times the detector count in a single telescope,
and will cover a field with four times the area of the \bicep2 field.
\spider~\citep[scheduled for Antarctic flight in 2014;][]{fraisse13}
is a balloon-borne telescope which also uses the same focal plane technology
as the \bicep/\keck\ family of experiments, with adaptations to
take advantage of the very low photon background available from a suborbital flight,
and with receivers at 100 and 150~GHz.
The excellent achieved sensitivity of \bicep2 demonstrates the successful
implementation of the enabling technology that is now being scaled up to higher
detector counts by successor experiments.

\section{Observing site and telescope mount}
\label{sec:obs_mount}
   \begin{figure*}
   \begin{center}
   \begin{tabular}{c}
   \def\svgwidth{15cm}
   \input{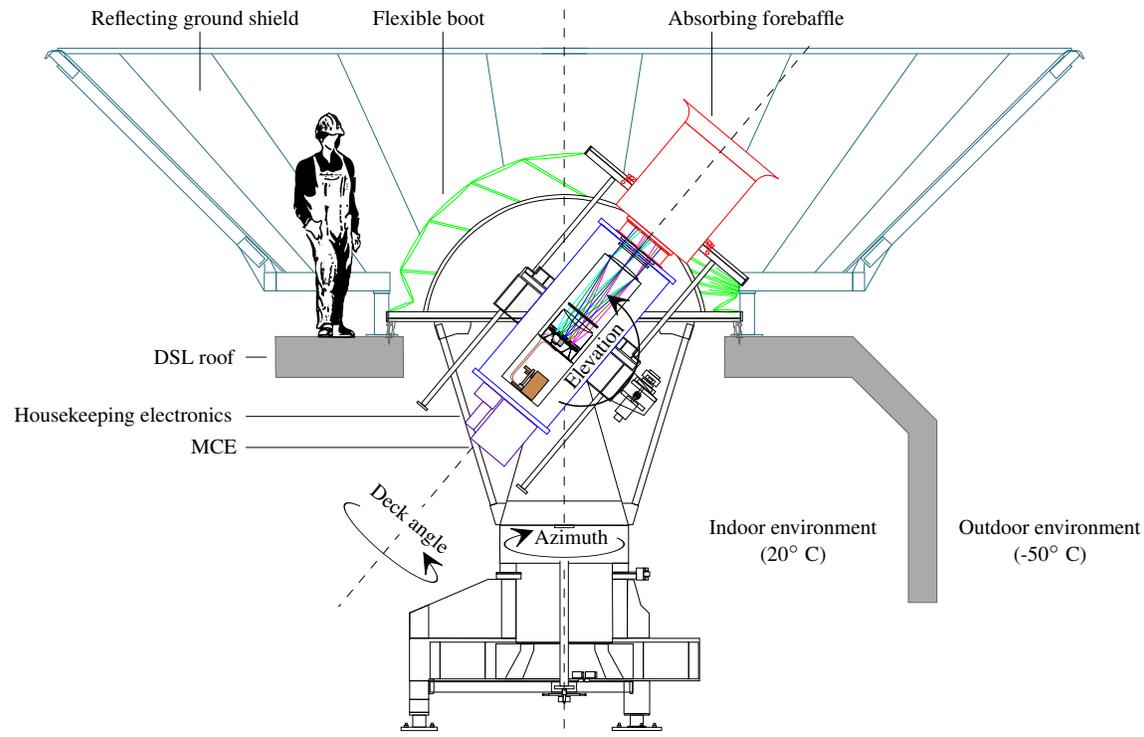}
   \end{tabular}
   \end{center}
   \caption[example]
   { \label{fig:mount}
     The \bicep2 telescope in the mount, looking out through the roof of the Dark Sector Laboratory (DSL)
     located 800~m from the geographic South Pole.  The three-axis mount allows for motion in azimuth,
     elevation, and boresight rotation (also called ``deck rotation'').
     An absorbing forebaffle and reflective ground screen prevent
     sidelobes from coupling to nearby objects on the ground.  A flexible environmental seal or ``boot'' maintains
     a room temperature environment around the cryostat and mount.  The telescope forms an insert within the liquid
     helium cryostat.  The focal plane with polarization-sensitive TES bolometers is cooled to 270~mK by a
     $^4$He/$^3$He/$^3$He sorption refrigerator.  The housekeeping electronics~(\S\ref{sec:housekeeping})
     and Multi-Channel Electronics (MCE,~\S\ref{sec:mce}) attach to the lower bulkhead of the cryostat.
   }
   \end{figure*}

\subsection{Observing site}

The South Pole is an excellent site for millimeter-wave observation from the
ground, with a record of successful polarimetry experiments including
\dasi, \bicep1, \QUAD\, and the South Pole Telescope.  Situated on the Antarctic Plateau, it has
exceptionally low precipitable water vapor~\citep{chamberlin1997}, reducing atmospheric noise
due to the absorption and emission of water near the $150~\mathrm{GHz}$
observing band.  The South Pole site also has very stable weather,
especially during the dark winter months, so that the majority of the
data are taken under clear-sky conditions of very low atmospheric
$1/f$ noise and low loading~\citep{stark02}.  The consistently low atmospheric loading is crucially important
because the sensitivity of the experiment is limited by photon noise, so that
low atmospheric emission is a key to high CMB mapping speed.

Finally, the Amundsen-Scott South Pole Station has hosted scientific
research continuously since 1958.  The station offers well-developed
facilities with year-round staff and an established transportation infrastructure.
\bicep1 and \bicep2 were housed in the Dark Sector Laboratory (DSL), which was built to
support radio and millimeter-wave observatories in an area $1~\mathrm{km}$
from the main station buildings and isolated from possible sources of
electromagnetic interference.

\subsection{Telescope mount and drive}
   \begin{figure}
   \begin{center}
   \begin{tabular}{c}
   \includegraphics[width=7.7cm]{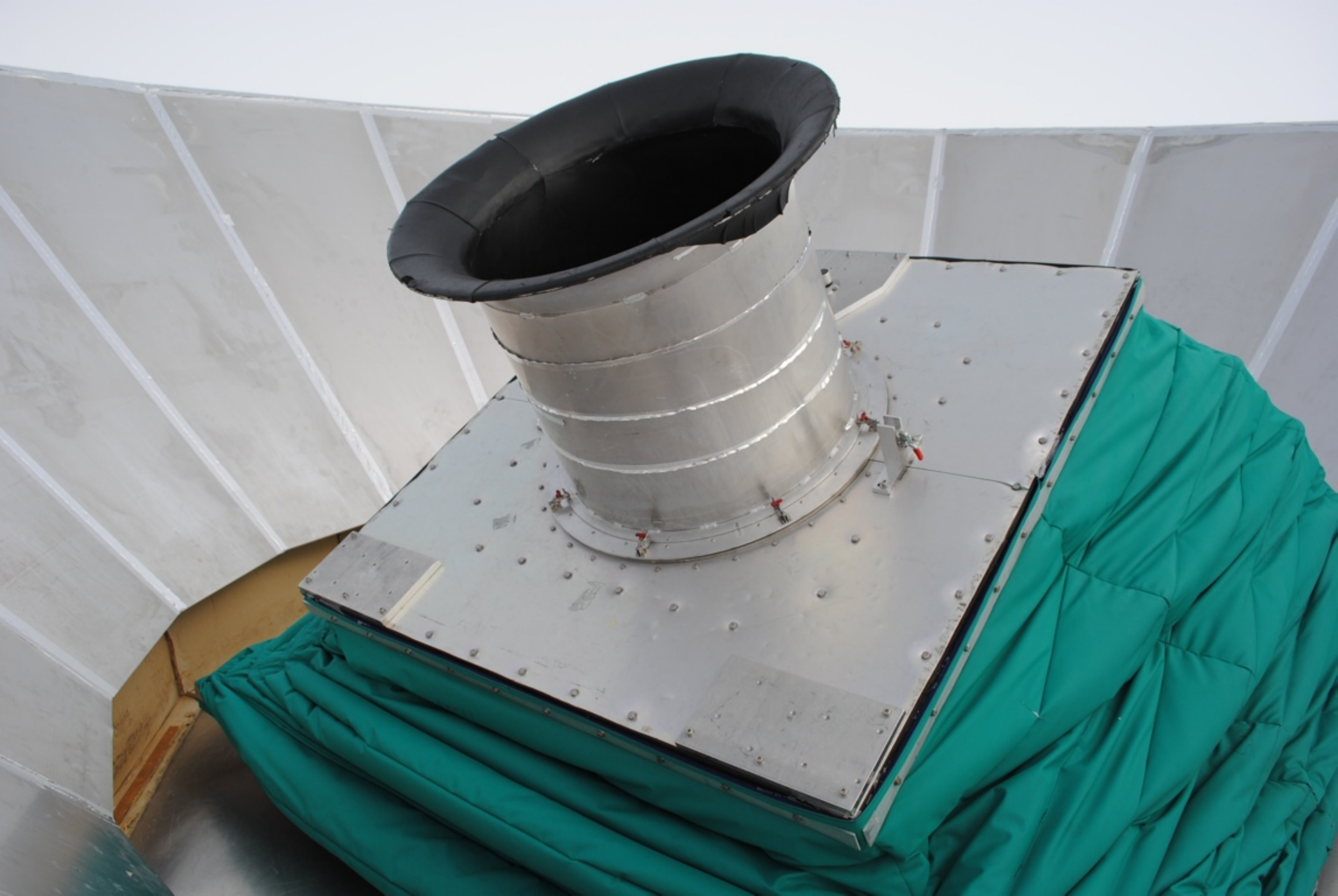}
   \end{tabular}
   \end{center}
   \caption[example]
   { \label{fig:ufbboot}
     \bicep2 absorbing forebaffle, flexible environmental seal (the ``boot''), and ground shield.
     The telescope and mount sat below the boot inside the Dark Sector Laboratory.
}
   \end{figure}

The telescope sat in a three-axis mount~(Fig.~\ref{fig:mount}) supported on a steel and wood
platform attached to the structural beams of the DSL building. The mount
was originally built for \bicep1 by Vertex-RSI\footnote{Now General Dynamics Satcom
Technologies, Newton, NC 28658,
\mbox{\url{http://www.gdsatcom.com/vertexrsi.php}}} along with a second, identical
mount that has remained in North America for pre-deployment testing.
The mount attached to a flexible environmental
shield or ``boot'' (Fig.~\ref{fig:ufbboot})
attached to the roof of the building, so that the cryostat, electronics, and drive
hardware were kept inside a climate-controlled, room temperature environment.

The mount moved in azimuth and elevation (which closely approximate
right ascension and declination when observing from the South Pole).  Its third
axis was a rotation about the boresight of the telescope, also known as
the ``deck angle''.
When installed in DSL its range of motion was
$50^\circ$--$90^\circ$ in elevation and $400^\circ$ in azimuth.  It was capable of
scanning at speeds of up to $5^\circ/\mathrm{s}$ in azimuth.  The major modification for \bicep2 was the
replacement of a slip ring with a cylindrical drum through which the readout
and control cables were fed.  This accommodates the much larger bundle of cables
needed for the \bicep2 housekeeping system~(\S\ref{sec:housekeeping}) while retaining a range of rotation
of 380$^\circ$ in boresight angle.  Our selection of
boresight angles for observing therefore remained unrestricted.

\section{Optics}
\label{sec:optics}
   \begin{figure}
   \begin{center}
   \begin{tabular}{c}
   \def\svgwidth{7cm}
   \input{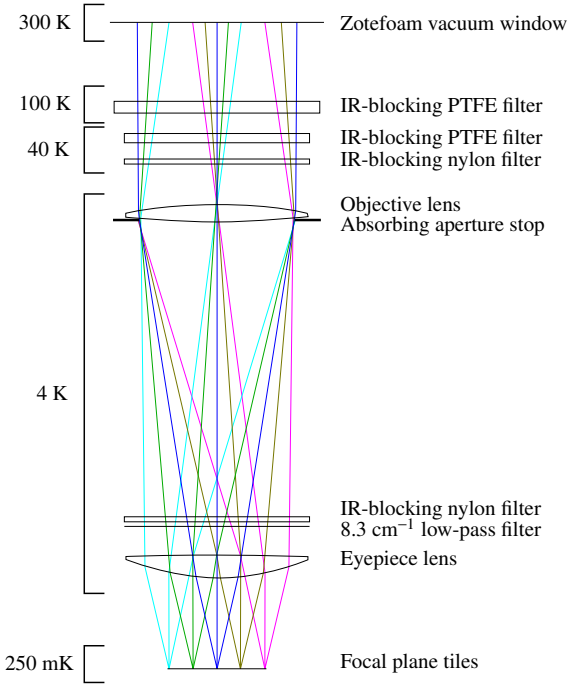}
   \end{tabular}
   \end{center}
   \caption[example]
   { \label{fig:optics}
The telescope optical system.  All components (except the
window) were anti-reflection coated to provide minimal reflection at
150~GHz.  All optics below the 40~K nylon filter were cooled to 4~K,
providing low and stable optical loading.  Due in large part to the
radially symmetric design, simulations predict well-matched beams for two
idealized orthogonally polarized detectors at the focal plane.
}
   \end{figure}

The \bicep2 telescope (Fig.~\ref{fig:optics}) was an on-axis refractor
similar to \bicep1~\citep{takahashi10}, with an aperture of $26.4~\mathrm{cm}$
and beams of width given by the Gaussian radius $\sigma\approx12'$.  The
relatively simple optical design~(Fig.~\ref{fig:optics}) and small aperture allowed \bicep2 to target
the predicted degree-scale peak of the inflationary $B$-mode signal while avoiding
reflective components that add expense and complexity and can have significant instrumental
polarization.  The telescope was efficient to
assemble and transport.   This design also allowed all optics to be cooled to 4~K for low optical
loading, and the beams to be measured in the far field ($>50~\mathrm{m}$)
using controlled optical sources on the ground.  The low loading and the ability to
extensively characterize the beams have been important for achieving high sensitivity
and control of beam systematics, respectively.

\subsection{Lenses and optical simulation}
\label{sec:lenses}
The telescope was designed to produce very well-matched beams for two
orthogonal linear polarizations coincident on the sky.  The two lenses were
made of high density polyethylene and were roughly $30~\mathrm{cm}$ in diameter.
The lens shapes and placement, along with other components
of the optical design, were guided by simulation of the beam properties
using the Zemax optical design software\footnote{ZEMAX Development Corporation,
Redmond, WA 98053, \mbox{\url{http://www.zemax.com/}}}.
We chose to place the first Airy null at the aperture stop for low internal
loading.  This  approximately satisfies the $2f\lambda$ criterion of \citet{griffin02}
for a wavelength $\lambda=2~\mathrm{mm}$.  The other constraints on the optimization
process were to minimize aberration and maintain telecentricity.  The resulting
$f/2.2$ configuration has an effective focal length of $587~\mathrm{mm}$ and a lens
separation of $550~\mathrm{mm}$.  Further details of the simulation and
optimization may be found in \citet{aikin10} and \citet{aikinthesis}.

Simulation of the selected design predicts a nearly ideal Gaussian
beam with width $\sigma=12.4'$ (FWHM=$29.1'$)
and cross-polar response below $5\times10^{-6}$.  The simulated beams for the two detectors
in each pair are the same to below $10^{-3}$ in ellipticity, $2\times10^{-3}$
in beam width, and $6\times10^{-3}$ in pointing (as a fraction of beam width).
These ideal parameters can be compared to the performance of the instrument
as built.   The polarization response was measured in far-field and near-field
calibration tests (\S\ref{sec:pol_response}), which found no
intrinsic cross-polar response detectable above the level of known instrumental
crosstalk~($\sim 0.5\%$).  The achieved beams have also been extensively measured
in the far field~(\S\ref{sec:beams}), allowing our analysis to fully account
for any departures from the ideal beams predicted by  the optics simulation.

\subsection{Vacuum window}

The vacuum window was 32~cm in diameter and 12~cm thick, made of four layers of
Propozote PPA30 foam\footnote{Zotefoams Inc., Walton, KY 41094,
\mbox{\url{http://zotefoams.com/}}} joined into a single piece by heat
lamination.  The PPA30 material is a closed-cell, nitrogen-filled polypropylene
foam with low scattering and high microwave transmission~\citep{tophat, runyan03}.
The window was sealed to its aluminum housing with Stycast 1266 epoxy.

\subsection{Optical loading reduction}
\label{sec:filters}

\begin{table}[t]
\caption{Modeled detector loading from elements in the optical path}
\label{tab:oploading}
\begin{center}
\begin{tabular}{lrrrr} 
\hline \hline
\rule[-1ex]{0pt}{3.5ex}  Element & $T_e$ [K] & Emissivity & Loading [pW] & $T_\mathrm{RJ}$ [K]\\
\hline
\rule[-1ex]{0pt}{3.5ex}  CMB & 3 & 1.00 & 0.12 & \\
\rule[-1ex]{0pt}{3.5ex}  Atmosphere & 230 & 0.03 & 2.0 & \\
\rule[-1ex]{0pt}{3.5ex}  Upper Forebaffle & 230 & 1.00 & 0.65 & \\
\rule[-1ex]{0pt}{3.5ex}  Window & 230 & 0.02 & 1.0 & \\
\rule[-1ex]{0pt}{3.5ex}  IR Blocker 1 & 100 & 0.02 & 0.45 & \\
\rule[-1ex]{0pt}{3.5ex}  IR Blocker 2 & 40 & 0.02 & 0.18 & \\
\rule[-1ex]{0pt}{3.5ex}  IR Blocker 3 & 40 & 0.02 & 0.18 & \\
\rule[-1ex]{0pt}{3.5ex}  IR Blocker 4 & 6 & 0.02 & 0.01 & \\
\rule[-1ex]{0pt}{3.5ex}  Lenses & 6 & 0.10 & 0.07 & \\
\rule[-1ex]{0pt}{3.5ex}  Total & & & 4.7 & 22\\
\hline
\end{tabular}
\end{center}
\end{table}

Optical loading contributes to photon noise, which sets the ultimate
sensitivity of the experiment.  We have therefore taken care to minimize
internal loading by ensuring that all microwave power reaching the detectors
comes only from the sky or cold surfaces.  This was accomplished by
intercepting stray radiation at a cold aperture stop and blackening
reflective surfaces.  The aperture stop, which defines the beam waist, was
an annular ring of $1.9~\mathrm{cm}$ thick Eccosorb
AN-74\footnote{Emerson \& Cuming Microwave Products,
Randolph, MA 02368. \mbox{\url{http://eccosorb.com/}}}
with inner diameter $26.4~\mathrm{cm}$.  It was placed on the lower surface
of the objective lens at $4~\mathrm{K}$ as shown in Fig.~\ref{fig:optics}.
Given the optical design parameters described above, we calculate that the 
aperture stop absorbed 20\% of total optical throughput.  The sides
of the tube supporting the optics and the magnetic shield~(\S\ref{sec:magshield})
were blackened using carbon-loaded Stycast 2850~FT epoxy
applied to a surface of roughened Eccosorb HR10.  This black surface
has very low reflectivity, and is especially effective in minimizing
specular reflection.
This textured black surface cycles cryogenically with minimal 
particulate shedding, and has very low reflectivity even at low angles 
of incidence.

Following an approach developed in \bicep1, we placed two
polytetrafluoroethylene (PTFE) filters in front of the objective lens to 
reduce thermal loading by absorbing infrared radiation.
These were heat sunk to $100~\mathrm{K}$ and $40~\mathrm{K}$.
We placed a $3~\mathrm{mm}$ thick nylon filter in front
of the objective lens, heat sunk to $40~\mathrm{K}$.
In addition, we placed a $5.2~\mathrm{mm}$ thick nylon
filter in front of the eyepiece lens, heat sunk to
$4~\mathrm{K}$.  We finally added a metal mesh low-pass edge
filter~\citep{ade06} with a cutoff at $8.3~\mathrm{cm}^{-1}$
($255~\mathrm{GHz}$) to reflect any coupling to submillimeter
radiation not absorbed in the plastic filters.
This filter was placed directly below the
nylon filter and was also cooled to 4~K.

We have modeled the expected loading for each optical component
and the atmosphere as shown in Table~\ref{tab:oploading}.
In the table, the emission temperature $T_e$ and estimated
emissivity are given for each optical element.  These are combined
with measured optical efficiencies for \bicep2~(\S\ref{sec:opteff})
The total loading is also expressed in units of Rayleigh-Jeans temperature $T_\mathrm{RJ}$.
Although the absorptive upper forebaffle had an emissivity of 1,
the aperture stop and blackening of the optics tube limited
sidelobes sufficiently that the forebaffle
only intercepted 1\% of the beam and contributed an acceptably low
loading power.  The 0.65~pW forebaffle loading in Tab.~\ref{tab:oploading}
is a measured value from tests with and without the forebaffle installed,
as described in Section~\ref{sec:sidelobes}.
The loading from internal components have been calculated in the model
with a total internal loading of 1.89~pW.  This is consistent with
laboratory test measurements~(\S\ref{sec:detprops}) that give an upper
limit of 2.2~pW.

\subsection{Antireflection coating}

Both lenses and the IR blocking filters have been coated
with an antireflection (AR) layer of porous PTFE (Mupor\footnote{Porex Corporation,
Fairburn, GA 30213, \mbox{\url{http://www.porex.com/}}}) optimized for
$150~\mathrm{GHz}$.
The PTFE thickness and density were chosen to minimize reflection
given the index of refraction of each optical element.
The AR layers were heat-bonded using a thin low-density polyethylene (LDPE)
film as a bonding layer.  In order to ensure uniform adhesion, the AR layer and LDPE film
were pressed against the surface
by enclosing each optical element in a vacuum bag during heat-bonding.

The metal mesh low-pass edge filter was separately coated with an antireflection
layer during its fabrication at Cardiff University.

\subsection{Membrane}

In front of the window was a 0.5~mil ($12.7~\mathrm{\mu m}$) transparent
membrane held tautly in place by two aluminum rings.  The membrane protected
the window from snow and created an enclosed space below, which was slightly
pressurized with dry nitrogen gas to prevent condensation on the Propozote foam.
Room-temperature air flowed through holes in the ring
onto the top of the membrane so that any outside snowfall sublimated away.

The initially deployed membrane was $0.5~\mathrm{mil}$ thick biaxially oriented
polyethylene terephthalate (Mylar), which is expected to have reflectivity of
only 0.2\% at 150 GHz.  During maintenance at the end of 2010 this was replaced
with a sheet of the same material and thickness, but held very taut within
the aluminum rings.  Vibrations of the new membrane caused intermittent
common-mode noise, strongly correlated across detectors.
We have verified that this noise does not significantly contaminate the pair-differenced
polarization maps, but as a precaution we remove the most affected data using
a cut on noise correlation~(\S\ref{sec:cuts}).
The membrane was replaced again on 2011 April 27
with less taut, $0.9~\mathrm{mil}$ thick biaxially oriented polypropylene~(BOPP),
while the pressure of the nitrogen gas purge was adjusted to minimize vibration.
After these changes the membrane noise signal was not seen in the remainder
of the 2011--12 data set.

\section{Telescope insert}
\label{sec:telescope}
\begin{figure}
   \begin{center}
   \begin{tabular}{c}
   \def\svgwidth{9cm}
   \input{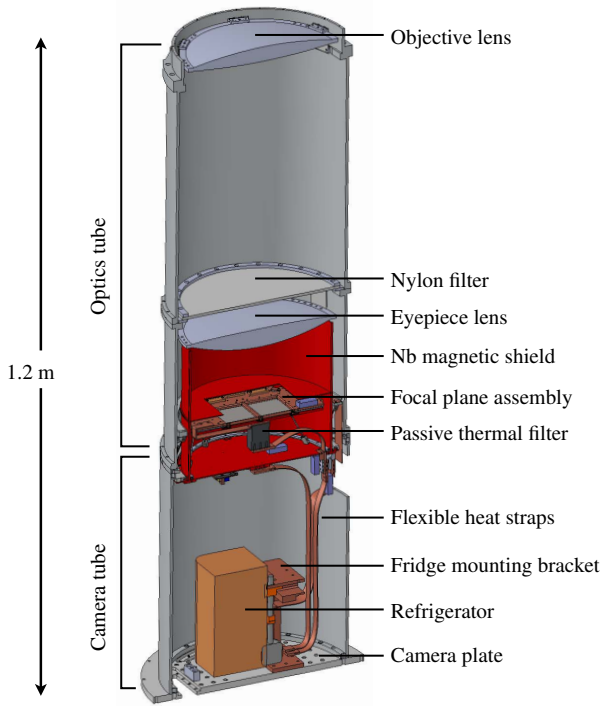}
    \end{tabular}
   \end{center}
   \caption[Telescope tube]
   { \label{fig:teletube}
     Cross-sectional view of the telescope insert.  The entire telescope insert assembly is cooled to $4~\mathrm{K}$ by a thermal link to a liquid helium bath.  The optics tube provides rigid structural support for the optical chain, including the lenses, filters, and aperture stop.  The camera tube assembly houses the sub-kelvin sorption refrigerator and the cryogenic readout electronics in a radiatively and thermally protected enclosure.  The sub-kelvin focal plane assembly sits within a superconducting Nb magnetic shield.  The focal plane is thermally connected to the fridge via a passive thermal filter.}
   \end{figure}

The entire telescope at $4~\mathrm{K}$ and colder formed a removable
insert that was installed into the cryostat (Fig.~\ref{fig:teletube}).  The upper
part of this insert was the optics tube, which contained the cold
lenses and the infrared-blocking filters. The bottom
section of the insert, called the camera tube, held the detector array, cold electronics,
and $^3$He/$^3$He/$^4$He sorption refrigerator. The
bottom plate of the insert was directly connected to the helium
bath. This plate provided sufficient cooling power at $4~\mathrm{K}$ to
cool the optics inside the telescope tube and to allow the refrigerator
to condense liquid $^4$He.

   \begin{figure*}
   \begin{center}
   \begin{tabular}{c}
   \def\svgwidth{13cm}
   \input{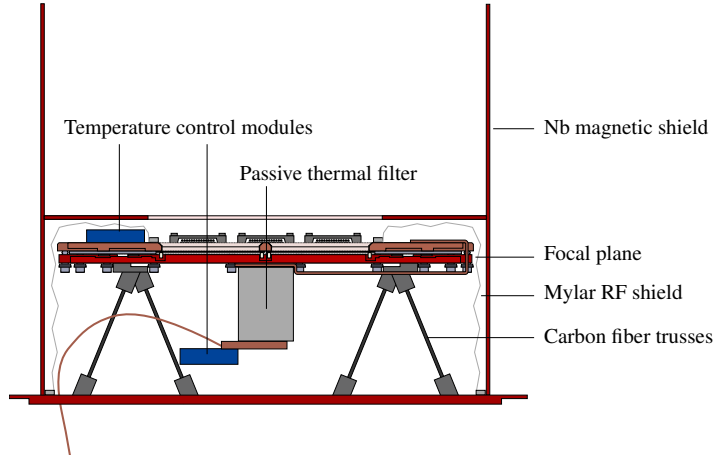}
    \end{tabular}
   \end{center}
   \caption[Focal plane assembled]
   { \label{fig:fpspittoon}
     Cross-sectional view of the sub-kelvin hardware.  The superconducting Nb magnetic shield
     is heat-sunk to $350~\mathrm{mK}$.  Within, the focal plane is isolated from thermal
     fluctuations by eight carbon fiber legs. A thin aluminized Mylar shroud
     extends from the top of the focal plane assembly to the bottom of the Nb magnetic shield
     to minimize radio frequency pickup.  Temperature stability is maintained through the
     combined use of active and passive filtering.  The passive thermal filter, on the
     bottom of the focal plane, serves to roll off thermal fluctuations at frequencies
     relevant to science observations, while active temperature control modules maintain
     sub-millikelvin stability over typical observation cycles.
   }
   \end{figure*}

\subsection{Carbon fiber truss structure}
\label{sec:truss}
\begin{figure*}
   \begin{center}
   \begin{tabular}{c}
   \includegraphics[width=7.7cm]{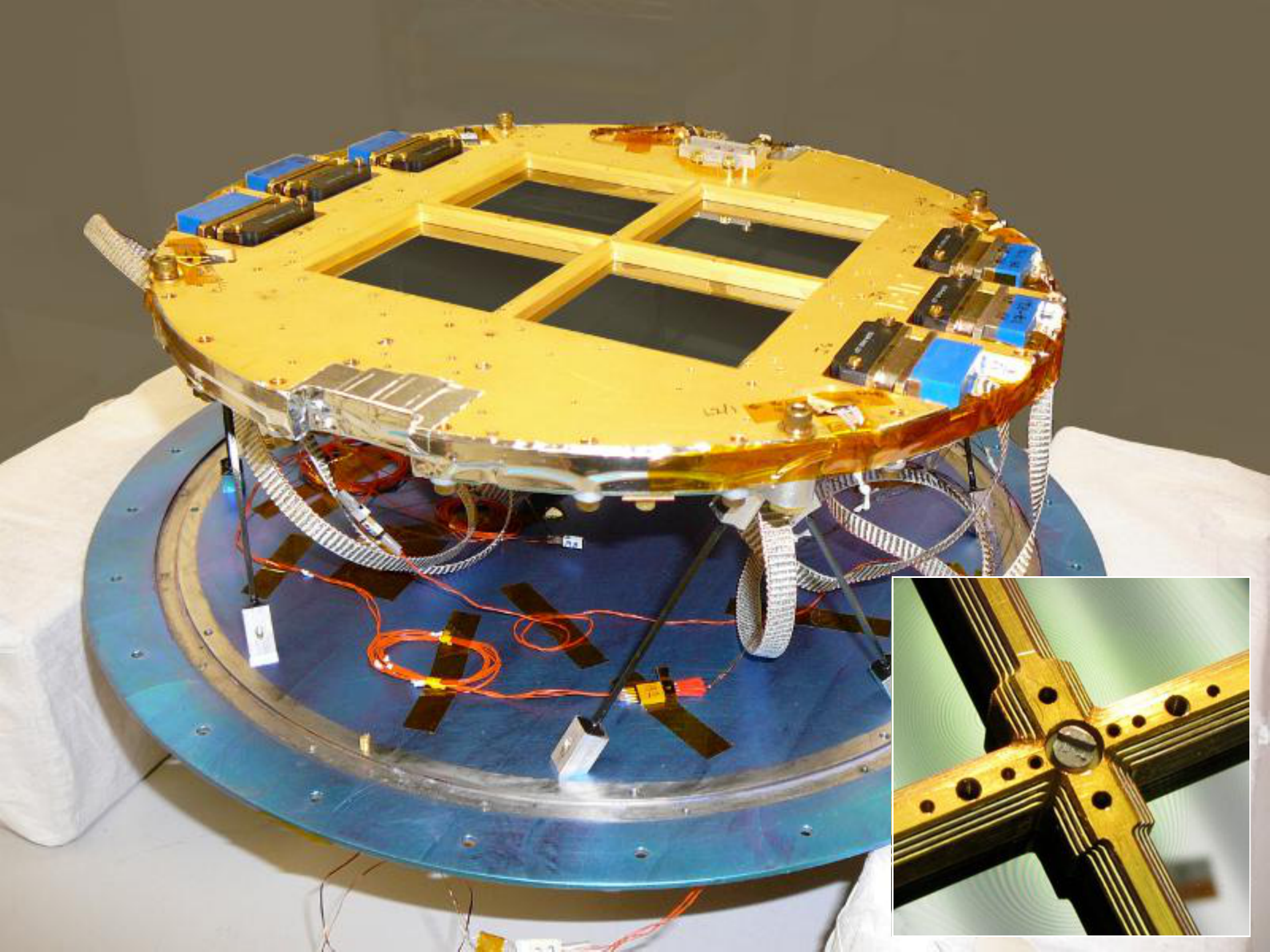}
   ~~
   \includegraphics[width=8.62cm]{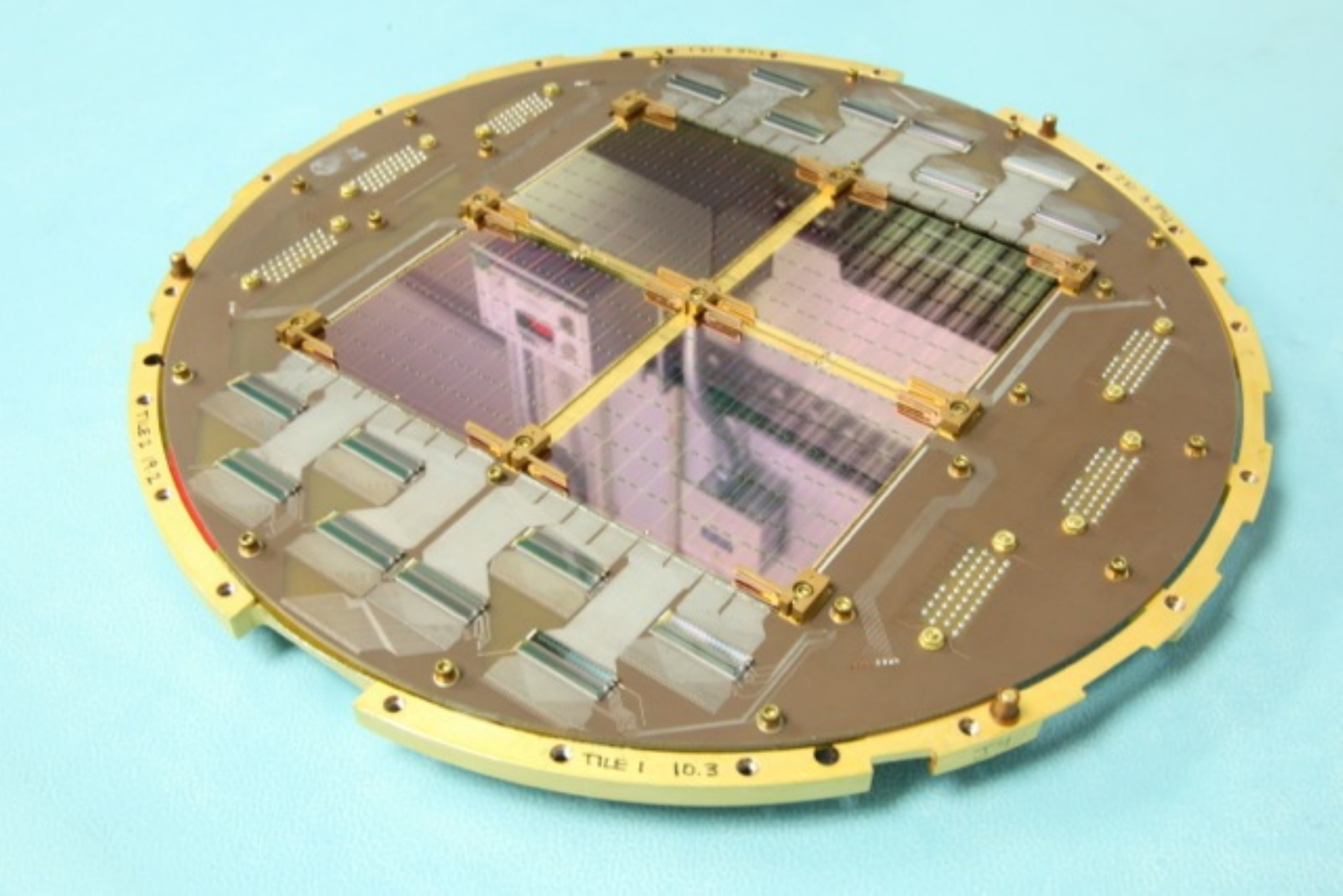}
    \end{tabular}
   \end{center}
   \caption[example]
   { \label{fig:fpu}
     The assembled focal plane on the carbon fiber truss structure and 350~mK Nb plate.
     The four anti-reflection tiles and detector tiles sit beneath square windows in the copper plate.
     This assembly will be covered in the aluminized Mylar radio frequency shield, with a square
     opening only above the detector tiles.
     \emph{Left:} Unshielded assembly; \emph{Left inset}: Corrugations in the edges of the copper
     plate next to the detector tiles;
     \emph{Right:} The underside of the focal plane Cu plate, with detector tiles and
     SQUID and Nyquist chips mounted.
   }
   \end{figure*}

The focal plane sat near the break between the camera tube
and the optics tube.  It required a compact, rigid support
structure with low thermal conductance to the walls of the
aluminum tube at 4~K.  This support was provided by
sets of concentric carbon-fiber truss structures
connecting the thermal
stages at $4~\mathrm{K}$, $2~\mathrm{K}$, $350~\mathrm{mK}$, and
$250~\mathrm{mK}$.  The trusses between the $350~\mathrm{mK}$ plate
and the $250~\mathrm{mK}$ focal plane are shown
schematically in Fig.~\ref{fig:fpspittoon}
and can also be seen in the left-hand panel of Fig.~\ref{fig:fpu}.
The carbon fiber has excellent mechanical properties and has
a very low ratio of thermal conductivity to strength at
temperatures below a few kelvin~\citep{runyan08}.

\subsection{RF shielding}
The detectors and cold SQUID readout electronics were enclosed in a
radio frequency (RF) shield depicted in Fig.~\ref{fig:fpspittoon}.
The RF shield began on the top of the focal plane,
just above the detector arrays. A square clamp held an
aluminized Mylar shroud (Fig.~\ref{fig:fpspittoon}) that extended from around the detectors
down to a circular clamp to the 350 mK niobium (Nb) plate.
A second Mylar sheet was used
to create a conductive path that surrounds the stages at different
temperatures without thermally linking them.  This sheet went
up from the $350~\mathrm{mK}$ ring to
a $2~\mathrm{K}$ ring, and then down to the $4~\mathrm{K}$ ring.
This ring connected to the aluminum walls of the optics and camera
tubes and the $4~\mathrm{K}$ base plate of the camera tube.
Filter connectors at the base plate protected the cold electronics from
RF interference picked up in wiring outside the cryostat.

\subsection{Magnetic shielding}
\label{sec:magshield}
The SQUIDs, TESs, and other superconducting components are sensitive to ambient
magnetic fields, including those of the Earth and of nearby electrical
equipment such as the telescope drive motors.
We attenuated the field in the vicinity of all sensitive elements by surrounding
them with passive magnetic shielding.
The final shielding configuration was chosen after simulation
using COMSOL Multiphysics software\footnote{COMSOL, Inc.,
Burlington, MA 01803, \mbox{\url{http://www.comsol.com/}}} and
experimentation with various options for each susceptible component.
This process led to the selection of superconducting and high-permeability
shielding materials according to their measured effectiveness in each
location.

   \begin{figure*}
   \begin{center}
   \begin{tabular}{c}
   \def\svgwidth{13cm}
   \input{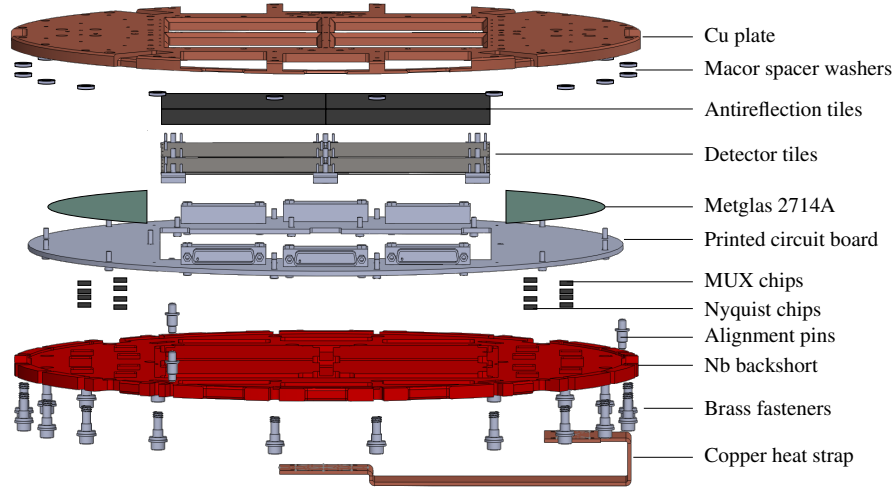}

    \end{tabular}
   \end{center}
   \caption[Focal plane layers]
   { \label{fig:fpexploded}
     Exploded view of the layers of the focal plane.  The Cu plate forms the
     substrate on which everything else is assembled.  The detector
     tiles are pressed against antireflection tiles and look out through
     four square cutouts in the Cu plate with corrugated edges.
     The TES detectors and antennas are on the bottom surface of the tile,
     so that radiation passes through the Si wafer before reaching the
     slot antennas.  A layer of Metglas magnetic shielding sits between the
     Cu plate and the printed circuit board (PCB).  The PCB layer routes
     electrical traces between the detector tiles, multiplexing (MUX) chips, and micro-D
     connectors, and acts as a base for wire-bonding the tiles.
     The MUX chips sit on alumina carriers that mate to the PCB.
     The Nb backshort is held at a distance of one quarter wavelength from the tiles by
     Macor spacers.  It is attached last to sandwich the circuit board,
     MUX chips, and tiles.
   }
   \end{figure*}

The focal plane assembly was surrounded to the greatest extent possible by a superconducting
shield shown in Fig.~\ref{fig:fpspittoon}.  This shield was composed of the
Nb plate at the 350~mK stage beneath the focal plane,
a Nb plate immediately in front of the focal plane, and
a cylindrical Nb shield that extends from the 350~mK plate upward. The Nb backshort immediately
behind the detector tiles provided additional shielding.

A cylinder of $1~\mathrm{mm}$~thick Cryoperm 10 alloy\footnote{Amuneal
Manufacturing Corp., Philadelphia, PA 19124, \mbox{\url{http://amuneal.com/}}}
was wrapped around the entire optics tube and held
at $4~\mathrm{K}$.  This high-permeability shield drew field lines into itself
so that they would not be trapped in the superconducting Nb shield around the
focal plane.

We placed sheets of Metglas 2714A\footnote{Metglas, Inc., Conway, SC 29526,
\mbox{\url{http://www.metglas.com/products/magnetic\_materials/}}}
behind the printed circuit board that housed the first and second-stage
SQUIDs~(Fig.~\ref{fig:fpexploded}).  In laboratory
comparisons this was found to give greater attenuation of applied fields than
Nb foil in this location.

Early tests showed that the instrument's magnetic sensitivity was dominated by
the SQUID series arrays (SSAs), which were located outside the focal plane
assembly, on the side of the refrigerator (Fig.~\ref{fig:teletube}).
The SQUID arrays were already enclosed in superconducting Nb shielding within
the SSA modules, and this shielding was greatly improved by wrapping several layers of Metglas 2714A
around the SSA modules. After this improvement the level of magnetic sensitivity from the SSAs
was much lower than that at other stages.

We characterized the remaining level of magnetic sensitivity
in laboratory tests by placing a Helmholtz coil in three orientations around
the cryostat, and \emph{in situ} by performing ordinary CMB observing
schedules with the TES detectors deliberately inactive.
We found that the shielding achieved an overall suppression factor of $\sim 10^6$,
leaving a residual signal from the Earth's magnetic field.
This had a median size corresponding to $\sim500~\mu\mathrm{K}_\mathrm{CMB}$,
or up to $5000~\mu\mathrm{K}_\mathrm{CMB}$ in the most sensitive channels.
The sensitivity was dominated by the first-stage
SQUIDs, which were especially sensitive in the MUX07a generation of hardware~\citep{stiehl11}.
The remaining signal has a simple sinusoidal form in azimuth and is ground-fixed,
so that it can be removed very effectively in analysis by 
low-order polynomial subtraction~(\S\ref{sec:polysub}) and ground-fixed signal
subtraction~(\S\ref{sec:groundsub}).

\section{Focal Plane}
\label{sec:FPU}

The focal plane unit (FPU) was constructed from several layers of different materials selected
to provide the stable temperature, mechanical alignment, and magnetic shielding required to
operate the camera.
The detector tiles must be held firmly in place while allowing for
differential thermal contraction and providing sufficient thermal conduction to the refrigerator.
The temperature of the focal plane must be kept very stable.  Sensitive components must be
further shielded from stray magnetic fields.  Finally, the optical backshort must be precisely
aligned at a quarter wavelength behind the detector tiles.  We have achieved these
goals using the focal plane components described below.

\subsection{Copper plate}

The focal plane was assembled around a gold-plated, oxygen-free high thermal conductivity copper (OFHC Cu) detector plate.
The detector tiles and most other focal plane components were mounted to its lower face.
The Cu plate with detector tiles and multiplexing components mounted can be seen in the
right-hand panel of Fig.~\ref{fig:fpu}, and an exploded view of all layers in the
assembly is shown in Fig.~\ref{fig:fpexploded}.
In the plate were four square windows that allowed radiation to reach the detectors.
To suppress electromagnetic coupling between the detector plate and the
antennas of pixels near the tile edges, we cut quarter-wavelength-deep
corrugations~(Fig.~\ref{fig:fpu} left, inset) into the edges of the windows~\citep{orlando10}.

\subsection{Niobium backshort}

A superconducting niobium (Nb) plate sat below the Cu at
a separation of $\lambda/4$ and served as an optical backshort.  It was held at the
correct distance by precision-ground Macor\footnote{Corning
Incorporated, Corning, NY 14831, \mbox{\url{http://www.corning.com/specialtymaterials/macor/}}}
washers, whose thermal contraction is negligible when cooled to millikelvin
temperatures.  The Nb backshort was supported at its perimeter
by a carbon-fiber truss and cooled at its center through a Cu foil strap (\S\ref{sec:thermal}).
This contact point ensured that the Nb backshort transitions
into a superconducting state from the center outwards so that it would
not trap flux as is possible with type-II superconductors.

\subsection{Printed circuit board}

An FR-4 printed circuit board (PCB) carried superconducting Al electrical
traces and served as a base for wire-bonding
the tiles and the SQUID chips.  Between the Cu plate and the PCB we
placed sheets of Metglas 2714A to create a low-field environment around the
SQUID chips.  The planar geometry between the Cu and Nb plates was especially effective in
lowering the normal field component to which the SQUID chips are most
sensitive.  The SQUIDs sat on alumina carriers on the PCB, giving sufficient
separation from the Metglas sheet to prevent magnetic coupling that could
cause increased readout noise.

\subsection{Assembly}
Each detector tile was stacked with a high-conductivity $z$-cut crystal
quartz anti-reflection (AR) wafer.  We attached the detector tiles and
AR wafers to the Cu plate in a way that provided precision alignment,
allowed for differential thermal contraction, and ensured sufficient
heat-sinking.  First precision-drilled holes and slots were made in the
detector tiles and AR wafers.  These registered to pins that were
press-fit in the Cu adjacent to each window.  The detector tile and
AR wafer stacks were clamped to the plate with machined tile clamps that
allow slipping under thermal contraction.
The weak clamping force was insufficient to effectively heat-sink the
tiles, so we further connected a gold ``picture frame'' around the tile edges
with gold wire bonds that made direct contact with the gold-plated Cu 
frame.  The thermal conductivity (limited by the Kapitza resistance
between the silicon substrate and the gold) was large enough to prevent
tile heating under thermal loading.

Additional wire bonds were used to electrically connect
mounted components to traces in the PCB.
The detector tiles had Nb pads on their back edges to be 
connected to the PCB traces with superconducting Al wire bonds.
The SQUID chips~(\S\ref{sec:muxing}) were similarly wire-bonded to the PCB,
as were NTD thermometers and heaters mounted directly on the detector tiles.

\section{Detectors}
\label{sec:det}

The focal plane was populated with integrated arrays of 
antenna-coupled bolometers.  This technology combines beam-defining planar
slot antennas, inline frequency-selective filters, and TES
detectors into a single monolithic package.  The JPL Microdevices Laboratory 
produces these devices in the form of square silicon tiles, each containing
an $8\times8$ array of dual-polarization spatial pixels (64~detector pairs or
128 individual bolometers). The \bicep2 focal plane had four of these tiles,
for a total of 500 optically coupled detectors and 12 dark (no antenna)
TES detectors.  The detector tiles were characterized at Caltech and JPL
during 2008--2009.  The rapid fabrication cycle of the Caltech-JPL
detectors made it possible to incorporate results of pre-deployment testing
into the final set of four tiles deployed in \bicep2.  Further details of
the detector design and fabrication will be presented in the Detector Paper,
which will report on improvements to the detector tiles
leading up to \bicep2 as well as further developments in subsequent generations
informed by \bicep2 testing.

   \begin{figure}
   \begin{center}
   \begin{tabular}{c}
   \includegraphics[width=7.7cm]{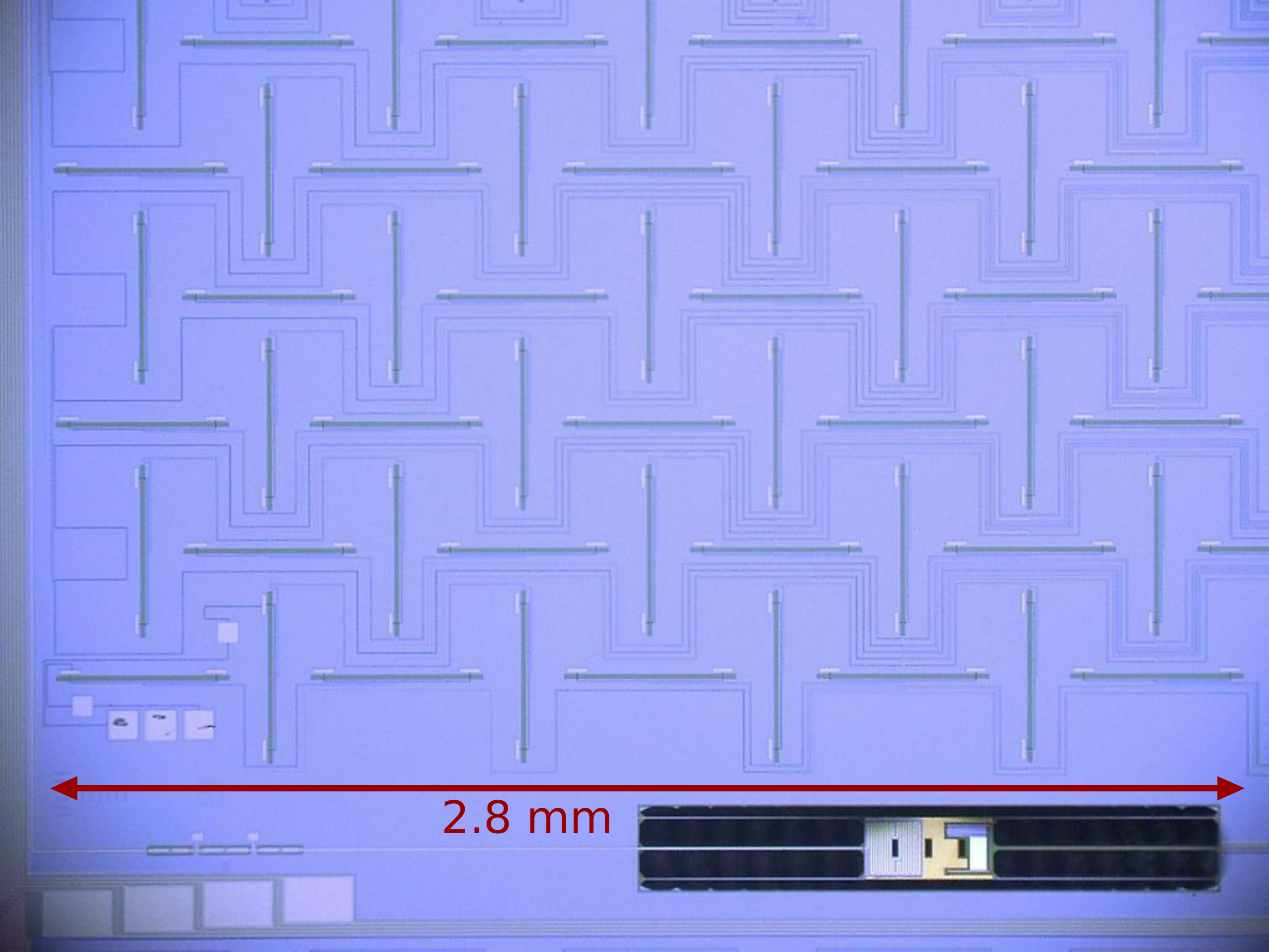}
   \end{tabular}
   \end{center}
   \caption[example]
   { \label{fig:antenna}
       Partial view of one \bicep2 dual-polarization pixel, showing the band-defining filter (lower left), TES island (lower right),
       and part of the antenna network and summing tree.  The vertically oriented slots are sensitive to horizontal polarization
       and form the antenna network for the A detector, while the horizontally oriented slots receive vertical polarization and
       are fed into the B detector.  In this way the A and B detectors have orthogonal polarizations but are spatially co-located
       and form beams that are coincident on the sky.  This view corresponds to a boresight angle of 90$^\circ$. At boresight
       angle of 0$^\circ$ the A detectors receive vertical polarization and the B detectors receive horizontal polarization.
  }
   \end{figure}

   \begin{figure}
   \begin{center}
   \begin{tabular}{c}
   \def\svgwidth{7.7cm}
   \input{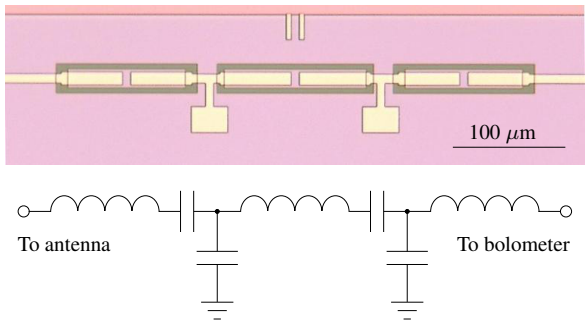}
   \end{tabular}
   \end{center}
   \caption[example]
   { \label{fig:bandfilt}
       150~GHz band-defining filter and equivalent circuit.  Each
       filter consisted of three inductors in series, coupled
       to each other through a T-network of capacitors.
   }
   \end{figure}

   \begin{figure}
   \begin{center}
   \begin{tabular}{c}
   \includegraphics[width=7.7cm]{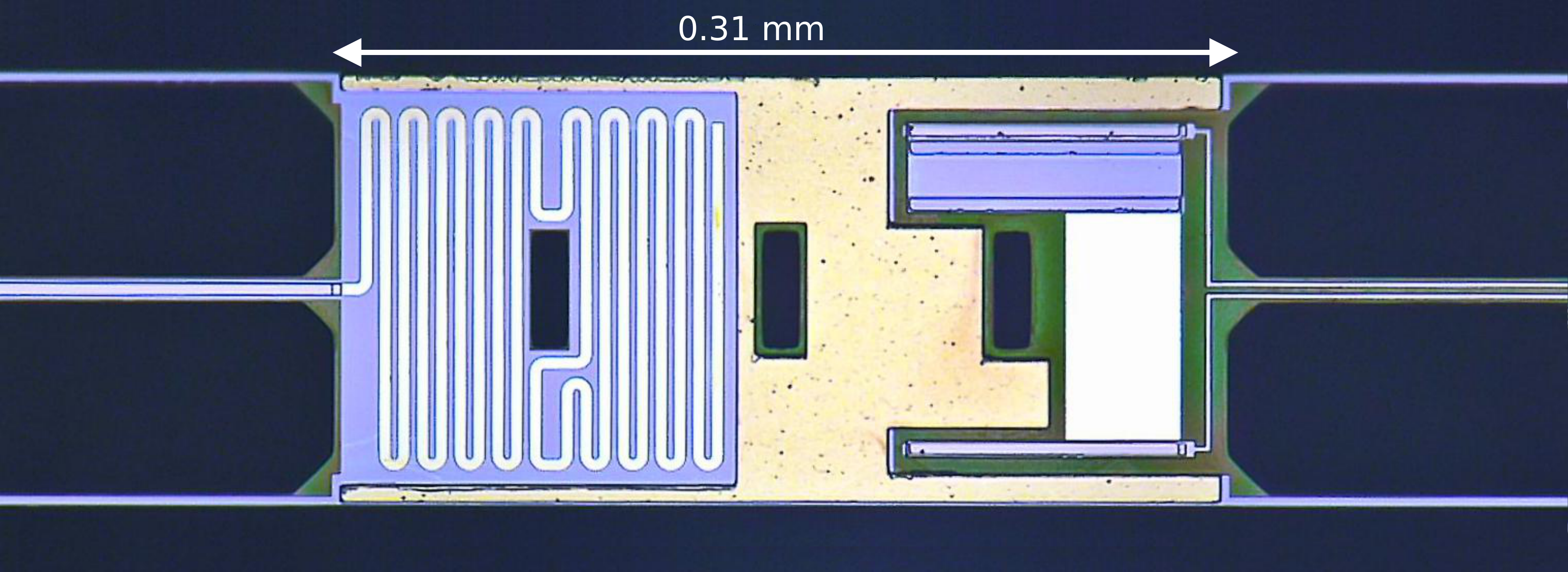}
   \end{tabular}
   \end{center}
   \caption[example]
   { \label{fig:island}
       TES island for a single \bicep2 detector. The island was supported
       by six lithographically etched legs.  Microwave power, entering from
       the left, terminated into a resistive meander.
       The deposited heat is measured as a decrease in electrical
       power (or current) dissipated in the titanium TES, which appears as
       a blue rectangle on the right of the island.
       The TES voltage bias was provided by two micro-strip lines
       at right.  To increase the dynamic range of the device, an aluminum
       TES (seen as a white rectangle below the titanium film) was deposited
       in series with the titanium TES, providing linear response across a
       wide range of background loading conditions.  The heat capacity of
       the island was tuned by adding $2.5~\mu\mathrm{m}$ thick
       evaporated gold, which is distributed across the remaining real estate of the island.
       This made the detector time constants~(\S\ref{sec:timeconst}) slow 
       enough for stable operation.
   }
   \end{figure}

\subsection{Antenna networks}
Optical power coupled to each detector through an integrated planar
phased-array antenna.  The sub-radiators of the array were slot dipoles etched
into a superconducting Nb ground plane.  The two linear polarization modes
were received through two orthogonal, but co-located, sets of 288 slots (for a
total of 576 slot-dipoles per dual-polarization spatial pixel).
Since the tiles were mounted in the focal plane with the detector side down, the antennas
received power through the silicon substrate.  A Nb backshort reflected the
back-lobe in the vacuum half-space behind the focal plane. The design of the 
slot antennas has gone through many iterations. The final design used in \bicep2, 
called the ``H" antenna after the arrangement of horizontal and vertical
slots~(Fig.~\ref{fig:antenna}), has exceptionally low cross-polar responses over $>30\%$
fractional bandwidth~\citep{kuo08}. 

Currents induced around the slots coupled to planar microstrip lines integrated
onto the backside of the antenna arrays.  The waves from the sub-radiators summed
coherently in a corporate feed network that accomplished the beam synthesis
traditionally handled by a feed horn.  Two
interleaved feed networks independently summed the two polarizations before
terminating at two different detectors.  Each pixel's antennas were
$7.8~\mathrm{mm}$ on a side, matching the $f/2.2$ optics such that the antenna
sidelobes terminated on the aperture stop or blackened surfaces inside the telescope tube.

\subsection{Band-defining filters}
Each microstrip feed-network contained an integrated filter (Fig.~\ref{fig:bandfilt}) to
define a frequency band centered at $150~\mathrm{GHz}$ and with 25\% fractional bandwidth
(defined at the 3~dB points).
The 3-pole filter contained lumped inductors made from short lengths
of coplanar waveguide.  Each of the three inductors coupled to its neighbor
through a T-network of capacitors.  The achieved bands are characterized in~\S\ref{sec:spectra}.

The band-defining filter was omitted in twelve detectors of the array to create dark TESs
with no connection to the antennas.  These
were used to characterize sensitivity  to signals such as temperature fluctuations
and RF interference.

\subsection{TES bolometers}
After passing through the band-defining filter, microwave power was carried to
a strip of lossy gold microstrip line on a released bolometer island~(Fig.~\ref{fig:island}).
The power thermalized in the gold resistor, heating the low-stress silicon nitride
(LSN) island.  The island was held by narrow LSN legs that formed a thermal weak link
to the rest of the focal plane with thermal conductance $G_c \approx 100~\mathrm{pW/K}$.
The leg conductivity was tuned~\citep{orlando10,obrient12} to optimize the
noise and saturation power, as described in \S\ref{sec:gtune}.

Each LSN island contained two TES detectors that changed in current in response to
changes in the temperature of the island.  A primary, titanium TES was designed to
operate under low loading conditions when observing the sky, with transition temperature
($T_c$) of $500$--$524~\mathrm{mK}$.  A second, aluminum TES was placed in series
with the primary TES.  The Al TES had a higher $T_c \approx 1.3~\mathrm{K}$ and higher saturation
power for use in the laboratory or when observing mast-mounted sources.
The sensitivity of the experiment depends crucially on the performance of the detectors.
Their optimization and characterization are reported in detail in Section~\ref{sec:det_performance}.

\subsection{Direct island coupling and mitigation}
\label{sec:islandcoupling}
In pre-deployment tests an earlier generation of detectors showed an unexpected,
small coupling to frequencies just above the intended band.
The out-of-band power detected was typically 3-4\% of the
total response and had a wide angular response.
We interpreted this response as power coupling directly to
the bolometer island.  This was reduced in the deployed
\bicep2 detectors through the addition of the metal mesh low-pass
edge filter to the optics stack~(\S\ref{sec:filters}) and several design changes
described in more detail in the Detector Paper.  We changed
the leg design to reduce the width of the opening in
the ground plane around the island and metalized the four outer
support legs with Nb to reduce the RF impedance to the island ground plane.
The dark island coupling was reduced to~0.3\% of the antenna
response in the experiment as deployed.

\subsection{Device yield}
Initial electrical testing of detector arrays checked for continuity across the
devices, with correct room-temperature resistance and no shorts.
This fabrication yield was extremely high, 99\% for the four
tiles in \bicep2.
When the detectors were integrated into the focal plane and
telescope there were additional losses from open lines in the readout,
further reducing the overall yield to 82\%.  The remaining 412 ``good light detectors''
are those that were optically coupled and had stable bias and
working SQUID readout.  A detector has been included in this count only if both
it and its polarization partner satisfy the same criteria.  The number
is reduced somewhat in analysis by data quality cuts on beam
shape and noise properties as described in~\S\ref{sec:cuts}.

\section{Cryogenic and thermal architecture}
\label{sec:cryodesign}

\subsection{Cryostat}
\label{sec:cryostat}
The telescope was housed within a 
Redstone Aerospace\footnote{Redstone Aerospace, Longmont, CO 80501, \mbox{\url{http://www.redstoneaerospace.com/}}}
liquid-helium cooled cryostat that was very 
similar to the \bicep1 dewar.  The major change was that the liquid nitrogen
stage of \bicep1 was replaced with two nested vapor-cooled shields, so
that liquid helium was the only consumable cryogen.  The helium reservoir had a capacity
of 100~L and consumed about 22~L/day during ordinary observing.

\subsection{Refrigerator}
\label{sec:fridge}

The detectors were operated at 270~mK in order to achieve photon-noise-limited
sensitivity.  Our focal plane and surrounding intermediate temperature
components were cooled using a closed-cycle, three-stage ($^4$He/$^3$He/$^3$He)
sorption refrigerator~\citep{duband99}.  The intermediate $^3$He stage
provided a 350~mK temperature used to heat-sink the niobium magnetic shield~(\S\ref{sec:magshield}),
while the final $^3$He stage provided a 250~mK base temperature.
The initial condensation of the $^4$He stage was performed by closing a heat switch
to thermally couple the fridge to the cryostat's liquid helium reservoir.
The condensed liquid was then pumped by a charcoal sorption
pump to pre-cool the next stage.

The refrigerator had an enthalpy of 15~J at the intermediate 350~mK stage,
and 1.5~J at the 250~mK stage.  The carbon-fiber truss structures~(\S\ref{sec:truss}),
along with other aspects of the thermal design, yielded very low
parasitic thermal loads.  The refrigerator was able to provide a stable
base temperature for more than 72 hours.  After the liquid reservoirs
were exhausted, they were replenished from the charcoal by performing
a five-hour regeneration cycle.  In order to allow for a margin of safety
and align with the \bicep2 observing pattern, we recycled the refrigerator
once within each observing schedule of three sidereal days, as described in
Section~\ref{sec:schedules}.

\subsection{Thermal architecture and temperature control}
\label{sec:thermal}

Several improvements were made in the thermal path between the
refrigerator and the focal plane relative to \bicep1, giving \bicep2 improved
stability and reduced vibrational pickup.  The coldest stage of the
refrigerator was linked to the focal plane through a thermal strap and a passive
thermal filter.  The thermal strap was designed as a flexible stack of
many layers of high-conductivity Cu foil, which reduces
the vibrational sensitivity relative to the stiffer linkages used in \bicep1.
The passive thermal filter was a rectangular
stainless steel block, $5.5~\mathrm{cm}$ in length and with a
$2.5~\mathrm{cm}\times 2.5~\mathrm{cm}$ cross-section.
The design approach for the passive filter was inspired by the distributed
thermal filter used in the \planck\ HFI instrument~\citep{planck03spie,planckfilter}.
The filter had high heat capacity and low thermal diffusivity in order to
achieve adequate thermal conduction with a sufficiently long time constant.
Stainless steel (316 alloy) was chosen as a readily available
material with suitable thermal and magnetic properties, though other materials, such as
holmium, have lower thermal diffusivity.
The filter effectively isolated the focal plane from thermal fluctuations on
time scales shorter than about $1300~\mathrm{s}$.

With no additional heating, the focal
plane achieved a base temperature of $\sim$250~mK.
Temperature control modules (TCMs) consisting of two NTD thermometers and one 
resistive heater were employed in a feedback
loop to control the temperature of the focal plane and the fridge side of the
thermal filter (as shown in Fig.~\ref{fig:fpspittoon}) to 280
and 272~mK respectively, well below the $500~\mathrm{mK}$
titanium TES transition temperature.
Temperature stability of the tile substrates was monitored using
NTD thermometers mounted on each detector tile and by dark 
TESs on the detector tiles.  The tile NTD data have been used to
demonstrate that the achieved thermal stability met the requirements
of the experiment~(\S\ref{sec:thermalstab}).

Temperatures were also monitored at critical points using Cernox resistive
sensors\footnote{Lake Shore Cryotronics, Inc., Westerville, OH 43082,
\mbox{\url{http://www.lakeshore.com/}}} and/or diode
thermometers. 

\subsection{Housekeeping}
\label{sec:housekeeping}
The AC signals from the NTD thermometers~\citep{rieke89} 
were read out using junction gate field-effect transistors that are housed at the $4~\mathrm{K}$ stage
(although self-heated to $\sim140~\mathrm{K}$) to
reduce readout noise~\citep{spirejfet}.
The NTD thermometers were read out differentially with respect to
fixed-value resistors, also cold, and each biased separately.
Resistor heaters provided control of the sorption fridge, a heat
source for temperature control of the cold stage, and
instrument diagnostics.

The warm housekeeping electronics were composed of two parts: a small
``backpack'' that attached directly to the vacuum shell of the cryostat~(Fig.~\ref{fig:mount})
and a rack-mounted ``BLAST bus'' adapted from the University of Toronto
BLAST system~\citep{wiebethesis}. The backpack contained
preamplifiers for readout channels and the digital-analog converters (DACs) for
temperature control and NTD bias generation, all completely enclosed within a
Faraday-cage conducting box. The BLAST bus contained the analog-digital converters (ADCs)
themselves, as well as digital components for the generation of the NTD
bias signals and in-phase readout of the NTDs. This split scheme was
designed to isolate the thermometry signals as much as possible from
pickup of ambient noise while keeping the backpack small enough to fit
within the limited space behind the scanning telescope.

The housekeeping system was upgraded after the first year of observing in order
to improve the noise performance of the NTD~readout. The upgraded
firmware allowed more effective use of the fixed resistors as a nulling
circuit to maximize the signal while maintaining linearity in response. The
frequency of the NTD bias was also increased from $55~\mathrm{Hz}$ to
$100~\mathrm{Hz}$ to improve noise performance. 

\section{Data acquisition system}
\label{sec:electronics}

\bicep2 used a multiplexed SQUID readout that allowed it to operate
a large number of detectors with low readout noise and acceptably low
heat load from the wiring.  We describe the NIST SQUIDs and other
cold hardware, the room-temperature Multi-Channel-Electronics (MCE)
system, and the custom control software that were used for data acquisition.

\subsection{Multiplexed SQUID readout}

\label{sec:muxing}

\bicep2 used the ``MUX07a'' model of cryogenic SQUID readout electronics provided by
NIST~\citep{dekorte03}.  These were designed for time-domain
multiplexing~\citep{chervenak99,irwin02}, in which groups of 33 channels
are read out in succession through a common
amplifier chain.  This scheme supports large channel counts with a small number
of physical wires so that the heat load on the cold stages remains low.

Each detector had its own first-stage SQUID, and the 33 first-stage
SQUIDs in one multiplexing column were coupled to a single second-stage SQUID
through a summing coil.  The first- and second-stage SQUIDs for one column of
detectors were packaged together into a single multiplexing (MUX) integrated circuit chip.
A second chip, the Nyquist
chip, contains the TES biasing circuitry, including a 3~m$\Omega$ shunt resistor
to supply a voltage bias for the $\sim$60~m$\Omega$ TES and a 1.35~$\mu$H inductor
to limit the detector bandwidth.  Both the MUX and Nyquist chips were bonded to alumina
carriers and mounted to the focal plane PCB layer (Fig.~\ref{fig:fpu} right-hand
panel; Fig.~\ref{fig:fpexploded}).
The PCB was connected to Nb/Ti twisted pair cables
running to the $4~\mathrm{K}$ stage, where SQUID series arrays (SSAs) were used for
impedance matching to room-temperature amplifiers.
This entire chain was operated in a flux-locked-loop mode
by applying a feedback signal to the first-stage SQUIDs.  This feedback ensured
that all SQUIDs operated very near their selected lock points and maintained
constant closed-loop gain.

\begin{table}[t]
\caption{Multiplexing parameters used by \bicep2}
\label{tab:mux}
\begin{center}
\begin{tabular}{lcc} 
\hline \hline
\rule[-1ex]{0pt}{3.5ex}  & 2010 & 2011--12 \\
\hline
\rule[-1ex]{0pt}{3.5ex}  Raw ADC sample rate & 50~MHz  & 50~MHz \\
\rule[-1ex]{0pt}{3.5ex}  Row dwell & 98~samples & 60~samples  \\
\rule[-1ex]{0pt}{3.5ex}  Row switching rate & 510~kHz & 833~kHz \\
\rule[-1ex]{0pt}{3.5ex}  Number of rows & 33 & 33 \\
\rule[-1ex]{0pt}{3.5ex}  Same-row revisit rate & 15.46~kHz & 25.25~kHz \\
\rule[-1ex]{0pt}{3.5ex}  Internal downsample & 150 & 140 \\
\rule[-1ex]{0pt}{3.5ex}  Output data rate per channel & 103~Hz & 180~Hz \\
\rule[-1ex]{0pt}{3.5ex}  Software downsample & 5 & 9 \\
\rule[-1ex]{0pt}{3.5ex}  Archived data rate & 20.6~Hz & 20.0~Hz \\
\hline
\end{tabular}
\end{center}
\end{table}

The SQUIDs and associated hardware are sensitive to ambient magnetic
signals.  This sensitivity was reduced by the gradiometric design of the
first-stage SQUIDs and further attenuated through magnetic shielding
(\S\ref{sec:magshield}), but the MUX07a model was particularly susceptible
to pickup at the first-stage SQUID~\citep{stiehl11}.
The multiplexed readout is also susceptible to several types of inter-channel
crosstalk~(\S\ref{sec:xtalk}), although development of the NIST hardware over several
generations has greatly reduced these effects.

\subsection{Warm multiplexing hardware}
\label{sec:mce}
The warm electronics for detector bias and multiplexed readout were the
Multi-Channel Electronics (MCE) system developed by the University of British
Columbia~\citep{mce08} to work with the NIST cold electronics.  The MCE is a 6U
crate that was attached to a vacuum bulkhead at the bottom of the cryostat as
in Fig.~\ref{fig:mount}.  It interfaced to the cold electronics through three
RF-filtered 100-pin micro-D metal (MDM) connectors and communicates with the control 
computers through two optical fibers (selected for their high data
rates and electrical isolation).  A third optical fiber connected the MCE to an
external synchronizing clock (``sync box''), which provided digital time
stamps used to keep the bolometer time streams precisely
matched to mount pointing and other data streams (see Section~\ref{sec:control}).

\subsection{Multiplexing rate}
\label{sec:muxoptimization}
The multiplexing rate was chosen to read out each detector frequently enough to
avoid noise aliasing while also waiting long enough between row switches to
avoid settling-time transients that could cause crosstalk.

Avoiding noise aliasing requires the readout rate to be sufficiently above the
knee frequency of the $LR$ circuit formed by the Nyquist inductor and the TES resistance.
For our typical device resistance ($R_\mathrm{TES}\approx 50\mathrm{m}\Omega$,
see~\S\ref{sec:detprops}) and $L_\mathrm{Nyq}=1.35~\mathrm{\mu H}$
the cutoff frequency is $R/L\approx 5\text{--}6~\mathrm{kHz}$.
At initial deployment \bicep2 used a row visit rate of $15.5~\mathrm{kHz}$,
which kept the level of crosstalk acceptably low but resulted in a significant
noise contribution from aliased TES excess noise (\S\ref{sec:noise}).

Additional studies of crosstalk and multiplexing rate were performed in late 2010,
resulting in SQUID tuning parameters that allowed a faster
row switching rate of $25~\mathrm{kHz}$ without a significant increase 
in crosstalk~\citep{brevik10}.
The $25~\mathrm{kHz}$ multiplexing parameters~(Table~\ref{tab:mux})
were adopted at the beginning of 2011, with an expected gain of $\sim 20\%$
in sensitivity.  The actual improvement in sensitivity is discussed
in~\S\ref{sec:mapspeed}.

\subsection{Control system}
\label{sec:control}
Overall control and data acquisition were handled by a set of Linux computers
running the Generic Control Program (\gcp), which has been used
by many recent ground-based CMB experiments~\citep{story12}.
The \bicep2 version of \gcp\ was based on the \bicep1 code base, with
changes to integrate with the MCE hardware and software.
It has been further adapted for use in the \keck.

\gcp\ provided control and monitoring of almost all components of the
experiment, including the telescope mount, focal plane temperature,
refrigerators, and detectors.  It provided a scripting language used
to configure observing schedules~(\S\ref{sec:schedules}).

\subsection{Digital filtering}
\label{sec:digifilt}

The TES detectors themselves had a very fast response, with typical
time constants of several milliseconds.  Given the scan pattern the band of
interest for science lay below $2.6~\mathrm{Hz}$~(\S\ref{sec:scanpat}).
In order to conserve bandwidth across the South Pole satellite data
relay we downsampled the data to 20 samples per second before archival.
This required an appropriate antialiasing filter, which was applied in
two stages.  The MCE firmware used a fourth-order digital Butterworth
filter before downsampling to 100 samples per second.  The second stage
was in the \gcp\ mediator, which applied an acausal, zero-phase-delay FIR
filter before writing data to disk.  As these were both digital filters,
their transfer functions are precisely known and do not vary.  The \gcp\
filter was designed using the Parks-McClellan algorithm~\citep{filtpm} with
a pass band at $0.6$ times the Nyquist frequency.  This Nyquist frequency
was set by the desired downsampling factor of $5\times$ (2010 data set)
or $9\times$ (2011-12 data set).  Both filters were modified at the end
of 2010 to accommodate the change from 15~kHz to 25~kHz multiplexing.
A small amount of March 2010 data used a more compact FIR filter with larger
in-band ripple.  This ripple is $<0.5\%$ with the earliest March 2010
settings, $<0.1\%$ with the settings used in the remainder of 2010,
and $<0.01\%$ with the 2011--12 settings.

\section{Detector performance and optimization}
\label{sec:det_performance}

We selected the parameters of the antenna-coupled TES detectors
for \bicep2 for low noise to maximize the instantaneous sensitivity of the experiment,
while also allowing a margin of safety for stable operation under typical loading conditions.
The noise in polarization (\ie~pair-differenced time streams) at low frequency was
dominated by photon noise, which was controlled by minimizing sensitivity
to bright atmospheric lines (\S\ref{sec:spectra}) and by reducing internal
loading~(\S\ref{sec:filters}).  The next largest noise component was
phonon noise from fluctuations in heat flow between the islands and the substrate.
This was kept low by tuning the leg thermal conductance (\S\ref{sec:gtune}).
Finally, we tuned the detector bias voltages to minimize aliased excess noise
(\S\ref{sec:biasoptimization}).

We extensively characterized the performance of the detector tiles as
fabricated, including the optical efficiency (\S\ref{sec:opteff}),
detector properties (\S\ref{sec:detprops}), time constants (\S\ref{sec:timeconst}),
and noise (\S\ref{sec:noise}).  After optimizations during the 2010 season,
the array has achieved an overall noise-equivalent temperature (NET)
of $15.8~\mathrm{\muK}_\mathrm{CMB}\sqrt{\mathrm{s}}$.

\subsection{Frequency response}
\label{sec:spectra}
\begin{figure}
   \begin{center}
   \begin{tabular}{c}
   \def\svgwidth{7.7cm}
   \input{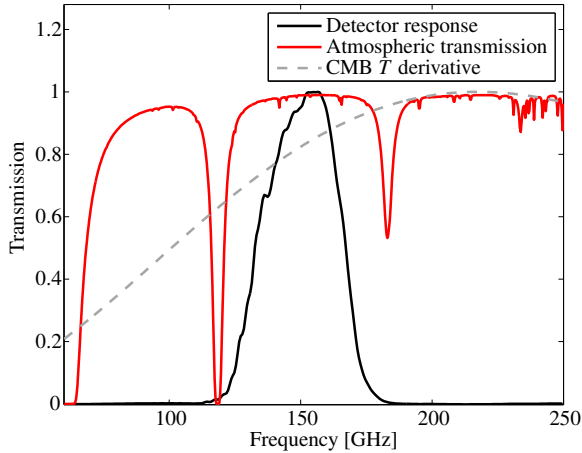}
   \end{tabular}
   \end{center}
   \caption[example]
   { \label{fig:spectra}
  The array-averaged frequency response spectrum (black solid line).  Atmospheric transmission from the South Pole 
  (red solid line) and the CMB anisotropy (gray dashed line) are also shown for comparison.
  The \bicep2 band center is 149.8~GHz and the bandwidth is 42.2~GHz~(28\%).  The detector response and
  CMB spectra are normalized to unit peak, and the atmospheric transmission spectrum is in
  units of fractional power transmitted.
}
   \end{figure}

The optics, antenna network, and lumped-element filters were tuned
for a frequency band at $150~\mathrm{GHz}$ with $\sim25\%$ fractional bandwidth.
The band was chosen to avoid to the spectral lines of oxygen at
$118.8~\mathrm{GHz}$ and water at $183.3~\mathrm{GHz}$
(red curve in Fig.~\ref{fig:spectra})
in order to reduce atmospheric loading, photon noise, and $1/f$ noise
from clouds and other fluctuations in the atmospheric brightness.

The achieved bands were characterized using
Fourier transform spectroscopy (FTS).  Measurements
were performed using a specially built
Martin-Puplett interferometer~\citep{martin82} designed to mount
directly to the cryostat window.  The spectrometer's output polarizing grid 
was attached to a rotation stage, which steered the output beam across the detector
array.
The stage also included a goniometer, a device for measuring the angular
orientation of the stage.
The FTS illuminated approximately a 4$\times$4 grid of
detectors per grid pointing, and multiple pointings were
combined to create the archival data set.  In order to probe 
measurement systematics, spectra were taken at
several boresight rotations and with several FTS
configurations.  The detector time streams were
combined with encoder readings from the mirror
stage to produce interferograms, or traces of
power as a function of mirror position.  The raw
interferograms were low-pass filtered, aligned on the white-light fringe
(zero path length difference) and Hann-windowed before
performing a Fourier transform to give the frequency response $S(\nu)$.
From the $S(\nu)$ for each detector's maximally illuminated data set
we compute its band center, defined as
\begin{equation}
\langle \nu \rangle = \int \nu S(\nu) d\nu,
\end{equation}
and its bandwidth,
defined as 
\begin{equation}
\Delta\nu = \frac{\left( \int S(\nu) d\nu \right)^2}{\int S^2(\nu) d\nu}.
\end{equation}
The \bicep2 array-averaged band center is $149.8
\pm 1.0$ GHz, and the array-averaged bandwidth is $42.2 \pm 0.9$ GHz.
Using this definition of the bandwidth, this corresponds to a fractional spectral 
bandwidth of 28.2\%.  The array-averaged frequency response is in Figure \ref{fig:spectra}.

A mismatch between the bandpasses of the two detectors in a pair can cause a
difference in gain that introduces a leakage of CMB temperature into polarization.
This is not fully corrected by the relative gain calibration~(\S\ref{sec:gaincal}),
which is based on an atmospheric signal with a different frequency spectrum
from the CMB.  We define the spectral gain mismatch for each detector pair
as in~\citet{bierman11}.
The array-averaged spectral mismatch is consistent with zero.  Because the
source is not fully beam-filling, the spectra for each detector vary somewhat
with pointing.  We have characterized this by calculating the spectral match
for several different pointings of the FTS.  We find that the pointing-dependent
systematic error on the spectral gain mismatch corresponds to a scatter of
1.7\%, so that the FTS measurement can only limit the root-mean-square spectral
mismatch per pair to be below this level.

Because a randomly distributed spectral mismatch at the level of 1.7\%
would introduce a significant false polarization, we have carried out
additional analysis to ensure that our polarization maps are not contaminated
by relative gain mismatch.  We apply the deprojection technique described in
the Systematics Paper, and we use simulations to show that leakage from
relative gain mismatch is suppressed to an acceptably small level.

\subsection{Optical efficiency}
\label{sec:opteff}

   \begin{figure*}
   \begin{center}
   \begin{tabular}{c}
   \def\svgwidth{14cm}
   \input{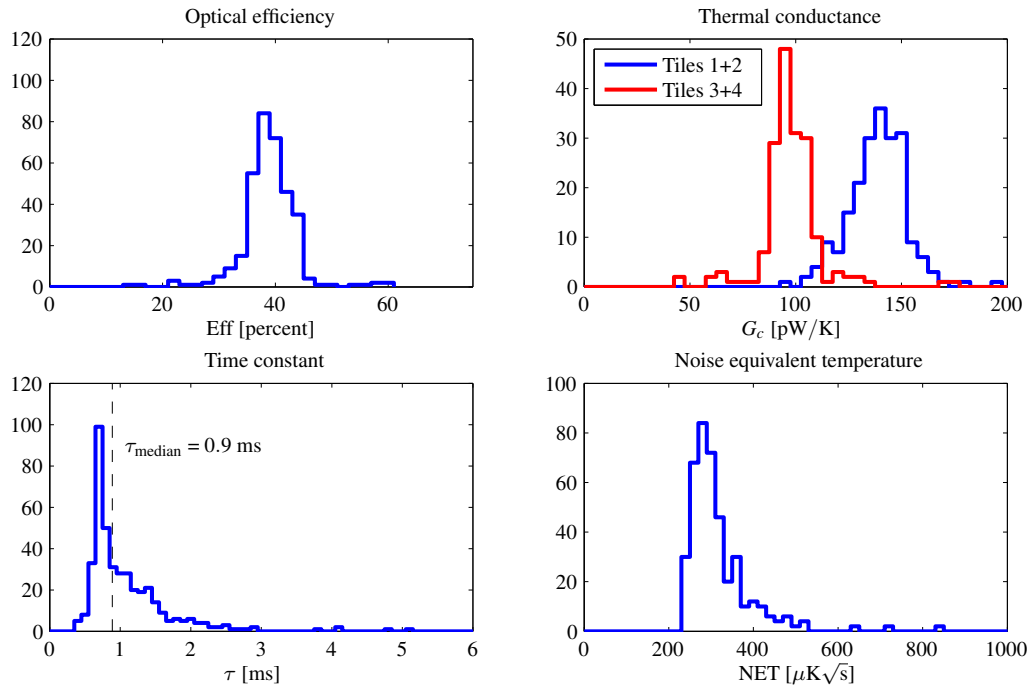}
   \end{tabular}
   \end{center}
   \caption[example]
   { \label{fig:dethists}
       Histograms of measured bolometer properties per detector.
       \emph{Top left:} Optical efficiency~(\S\ref{sec:opteff}).  This measurement was
       taken in lab with a beam-filling source.  It was converted to an efficiency
       number using the measured spectral bandwidth of 42~GHz (\S\ref{sec:spectra}).
       \emph{Top right:} Thermal conductance of the legs~(\S\ref{sec:gtune}).  Two of the tiles
       have $G_c\approx100~\mathrm{pW}/\mathrm{K}$ and two have higher
       $G_c\approx140~\mathrm{pW}/\mathrm{K}$.
       \emph{Bottom left:} Time constants with 2011--12 biases~(\S\ref{sec:timeconst}).
       The vertical dashed line shows the median of the distribution, 0.9~ms.
       These time constants are taken from raw-mode data in which the MCE and
       \gcp\ digital filters have not been applied.
       \emph{Bottom right:} Noise equivalent temperature~(\S\ref{sec:noise}) per detector,
       in units of CMB temperature.  NET is shown for the 2011--12 operating parameters.
  }
   \end{figure*}

The optical efficiency is the fraction of input light that the detectors 
absorb.  It is dependent on the losses within the optics, the antennas, the band 
defining filters and the detectors.  Higher optical 
efficiencies increase the responsivity and the bottom line 
sensitivity numbers, but also increase the optical loading and the photon noise.  
For a beam-filling source with a blackbody spectrum, 
the power deposited on a single-moded polarization-sensitive detector is
\begin{equation}
\label{eq:popt}
P_\mathrm{opt} = \frac{\eta}{2} \int \mathrm{d}\nu \lambda^2 S(\nu) B(\nu,T),
\end{equation}
where $\eta$ is the optical efficiency, $B(\nu)$ is the Planck blackbody spectrum,
 and $S(\nu)$ is the detector response 
in frequency space as defined in \S\ref{sec:spectra}.  Here we choose the
normalization condition
\begin{equation}
\frac{\int S^2(\nu) \mathrm{d}\nu}{\int S(\nu) \mathrm{d}\nu}=1.
\end{equation}
In the Rayleigh-Jeans limit
($h\nu\ll kT$), Eq.~\ref{eq:popt} reduces to
\begin{equation}
P_\mathrm{opt} = kT\eta \int \mathrm{d}\nu S(\nu) = kT\eta \Delta \nu.
\end{equation}

The optical efficiency was measured in the laboratory using a beam-filling,
microwave-absorbing load at both room temperature and liquid nitrogen
temperature. This end-to-end measurement, including losses from all optics
and using bandwidth of~42~GHz,
yielded per-detector optical efficiencies as shown in the upper left
histogram of Fig.~\ref{fig:dethists}, with an array average of 38\%.

\subsection{Thermal conductance tuning}
\label{sec:gtune}
After photon noise, the next largest noise contribution was phonon noise,
corresponding to random heat flow between the island and substrate
through the isolation legs.  The noise-equivalent power (NEP) from this source 
is proportional to the island temperature and the square root of the leg 
thermal conductance $G$ \citep[see \eg~][]{hiltonirwin2005}:

  \begin{equation}
		\label{eqn:phonon}
    \mathrm{NEP}=\sqrt{4k_BGT^2F}.
  \end{equation}
  
Here $F$ is a numerical factor (typically $\sim0.5$ for these devices) 
accounting for the finite temperature gradient across the isolation legs.
Reducing the thermal conductance lowers the phonon noise power and
lengthens the detector time constants.  It also decreases the 
detector's saturation power, the amount of optical loading required to drive 
the detectors out of transition and into the normal state.  If the saturation power 
is too low, it may not be possible to operate the detectors during all weather conditions.
The selection of $G$ is thus a balance between the requirements for
low noise and sufficient saturation power.

For \bicep2 we expect edoptical loading of $4\text{--}6~\mathrm{pW}$ during
representative weather conditions~(\S\ref{sec:filters}).
We chose to make the optical power and Joule power approximately equal.  This
gave a saturation power of about twice the ordinary optical
loading for a safety factor of two, so that the detectors could
operate in almost all weather conditions without saturating.
We thus required a saturation power of $10~\mathrm{pW}$.
For a TES bolometer with thermal conductance $G \propto T^{n}$,
the saturation power is given by
  \begin{equation}
		\label{eqn:satpow}
    P_\mathrm{sat}=G_0 T_0 \frac{(T_c/T_0)^{n+1}-1}{n+1}.
  \end{equation}
With a typical thermal conductance exponent $n=2.5$,
transition temperature $T_c=500~\mathrm{mK}$ and substrate temperature
$T_0=250~\mathrm{mK}$, this gives a thermal conductance
$G_0=14~\mathrm{pW/K}$ at substrate temperature or
$G_c=80~\mathrm{pW/K}$ at $T_c$.
We have used the latter as the fabrication target for \bicep2
detectors.

\subsection{Measured detector properties}
\label{sec:detprops}

\begin{table}[t]
\caption{Measured detector parameters}
\label{tab:detparams}
\begin{center}
\begin{tabular}{lc} 
\hline \hline
\rule[-1ex]{0pt}{3.5ex}  Detector Parameter & Value \\
\hline
\rule[-1ex]{0pt}{3.5ex}  Optical efficiency, $\eta$ & 38\% \\
\rule[-1ex]{0pt}{3.5ex}  Normal resistance, $R_N$ & 60--80 m$\Omega$ \\
\rule[-1ex]{0pt}{3.5ex}  Operating resistance, $R_\mathrm{op}$ & 0.75~$R_N$ \\
\rule[-1ex]{0pt}{3.5ex}  Saturation power, $P_\mathrm{sat}$ & 7--15~pW \\
\rule[-1ex]{0pt}{3.5ex}  Optical loading, $P_\mathrm{opt}$ &4--5.5~pW \\
\rule[-1ex]{0pt}{3.5ex}  Thermal conductance, $G_c$ & 80--150 pW/K\\
\rule[-1ex]{0pt}{3.5ex}  Transition temperature, $T_{c}$ & 505--525~mK\\
\rule[-1ex]{0pt}{3.5ex}  Thermal conductance exponent, $n$ & 2.5\\
\hline
\end{tabular}
\end{center}
\end{table}

The detector properties were measured in the laboratory and on the sky to be close to the 
design values. Table~\ref{tab:detparams} summarizes these properties.  The detectors were 
fabricated at JPL in two separate batches, and the differences between these two batches
account for the majority of the variation in detector properties, particularly
the thermal conductance $G_c$ and the saturation power $P_\mathrm{sat}$.

The thermal conductance can be measured by taking sensor current-voltage characteristics or
``load curves'' in which we sweep the bias voltage and measure the output current.
This was repeated at several focal plane temperatures to give a measurement of $G$
as shown in the upper right panel of Fig.~\ref{fig:dethists}.
We found $G_c$ in the range $80$--$150~\mathrm{pW/K}$, with the detectors on two of the
tiles (Tiles 1 and 2) matching the design characteristic of $80~\mathrm{pW}/\mathrm{k}$
and a higher $G_c$ on the other two tiles (Tiles 3 and 4).
The transition temperature was measured from the same load curve data, with
$T_c=505$--$525~\mathrm{K}$.

Since the saturation power is directly related to the 
thermal conductance~(Eq.~\ref{eqn:satpow}), the fractional variation in $P_\mathrm{sat}$
is similar to that in $G_0$.
With the telescope pointed at the center of the CMB observing field at $55^\circ$ 
elevation, the saturation power for the light detectors was $7$--$15~\mathrm{pW}$. 

The contributions of Joule heating power and optical power to the total can be
determined by calculating the Joule power from known $G_c$ and $T_c$~(Eq.~\ref{eqn:satpow})
or by using the dark detectors, which have no optical power.
Both techniques show the \bicep2 optical loading to be $4$--$5.5~\mathrm{pW}$, or $18$--$25~\mathrm{K}_\mathrm{RJ}$.

The optical loading can further be separated into internal loading and atmospheric 
loading by measuring the saturation power of the detectors with a mirror placed at the aperture.
Because the flat mirror reflects some radiation from the filters, lenses, window, and optics
tube, the loading in the mirror test is an upper limit on the internal loading.  For detectors near
the center of the focal plane, where the reflected radiation is low, the mirror test loading is around
2.2~pW or 10~K$_\mathrm{RJ}$.  This is similar to the 1.89~pW calculated from the optical loading using
temperatures and emissivities of the receiver components as
described in Section~\ref{sec:filters} and Table~\ref{tab:oploading}. Roughly half of \bicep2's optical
loading was from the atmosphere and half from internal loading.

\subsection{Detector bias}
\label{sec:biasoptimization}

   \begin{figure}
   \begin{center}
   \begin{tabular}{c}
   \def\svgwidth{7.7cm}
   \input{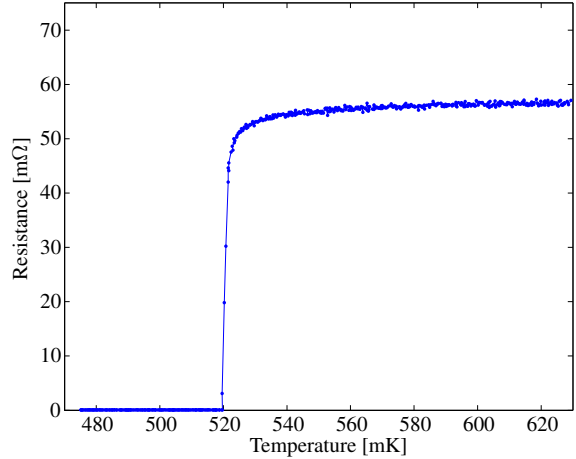}
   \end{tabular}
   \end{center}
   \caption[example]
   { \label{fig:r_vs_t}
   Example resistance vs. temperature characteristic for a JPL TES detector.
   The resistance rises from zero to 90\% of the normal state resistance within
   $5~\mathrm{mK}$.
}
   \end{figure}

The choice of TES bias voltage affects the noise level and stability of
the detectors and their safety margin before saturation.  We have taken noise
data at a range of biases under low loading conditions during winter 2010,
in order to choose the settings that give the lowest noise and greatest
sensitivity.  The optimization is described in detail in~\citet{brevik10}.

\begin{figure}
   \begin{center}
   \begin{tabular}{c}
   \def\svgwidth{7.7cm}
   \input{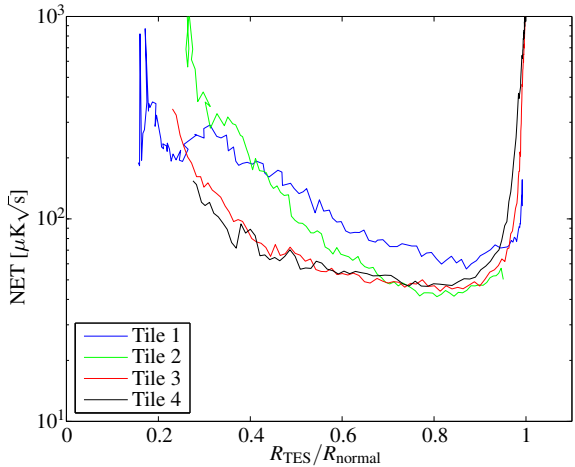}
   \end{tabular}
   \end{center}
   \caption[example]
   { \label{fig:net_vs_bias}
       Per-tile noise equivalent temperature (NET) in units of CMB temperature
       as a function of the detector resistance.  These data were taken under
       conditions of low atmospheric loading during the winter of 2010 and used to
       select new TES bias values to improve the instrumental sensitivity.
       NET sharply increases at the top of the superconducting-normal transition
       (high $R_\mathrm{TES}/R_\mathrm{normal}$) as the detectors saturate.
       In \bicep2 the NET also increases in the middle and lower part of the
       transition because of TES excess noise.  The excess noise increases
       with increasing transition steepness $\beta$, which is larger at
       low fractional resistance~(Eq.~\ref{eqn:transitionbeta}, Fig.~\ref{fig:r_vs_t}).
       Note that the minima of the NET curves shown here do not directly represent the
       final noise level of \bicep2 after optimization, for several reasons.
       The NET values have been approximately converted to CMB temperature units assuming
       a typical value of the sky temperature at zenith.  Variation in sky temperature
       will therefore affect the minimum NET as plotted, but does not impact the selection
       of optimal bias point.  The NET as optimized is somewhat better than shown because
       the TES bias is configured per column rather than using a single value for each tile.
       Finally, the 2010 data shown here do not use the improved 2011--12
       multiplexing configuration~(\S\ref{sec:muxoptimization}).
       
  }
   \end{figure}

The optimal bias voltage for a given TES detector depends on its responsivity
(\ie~the shape of the transition, or $R$ vs. $T$ curve, as in Fig.~\ref{fig:r_vs_t})
and on its noise properties.
Fig.~\ref{fig:net_vs_bias} shows the noise as a function of bias point
in the same 2010 noise data set that was used to optimize the TES biases.
For \bicep2 detectors the responsivity was highest in the lower portion
of the transition, when the fractional resistance $R/R_\mathrm{normal}<0.5$.
When the detector was very low in the transition, with $R/R_\mathrm{normal}$ close
to zero, the detector could enter a state of unstable electrothermal feedback.
Higher in the transition, the responsivity decreased and the detector could
saturate or have a gain that varies with atmospheric temperature.  There was a
broad region in the middle of the transition with suitably high and stable
responsivity.

Some components of noise also depend on the TES bias voltage.  The \bicep2
noise data showed TES excess noise~(\S\ref{sec:noise}) aliased into
the low-frequency region $<2~\mathrm{Hz}$.  The TES excess noise
generically increases with increasing transition steepness parameter
  \begin{equation}
    \beta=\left(R/I\right)\left(\partial I/\partial R\right)|_{T}
    \label{eqn:transitionbeta}
  \end{equation}
For our detectors
$\beta$ was largest low in the transition, so the excess noise was minimized
and the sensitivity was highest when the bias point was toward the high end.

The 32 TESs in a multiplexing column shared a common
bias line, so this optimization was performed column-by-column to maximize
the array sensitivity.  At the optimal bias some detectors could be saturated
(high bias) or unstable (low bias).  This was an acceptable price
for maximizing the overall sensitivity.

Before the mid-2010 noise data were taken, we used an initial set of
biases chosen based on noise data taken during summer, with higher
optical loading.  These were deliberately chosen to be conservative,
with lower bias for greater margin of safety against saturation.
We switched to the optimized detector biases on 2010 September 14 and continued to use
them throughout the remainder of the three-year data set.  They gave an
improvement of 10--20\% in mapping speed~(\S\ref{sec:mapspeed}).

\subsection{Detector time constants}
\label{sec:timeconst}

The TES detectors had a thermal time $\tau$ constant determined only
by the heat capacity $C$ of the island and the thermal
conductance $G_c$ of the legs.
The heat capacity was dominated
by the electronic heat capacity of the $0.3~\mathrm{\mu g}$ of added gold,
$C_\mathrm{Au}\approx0.3\text{--}0.5~\mathrm{pJ}/\mathrm{K}$.
The conductance varied between $80$ and $150~\mathrm{pW}/\mathrm{K}$
(\S\ref{sec:gtune},~\S\ref{sec:detprops}).  These combined to give thermal time
constants of $\tau=C/G_c\approx4~\mathrm{ms}$, with some variation from detector to detector
because of nonuniform $G$.  The time constants were faster when the detectors
were operated in negative electrothermal feedback~\citep{hiltonirwin2005},
so that the effective time constant for a typical detector was well below the
4~ms thermal time constant.

Because the frequencies of interest for $B$-mode science are much lower,
$f<2~\mathrm{Hz}$~(\S\ref{sec:scanpat}), the detector transfer functions are to
a good approximation perfectly flat.  This holds
as long as the detectors were biased sufficiently low in the transition, with a narrow
transition (high $\beta$) and strong electrothermal feedback.  If a detector was near
saturation, its time constant would become slower.

   \begin{figure*}
   \begin{center}
   \begin{tabular}{c}
   \def\svgwidth{14cm}
   \input{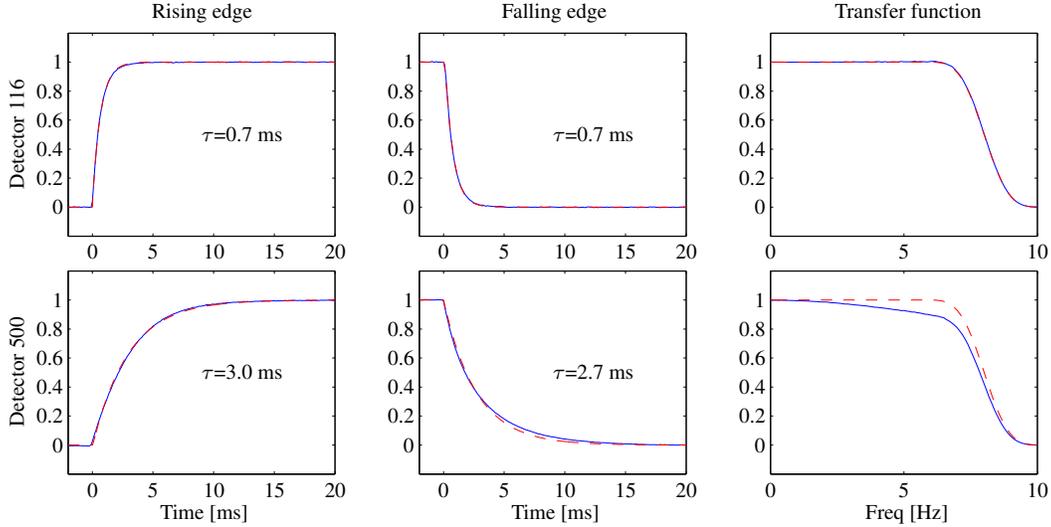}
   \end{tabular}
   \end{center}
   \caption[example]
   { \label{fig:timeconst}
   Measured time constants and transfer functions for a typical detector {\it (top)}
   with time constant $\tau<1~\mathrm{ms}$ and a slow detector {\it (bottom)}
   with time constant $\tau\approx 3~\mathrm{ms}$.  The three panels show a rising
   edge, falling edge, and a transfer function measured with a square-wave-modulated
   microwave source as described in the text.  Step responses are normalized
   to the step size, and transfer functions are normalized to unit gain at
   the 10~Hz modulation frequency of the data.
   The blue curves are data.  For the time constant panels, the red dashed
   lines are a fit to a single-exponential rise and fall with time constants
   as indicated.  For the transfer function panels, the red dashed curve is
   the transfer function of the MCE and \gcp\ digital filters alone~(\S\ref{sec:digifilt}).
   For fast detectors such as the one in the top panel, the data match this
   profile to within $0.5\%$, with no evidence of additional time constants.
   The MCE and \gcp\ digital filters have not been applied to the raw-mode 
   time constant data shown in the left and center panels.
   All data shown use the 2011--12 bias and multiplexing settings.
}
   \end{figure*}

We measured the time constants and end-to-end transfer functions of the detectors in
special-purpose calibrations during two of the austral summers.
The telescope was illuminated with a broad-spectrum
noise source chopped by a PIN diode to a square wave.  Metal washers were
inserted into a sheet of Propozote foam that was placed over the telescope window to
scatter the radiation and uniformly illuminate the focal plane.
For time constants the data were taken with $1~\mathrm{Hz}$ modulation and no
multiplexing, without applying any digital filters.  For transfer functions the
data were taken with a $10~\mathrm{mHz}$ square wave, applying the MCE and \gcp\
digital filters as in standard observing.
(This frequency was chosen to match the modulation of the atmospheric signal in
the \elnods\ used for relative gain calibration~(\S\ref{sec:gaincal}).  The
resulting transfer functions could then be used to verify that the relative gains
from \elnods\ also held within the full science band.)
The transfer function data used a standard data-taking
configuration including the MCE and \gcp\ filters~(\S\ref{sec:digifilt});
the detector time streams were Fourier transformed to give the transfer functions.
The response of most detectors was fast enough that the
results were indistinguishable from the transfer functions of the
digital filters applied by the data acquisition system.  A small number of
detectors were biased high in the transition and as a result had slower transfer functions.
These detectors showed faster transfer functions under lower optical loading, so that
the time constants measured with the bright noise source represent worst-case
performance.  The calibration data are shown in Fig.~\ref{fig:timeconst}
for two detectors, one of which had typical fast response, and one of which had
a slow response under the bright illumination of the noise source.  The typical
detectors' time constants were sufficiently fast that their transfer functions match
the model from the MCE and \gcp\ filters to within $0.5\%$.  These tests
were repeated for all detectors using the 2010 and 2011--12 TES bias and filter
settings.  The distribution of time constants across the array is shown in the lower left panel of Fig.~\ref{fig:dethists}.

The time constants are relevant not only to the time stream noise and resulting
instrumental sensitivity, but also to the systematics budget of the experiment.
Our data analysis deconvolves only the digital filters (\S\ref{sec:deconv}).
Following the general strategy for systematics control described in \S\ref{sec:performance}
we have performed simulations to show that the flatness of the achieved transfer functions,
and in particular the consistency between the A and B detectors in a pair, are
sufficient to ensure that the small departures from non-ideality do not significantly
impact our results.  We confirm this
conclusion using the difference map (jackknife test) of left-going and right-going scans.
These constraints on the contamination of $B$-modes
from detector time constants can be found in the Systematics Paper.

\subsection{Time stream noise}
\label{sec:noise}

   \begin{figure*}
   \begin{center}
   \begin{tabular}{c}
   \def\svgwidth{18cm}
   \input{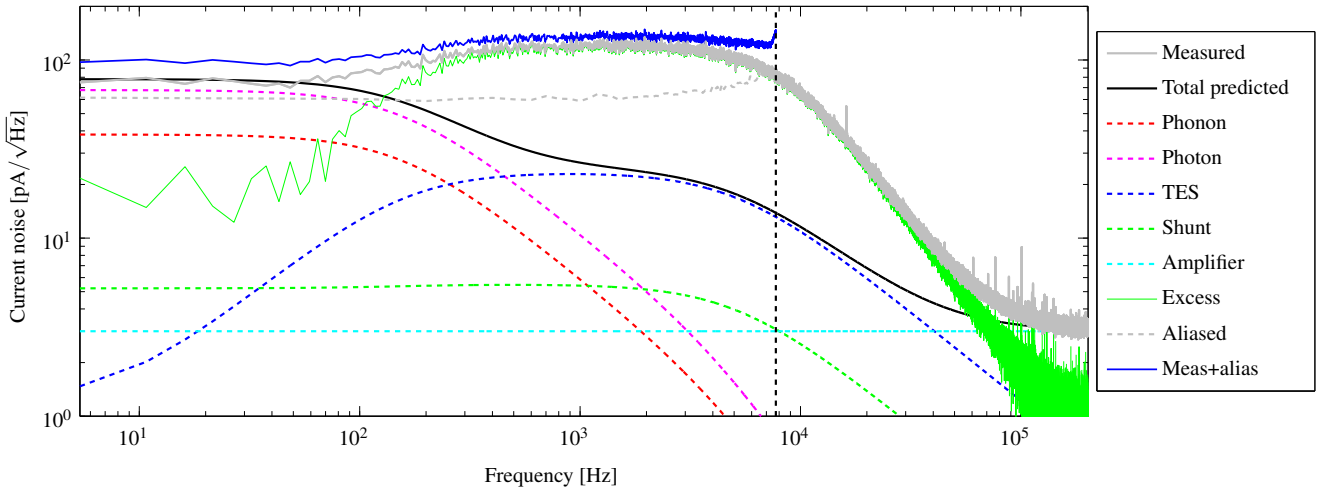}
   \end{tabular}
   \end{center}
   \caption[example]
  { \label{fig:noise_psd}
   Measured noise spectrum of a characteristic detector under typical observing conditions and with 2010 readout settings.
   Also shown are modeled noise components estimated from measured detector parameters.  The measured
   noise is well described by the sum of individual noise model components at low frequencies, where
   the scan-modulated science signal peaks.  Due to the limited bandwidth of the detector readout
   (indicated by the dashed vertical line), aliased excess noise contributes non-negligibly to the
   achieved noise performance at low frequencies.
   For the 2011 and 2012 observing seasons, the increase in multiplexing rate from 15~kHz to
   25~kHz~(\S\ref{sec:muxoptimization}) reduced the aliased noise to approximately half its original level.
}
   \end{figure*}

The noise level in the detectors has been previously documented in~\citet{brevik10,brevik11}.
The noise was characterized in special-purpose data
taken at a fast readout rate of 400~kHz by skipping the multiplexing step, allowing aliased
noise to be studied separately from intrinsic noise at low frequency. Although
degree-scale CMB anisotropies correspond to frequencies of 0.05--1~Hz~(\S\ref{sec:scanpat}),
the noise at much higher frequencies can become relevant through aliasing.  This
was especially true for the 2010 season, which used a slower multiplexing rate
of 15~kHz rather than 25~kHz as in 2011--12.

The noise is broken down by component in Figure~\ref{fig:noise_psd}, for
the 2010 readout settings.  At low frequencies it was dominated by photon noise.  The NEP
from the photon noise was a combination of the Bose and shot noise 
(see \eg~\citet{hiltonirwin2005}):
  \begin{equation}
    \mathrm{NEP}^2_{\mathrm{photon}}=2h\nu P_{\mathrm{load}}+\frac{2P_{\mathrm{load}}^2}{\nu (\Delta\nu/\nu)}
  \end{equation}
where $\nu$ is the band center, $\Delta\nu / \nu$ is the fractional bandwidth, and $P_{\mathrm{load}}$ 
is the photon loading.  For 4--5.5~pW of loading, as measured in ~\S\ref{sec:detprops}, 
the photon noise contributed 41--56~aW/$\sqrt{\rm{Hz}}$.  

The next largest contribution to noise at low frequencies was the phonon noise from thermal
fluctuations across the SiN legs.  The NEP contribution~(Eq.~\ref{eqn:phonon})
was 27~aW/$\sqrt{\rm{Hz}}$.  All other noise contributions, such as Johnson and
amplifier noise, were subdominant in the low-frequency region.

However, at frequencies of 1~kHz, the TES Johnson noise and the TES 
excess noise both contributed substantially.
The excess noise~\citep{galeazzi11} increased at lower TES biases and had a
power spectral density similar to Johnson noise.  The 15~kHz multiplexing 
rate used in 2010 (shown as a vertical line in Figure \ref{fig:noise_psd})
aliased that noise into the low-frequency region.  The increased multiplexing
speed of 25~kHz in 2011--2012 reduced that aliasing amount.  The total noise, 
including aliasing effects, was 67--78~aW/$\sqrt{\rm{Hz}}$
with 2010 settings and 56--64~aW/$\sqrt{\rm{Hz}}$ for 2011--12 configuration.  

Combining the noise, optical efficiency, optical loading, and yield
using the method described in~\citet{kernasovskiy12}, and converting to CMB
temperature units, \bicep2 as a whole is predicted to have an NET of
15~$\mu$K$_\mathrm{CMB}\cdot \sqrt{\rm{s}}$ with the 2011--12 settings.
The actual detector performance was evaluated using the noise in the range
0.1--2~Hz in a subset of 2012 CMB data, giving
316~$\mu$K$\cdot \sqrt{\rm{s}}$ per detector and 
15.9~$\mu$K$\cdot \sqrt{\rm{s}}$ for the array~\citep{brevik11}.
The per-detector 
distribution is shown in the lower right histogram of Fig~\ref{fig:dethists}.  
The NET as calculated from the time streams agrees well with the results
of a separate calculation from coadded maps, which gives 15.8~$\mu$K$\cdot \sqrt{\rm{s}}$.

\section{Instrument performance}
\label{sec:performance}
While the previous section focused on detector properties that affect the
sensitivity of the experiment, the instrumental performance characteristics
described in this section contribute to the systematics budget.  We have
extensively measured these characteristics in both pre-deployment tests and
post-deployment calibration measurements.  The results in this
section combine results from laboratory tests, \textit{in situ} calibrations,
and (in some cases) the CMB data set itself.

In general, we have not relied on meeting predetermined benchmarks in these
properties to guarantee adequate control of systematics.  Instead, we use
the results of tests and calibration data as inputs to detailed simulations
that we use to calculate the contribution of each effect given the actual
performance of \bicep2, its observing pattern and noise levels, and
the same analysis pipeline that we use to prepare maps and angular power
spectra from real data.

The Systematics Paper will present the set of simulations and the
powerful analysis technique of deprojecting instrumental effects.
The Beams Paper will apply these same methods to the important
class of systematics related to beams.  It will present a set of simulations
made from observed high signal-to-noise beam maps for each detector, with no assumption
of Gaussianity or ideality.  In the current paper we describe the calibration measurements
including the high-quality beam maps, and note that the simulation campaign
has shown that the instrumental performance as reported here meets the requirements
for \bicep2 to remain sensitivity limited rather than systematics limited.

\subsection{Mast-mounted source calibrations}
\label{sec:mastcal}

Many of the calibration measurements at the South Pole involved observation of a
millimeter-wave source in the optical far field.  We
mounted sources on a $12.2~\mathrm{m}$ high mast on the Martin A. Pomerantz
Observatory (MAPO) at a distance of $195~\mathrm{m}$ from the telescope.  The
source then appeared at an elevation of about $3^\circ$ above the horizon as
seen from \bicep2.  The ground shield and roof penetration did not allow \bicep2
to directly observe at elevations below $50^\circ$, so far-field source
calibrations were made with the aid of a $1.6~\mathrm{m}\times 1.1~\mathrm{m}$ flat mirror mounted to the
front of the telescope.  The far-field flat mirror was also used for occasional
observations of the Moon and Venus.  Because observations of terrestrial and
astrophysical compact sources all required the flat mirror to be installed, these
measurements could only be made during summer calibration work.

\subsection{Beams}
\label{sec:beams}

The far-field optical response of each detector was measured before and after each observing
season at the South Pole.  The far-field beam mapping campaign and full beam properties will be presented
in the Beams Paper along with simulations that establish stringent limits on the level at which
beam systematics enter our $B$-mode results.  Here we briefly describe the beam maps used in
this analysis and summarize the overall properties of the beams individually and in pair difference.

\begin{figure}[ht]
   \begin{center}
   \begin{tabular}{c}
   \def\svgwidth{7.7cm}
   \input{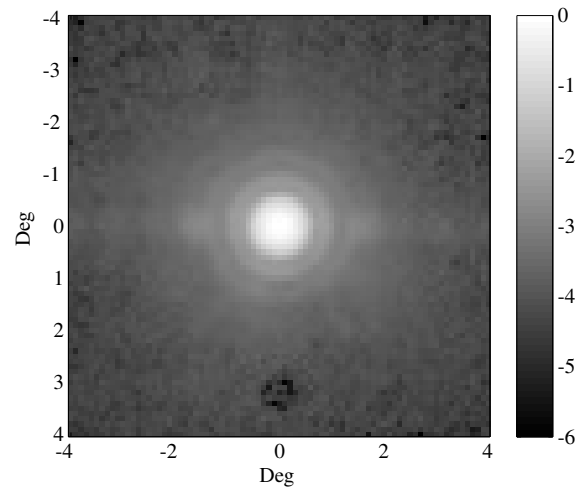}
   \end{tabular}
   \end{center}
   \caption{ \label{fig:stackedBeams}  A map of the \bicep2 far-field response made with the thermal
     source in units of $\log_{10}(\mathrm{power})$, showing dynamic range of more than
     six decades. Beam maps for individual detectors are shifted to align the peaks and coadded
     over all operational detectors.  The measured shape of the main beam and Airy ring structure are
     well matched by simulations, as shown in the Beams Paper.  Crosstalk features are apparent as small additional beams to the
     left and right, at a low level relative to the main beam strength.  The dark feature near the
     bottom is negative-going crosstalk found only in a subset of channels.
   }
   \end{figure}

The characterization of the shape and position of each detector's beam was performed by mapping the 
optical response to a chopped thermal source mounted on the MAPO mast (in the optical far field).
A chopper wheel modulated between the cold sky ($\sim15$~K) and ambient
temperature ($\sim250$~K) at a rate of $18$~Hz.  This largely unpolarized blackbody source was 
well suited for measuring the spectrally-averaged optical response of the instrument.  
The quality of this data set is illustrated in Fig.~\ref{fig:stackedBeams},
which shows a composite beam map that has been
centered and co-added over all operational channels.
The Beams Paper will show the results of a set of optical simulations which give a very good
match to the observed main beam and Airy ring structure.

\begin{figure*}[ht]
   \begin{center}
   \begin{tabular}{c}
   \def\svgwidth{14cm}
   \input{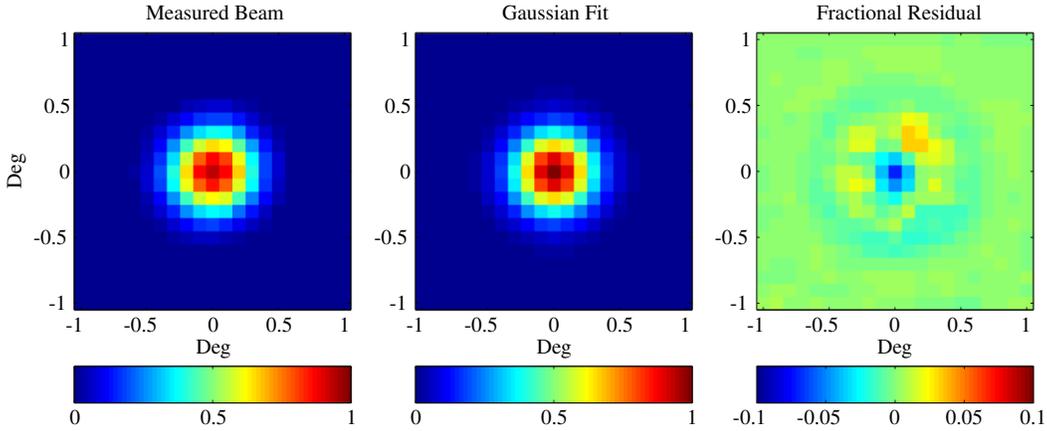}
   \end{tabular}
   \end{center}
   \caption{ \label{fig:farFieldBeam}
     Results of far-field beam characterization with a chopped thermal source.
     \emph{Left:} Typical measured far-field beam on a linear scale. \emph{Middle:} The Gaussian fit to the measured beam pattern. 
     \emph{Right:} The fractional residual after subtracting the Gaussian fit.  Note finer color scale in the right-hand
     differenced map.
   }
 \end{figure*}

We use elliptical Gaussians as a convenient way to parameterize beam shapes in Table~\ref{tab:beam},
although our analysis does not rely on an assumption of Gaussianity.
Fig.~\ref{fig:farFieldBeam}
shows an example map of a typical detector, the elliptical Gaussian fit to the beam,
and the fractional residual remaining after subtracting the fit.
We extract five parameters for each detector: position (in two directions),
beam width, and ellipticity (two parameters).  Two parameters
are required to fully specify ellipticity: these could be ellipticity and orientation,
but we use two orthogonal components known as the ``plus'' and ``cross'' orientations,
which are analogous to Stokes parameters for polarization.
These five parameters are defined in terms of the semimajor and semiminor axes
$\sigma_\mathrm{maj}$ and $\sigma_\mathrm{min}$, and the rotation angle $\theta$ of the major axis.
Table~\ref{tab:beam} lists the mean value for each quantity along with the scatter
among detectors.
The beam width $\sigma=\sqrt{\left(\sigma_\mathrm{maj}^2+\sigma_\mathrm{min}^2\right)/2}$
has an average value of $13.3'$ with standard deviation of
$0.3'$ across detectors ($0.221^\circ\pm0.005^\circ$).
This is close to the $12.4'$ predicted by optics
simulations~(\S\ref{sec:optics}).

\renewcommand{\tabcolsep}{4pt}
\begin{table*}[t]
\begin{center}
\caption[Measured per-detector beam parameters]{Measured per-detector beam parameters \label{tab:beam}}
\begin{tabular}[c]{lcccc}
\hline
Parameter\footnote{The per-detector parameters are calculated as an inverse-variance weighted combination of the elliptical Gaussian fits to 24 beam maps
  with equal boresight rotation coverage.} 
     & Symbol
     & Definition
     & Mean\footnote{Mean across all detectors used in science analysis.}
     & Scatter\footnote{Standard deviation across all good detectors.  The uncertainty of the measurement for each detector is small compared
              to the true variation from detector to detector.} \\
\hline\hline
Beam width & $\sigma_i$  & $((\sigma_\mathrm{maj}^2 + \sigma_\mathrm{min}^2)/2)^{1/2}$ & $13.3'$ & $0.3'$ \\
Ellipticity Plus ($+$) & $p_i$  & $\left(\frac{\sigma_\mathrm{maj}^2 - \sigma_\mathrm{min}^2}{\sigma_\mathrm{maj}^2 + \sigma_\mathrm{min}^2}\right)\cos 2 \theta $ & 0.013 & 0.03 \\
Ellipticity Cross ($\times$)  & $c_i$   & $\left(\frac{\sigma_\mathrm{maj}^2 - \sigma_\mathrm{min}^2}{\sigma_\mathrm{maj}^2 + \sigma_\mathrm{min}^2}\right)\sin 2 \theta $ & $0.002$ & 0.03 \\
\hline
\end{tabular}
\end{center}
\end{table*} 

\begin{table*}[t]
\begin{center}
\caption[Differential beam parameterization]{\label{tab:diffbeam}Differenced beam parameters}
\begin{tabular}[c]{lcccc}
\hline
     Parameter\footnote{Differential parameters are calculated by differencing measured beam parameters for detectors $A$ and $B$ within a polarized pair.}
     & Definition
     & Mean\footnote{Mean across all detector pairs used in science analysis.}
     & Scatter\footnote{Standard deviation across all detector pairs used in science analysis, dominated by true pair-to-pair variation.}\\
\hline\hline
Differential Pointing, $ \delta x$ & $x_A - x_B $  & $0.80'$ & $0.38'$ \\
Differential Pointing, $ \delta y$ & $y_A - y_B $  & $0.80'$ & $0.42'$ \\
Differential Beam Width, $\delta \sigma$ & $\sigma_A - \sigma_B$  & $-0.02'$ & $0.1'$ \\
Differential Ellipticity, $ \delta p$ & $p_A - p_B $   & $-0.002$ & 0.016 \\
Differential Ellipticity, $\delta c$  & $c_A - c_B$  & $-0.004$ & 0.014 \\
\hline
\end{tabular}
\end{center}
\end{table*} 

The differential beam parameters for a pair of co-located orthogonally polarized
detectors are calculated by taking the difference
between the main beam parameters for each detector within the pair.
These differential beam parameters are shown in Table~\ref{tab:diffbeam}.
The measured differential pointing per pixel for \bicep2 was larger than that observed in \bicep1 and much 
larger than optical modeling of the telescope predicted (see~\S\ref{sec:lenses}).
Subsequent detector testing has shown that this differential pointing
is related to contamination in the Nb films and crosstalk within the
microstrip lines of the antenna array feed networks.  Design and fabrication
changes described in~\cite{obrient12} and the upcoming Detector Paper have addressed these two issues
to reduce differential pointing for subsequent devices used in the \keck\
and \spider.

The effects of differential pointing, differential beam width, and differential ellipticity have been
strongly reduced through the adoption of the deprojection technique described in the Beams
and Systematics Papers.  We find that no other modes of beam mismatch are present at a sufficiently
large level to justify the use of deprojection.

We calculate the ultimate level at which beam imperfections affect our polarization maps by performing
simulations with the measured thermal source beam maps as inputs.  The simulation pipeline is run with
the observed beam map for each detector rather than a Gaussian or other approximation.  This technique
allows us to include the effects of all possible beam imperfections, not just those that can be represented
in terms of the elliptical Gaussian parameters or modes of the deprojection method.  We find that the
level of contamination from beam shape mismatches is below the noise-limited sensitivity of the experiment.

\subsection{Far sidelobes}
\label{sec:sidelobes}
Far sidelobes of the \bicep2 telescope could potentially see the 
bright Galactic plane as well as radiation from the ground or nearby buildings.  
To mitigate far sidelobe contamination in CMB observations, \bicep2 implemented a combined
ground shield and forebaffle system~(shown in Fig.~\ref{fig:mount}) similar to \polar~\citep{polar03}.  The first stage
was a large, ground-fixed reflective screen that removed a direct line of
sight between the telescope and the ground and redirects any far sidelobes
to the cold sky, lowering loading and preventing spurious signals. 

The second stage was a co-moving absorptive baffle~\citep{keating01}
that rotated with the telescope around its boresight and was designed to intercept 
the farthest off-boresight beams at $\sim15^{\circ}$ from beam center.  
It was constructed from an aluminum cylinder with a rolled lip lined with 10~mm thick 
sheets of Eccosorb HR.
The Eccosorb was coated with Volara foam\footnote{Sekisui Voltek,
Lawrence, MA 01843, \mbox{\url{http://www.sekisuivoltek.com/products/volara.php}}}
to prevent snow accumulation and disintegration of the Eccosorb in the Antarctic climate.

The system was designed such that at the lowest CMB observation 
elevation angle ($~55^\circ$) rays from the telescope must
diffract twice (once past each stage of the shielding system) before they hit the ground.
This is an identical strategy to \bicep1, described in~\citet{takahashi10} 
and shown in Fig.~\ref{fig:mount}.

We have two different methods for measuring the far sidelobes: one finds the total power coupling to
the detectors from outside the main beam, and the second maps the angular pattern of the sidelobes.
Additional information on these measurements can be found in the Beams Paper.

We have used the loading from the absorptive forebaffle as a measure of the total power in
far sidelobes.
As we removed and reinstalled the forebaffle with the telescope pointed at the zenith,
the change in loading per detector corresponded to $3$--$6~\mathrm{K}_\mathrm{CMB}$.
This is higher than the measured \bicep1 value.  The origin of this coupling is attributed to a combination of scattering from the 
foam window, shallow-incidence reflections off the inner wall of the telescope tube, and residual 
out-of-band coupling~(\S\ref{sec:islandcoupling}).  The forebaffle loading was highest for pixels located
near the center of the focal plane, because these have the largest fraction of their sidelobes
terminate on the forebaffle rather than internal surfaces or the sky.

The angular pattern in far sidelobes was mapped using a broad-spectrum noise source with fixed 
polarization and modulated with a chopper wheel.  We made maps at two different levels of source brightness to achieve a high signal-to-noise ratio
over a $\sim70$~dB dynamic range.  This allowed us to measure both the main beam and dim, outlying features
without significant gain compression.
We found no sharp features in the far sidelobe regime. For a typical detector,
less than 0.1\% of the total integrated power remained beyond $25^\circ$ from the main beam.

In order to verify that the total power in far sidelobes matches the integrated power in the angular pattern
measured with the broad-spectrum noise source, we made maps of the far sidelobe response with and
without the forebaffle installed.  The results were consistent with the other far sidelobe measurements.
The fractional amount of loading intercepted by the forebaffle averaged across the focal plane 
was 0.7\%, which corresponds to $3~\mathrm{K}_\mathrm{CMB}$.

\subsection{Polarization response}
\label{sec:pol_response}
The polarization response of the detectors was measured in two types of calibration
tests.

The first technique used a dielectric sheet as described in~\citet{takahashi10}.
The dielectric sheet calibrator worked as a partially
polarized beam splitter, directing one polarization mode to the cold sky and the
orthogonal mode to a warm microwave absorber at ambient temperature~\citep{polar03}.
Because of this temperature contrast, the arrangement acted as a polarized beam-filling
source.  By rotating the telescope about its boresight beneath this source we
obtained a precise measurement of the polarized response of each detector as a
function of source angle.  This technique
was fast and precise but also sensitive
to the exact alignment of the calibrator.  

\begin{figure}
   \begin{center}
   \begin{tabular}{c}
   \def\svgwidth{7.7cm}
   \input{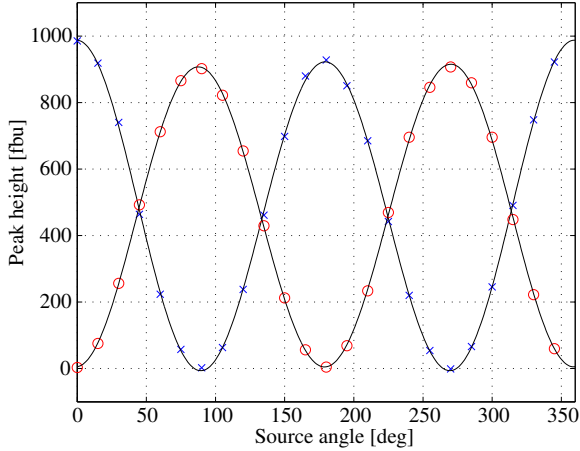}
   \end{tabular}
   \end{center}
   \caption[example]
   { \label{fig:rps_mod}
   Polarization response of a detector pair from a rotating polarized source measurement.
   The $x$-axis shows the source polarization angle relative to the vertical.  The A
   detector (blue points) responds to vertical polarization and the B detector (red points)
   responds to horizontal polarization at a boresight angle of $0^\circ$ (see also
   Fig.~\ref{fig:antenna}).  The cross-polar signal is $\sim0.5\%$, consistent with
   the level of crosstalk.  The small deviation from a sinusoidal form is caused by variation in
   source illumination of the telescope, included in the fit curves.
}
   \end{figure}

The second technique used beam maps of a rotatable polarized broad-spectrum noise 
source mounted on the MAPO mast~\citep{bradfordthesis}.  
To map the response of every detector as a function of polarization angle
incident on the detector, we set the polarized source to a given polarization angle
and rastered in azimuth over the source over a tight elevation range to obtain beam maps of
one physical row of detectors on the focal plane at one polarization angle.  
We then repeated this measurement in steps of 
$15^\circ$ in source polarization angle over a full $360^\circ$ range.  
After completing all source polarizations for a given row of detectors,
we moved to the next row of detectors and repeated the sequence.  We repeated the entire set of measurements 
at two distinct boresight rotation angles as a consistency check.
The response of a single detector to
rotation of the polarized source is shown in Fig.~\ref{fig:rps_mod}.

Both the dielectric sheet and rotating polarized source calibrators found a very low cross-polar response, $\sim0.5\%$.
This is consistent with the known level of crosstalk (\S\ref{sec:xtalk}) between the two detectors in each polarization pair.
The cross-polar response enters the analysis only as a small adjustment to the overall gain of the $E$ and $B$
polarization, but cannot create any false $B$-mode signal.

The primary $B$-mode target of \bicep2 requires only modest precision
in the measurement of the absolute angles of polarization response.
We have adopted the per-detector polarization angles from the
dielectric sheet calibrator for use in making polarization maps, as they have
low statistical error $<0.2^\circ$.  However, a coherent rotation of polarization
angles for all detectors is less strongly constrained because of possible systematics
in the alignment and material of the sheet calibrator, and the alignment of the source
in the case of the rotating polarized source.  We estimate the overall rotation angle from
the $TE$ and $EB$ correlations of the CMB using a self-calibration procedure~\citep{polselfcal}.
This indicates a coherent rotation of $\sim 1^\circ$, which is included as an adjustment
in the $B$-mode analysis.  We also simulate the effect of a similar overall offset to
show that the contribution at low $\ell$ is small even for an angle of $1^\circ$.

\subsection{Crosstalk}
\label{sec:xtalk}

The use of a multiplexer in \bicep2 presented several potential sources of
crosstalk that did not exist in the single-channel readout used in \bicep1.
For a full treatment of the crosstalk mechanisms see~\citet{dekorte03}.  The
crosstalk arises due to the use of common components for rows or columns of
detector in the multiplexer to reduce wiring count and due to the close
proximity of magnetically sensitive components.

The largest crosstalk mechanism in \bicep2 was inductive crosstalk between detectors
that are nearest neighbors within a multiplexing column, for which first-stage SQUIDs and
input coils were in close proximity.  A second mechanism was settling-time crosstalk caused
by the finite recovery time of the electronics after the multiplexer switched rows.
This crosstalk mechanism depends on the dwell time per row.  It was extensively tested
in 2010 before the increase in readout rate (\S\ref{sec:muxoptimization}) and measured
at a level of $-36~\mathrm{dB}$ (2010 settings) or $-34~\mathrm{dB}$ (2011--12 settings).

Crosstalk has been assessed from maps acquired by scanning across
both the broad-spectrum noise source and the chopped thermal source.
These measurements
used the far-field flat mirror and the aluminum transition of the bolometers
(which could operate under high
loading).  The signal-to-noise ratio in these maps was generally adequate to probe crosstalk to a
level of around $-40~$dB.  These large-signal beam maps could be sensitive to nonlinear
and threshold-dependent crosstalk mechanisms that do not affect standard CMB
observations on the titanium transition.  We have made a second calculation of
crosstalk levels using cosmic ray hits during CMB data taking.  When a cosmic
ray interacts in a TES island, the deposited energy thermalizes and raises the
temperature of the island.  This appears as a spike in the time stream exactly as if
it were an instantaneous spike in optical power, and is subject to the same forms
of crosstalk.  We have selected cosmic ray hits of moderate amplitude (equivalent to
15--300~mK$_\mathrm{CMB}$), finding around 100 such events per detector in the full
data set.  We excluded events in which multiple detectors see a large amplitude in order
to exclude showers.  These were stacked, and the crosstalk level was read from the corresponding
samples in all other channels.

The thermal source beam maps show crosstalk at a level of $0.25\pm0.16\%$
between detectors that are nearest neighbors in the multiplexing sequence.  The 
scatter around the mean is dominated by detector variations; the noise 
per detector is lower at $0.03\%$.  The next-nearest neighbors see $0.03\pm0.10\%$ and 
all other channels are below the noise
level.  This is consistent with the 0.25\% reported by 
NIST-Boulder~\citep{dekorte03} for an earlier version of the multiplexing chip.  
This cosmic ray analysis is also consistent with the beam maps, measuring a 
nearest-neighbor crosstalk level of $0.38\pm 0.23\%$.  This confirms that the 
crosstalk levels were consistent between the large-signal, aluminum transition 
data taking mode of the beam maps and the normal 
CMB data taking mode of small signals on the titanium transition. 

Atypical crosstalk between several multiplexer rows was discovered in beam maps
taken during instrument commissioning. Single-channel maps showed multiple
beams with amplitudes comparable to the expected main beam.
The SQUID tunings of the affected channels also showed an unusual flux response.
The problem was traced to wiring shorts between the bias lines of several multiplexer
rows, which caused first-stage SQUIDs to be inadvertently biased and read
out during the wrong part of the multiplexing cycle.  This crosstalk was mitigated
by reducing the bias levels for the shorting rows until they were low enough to
prevent turn-on of multiple SQUIDs at the same time.  The affected channels
are excluded from mapmaking for the period of time before the fix by channel
selection cuts~(\S\ref{sec:cuts}).

\subsection{Glitches and unstable channels}
\label{sec:unstable}

Whenever the signal in one detector underwent a large step or glitch (usually a cosmic ray hit),
there were additional crosstalk considerations beyond the
nearest-neighbor mechanisms described in~\S\ref{sec:xtalk}.
There was a coupling to all other channels in the same multiplexing column
at a lower level than the nearest-neighbor crosstalk.  There was also
a coupling to other channels in the same multiplexing row (channels read
out at the same time) through the common ground of the ADCs and DACs in the MCE.
These mechanisms could introduce small steps in many channels coincident
with a glitch or flux jump in a single channel.  These are handled conservatively in
analysis by deglitching and cutting all channels that might be affected by crosstalk
from a glitch event (\S\ref{sec:deglitch}).

A small number of channels had readout hardware defects that caused their
raw amplifier signal to very frequently undergo spontaneous jumps.
A small number of other detectors had unstable TES bias points and
sometimes entered a state of electrothermal oscillation.  Either class
of pathology could cause localized transient signals or step offsets on
other detectors in the same multiplexing row and column, which then lost
livetime to deglitching~(\S\ref{sec:deglitch}) and cuts~(\S\ref{sec:cuts}).
We found that it was useful to disable SQUID flux feedback only for those
detectors that showed no optical response in the time streams.
For channels that were severely unstable but had some optical response, we kept
the feedback servo active to minimize disruption to neighboring channels.

\subsection{Thermal stability}
\label{sec:thermalstab}

\bicep2 maintained a stable focal plane temperature through passive filtering
and active thermal control as described in~\S\ref{sec:cryodesign}.
We have measured the performance of both these components to show that the
achieved stability met the requirements of the experiment.

To characterize the passive thermal filter, a heater near the sub-kelvin fridge
was turned on and off in a square-wave pattern with an amplitude of 7 mK.  The
frequency of the square wave was varied while temperatures were monitored on
the fridge and focal plane sides of the passive thermal filter.  The filter's
low-frequency performance ($<$0.1~Hz) can be modeled as a continuous-pole
low-pass filter with a characteristic frequency of 0.291~mHz.  Measurement of
the filter's response in our science band was limited by crosstalk between the
thermometers.  The thermal filter suppressed thermal fluctuations originating
from the refrigerator by at least a factor of $10^{4}$ for $f<2~\mathrm{Hz}$.

In CMB units, the spurious polarization signal caused by temperature fluctuations is
proportional to the matching of $G_0/\eta$ in a detector pair, where $G_0$ is the thermal
conductance of the bolometer at the base temperature and $\eta$ is the optical efficiency.
Because \bicep2 had higher optical efficiency and lower thermal conductance than \bicep1,
this ratio was much lower for \bicep2 and the sensitivity to thermal fluctuations was
correspondingly reduced.
We calculate the required level of thermal stability using simulations with measured
detector thermal responsivities, as was done for \bicep1.
For a target of $r$ = 0.01 we found the required stability is 6.0~nK$_\mathrm{FPU}$ at 
$\ell=100$. For comparison, \bicep1's thermal stability 
benchmark was 3.2~nK$_\mathrm{FPU}$ for a target of $r$ = 0.1~\citep{takahashi10}.

To quantify \bicep2's achieved thermal stability, angular power spectra
were calculated from the NTD thermometer channels instead of the TES bolometer
channels.  These NTD thermometers were not optically coupled and provide
a measure of the temperature fluctuations on the focal plane.  To produce
spectra, the thermometer time streams were processed into maps using the
standard analysis pipeline (see~\S\ref{sec:pipeline}).
The NTD maps are noise-dominated, so they provide only an upper limit on
the achieved thermal stability of the experiment.  This upper limit
gives temperature fluctuations of $0.4~\mathrm{nK}$ at $\ell=100$, averaged
over a year of data.  This is well below the $6~\mathrm{nK}$ requirement.
Further details of this calculation can be found in the Systematics Paper.

\subsection{Electromagnetic interference}
\label{sec:satcom}
The telescope was sensitive to the $2.0~\mathrm{GHz}$ signal of the
Amundsen-Scott South Pole Station's S-band uplink to the GOES satellite, which provides
seven hours of telephone and network connectivity to the station each day.
When the ground station transmitter was powered on it appeared in \bicep2 data as a ground-fixed
signal that had a characteristic pattern in azimuth and did not vary strongly
with time.  The level of sensitivity varied from detector to detector, with
most seeing a small signal equivalent to $1~\mu\mathrm{K}_\mathrm{CMB}$ and a few
seeing pickup as large as $300~\mu\mathrm{K}_\mathrm{CMB}$.  The amplitude of this
signal scaled linearly with the power at the transmitter, and it did not
appear in open-input SQUID channels or in tests with unbiased TESs.  These
results suggest that power from the ground station transmitter was eventually
thermalized in the TES island.  Further tests with the \keck\ have shown that the
2.0~GHz radiation entered through the window rather than through the electronic
feedthroughs on the back of the telescope.
We have confirmed that all other frequency bands used for satellite communication
at the South Pole have much lower power levels.

Because the signal had a fixed pattern in azimuth, it can be effectively removed from the bulk of our data by the ground-subtraction step in mapmaking.
The exceptions are the beginning and end of each satellite pass, when the transmitter changed state during a scan.
The pickup signal is largest only
when the telescope pointed toward the ground station.  The \bicep2 observing schedule and the
satellite communications schedule were both based on sidereal time, and they fortuitously aligned such that
the telescope always pointed away from the ground station during scheduled GOES communications.  The
small remaining signal had a negligible impact on CMB maps, as will be shown in the Systematics Paper.

The case is somewhat different for observations of the Galaxy~(\S\ref{sec:fields}).
During these parts of the schedule \bicep2 regularly pointed toward the actively transmitting ground station.
Past analysis of the Galactic field (as in \bicep1,~\citet{bierman11}) did not use
the ground subtraction analysis technique~(\S\ref{sec:groundsub}).  Studies of the \bicep2 Galactic data must therefore ensure
that the satellite pickup signal is adequately removed either through cuts, ground template subtraction, or some other
technique.

After these transmissions were detected in \bicep2 data, an RF-absorbing barrier was installed on the wall of
the GOES radome to attenuate the spillover of power toward the Dark Sector.  This will greatly reduce the S-band power
incident on the \keck\ and other experiments beginning with the 2014 observing season.

\subsection{Pointing}
\label{sec:pointing}

The pointing of the telescope can be analyzed in two parts.  The first is the
pointing of the central boresight axis of the cryostat, and the second is the pointing
of each detector's beam relative to this axis.
The boresight of the telescope was defined as the line in space that remained fixed as the mount
rotates in its third axis.  All other pointing directions were defined as offsets from
the boresight.

The boresight pointing is complicated by the fact that the building
sits on a packed-snow foundation over an ice sheet that moves at
$\sim 10~\mathrm{m}/\mathrm{yr}$.
The intrinsic precision of the mount and control electronics, combined with
the short-term stability of the platform, were sufficient to give
blind pointing accuracy below $20''$~\citep{yoon06}.
However, the movement and settling of the building caused pointing
drifts of $\sim$1' per month.  We accounted for these shifts by taking star observations
at three boresight angles as often as once every six days~(\S\ref{sec:starpoint}).
The star-pointing data were used to fit a seven-parameter model for the
orientation of the mount and the alignment of its axes.  In offline
data analysis the nearest star pointing fits are used with the pointing model
to convert raw archived encoder readings into boresight pointing in
horizontal coordinates.

The pointing of each beam relative to the boresight was determined by making temperature maps
for each detector as in~\S\ref{sec:pipeline}, but omitting the pair difference step.  The maps were made 
separately for each of the four boresight rotation angles, for left-going and right-going scans,
for each detector across the 2010--11 data set.  Each of these eight maps was then cross-correlated
with the temperature map from the five-year \wmap\ W band data set~\citep{wmap5yr}.  The external
temperature map had the \wmap\ beams deconvolved and was Gaussian-smoothed to the \bicep2
beam size before cross-correlation.  The offset that maximized the cross-correlation was taken as
a correction to the ideal detector pointing that had been used in forming the single-detector map.
From comparison among the eight maps for each detector, we estimate that this procedure gives
beam centers accurate to $2'$ rms.  The precision of the fitted beam centers is limited primarily
by the $0.25^\circ$ step in elevation of the CMB observing pattern, which makes the cross-correlation
more weakly sensitive in this direction.
We have simulated the effect of cosmological $TE$ correlations
as a bias on the beam centers and find it well below $5''$.
The same beam-fitting procedure has been repeated with \planck\ $143~\mathrm{GHz}$
maps~\citep{planckI,planckVI} instead of \wmap\ templates.  The results are identical to within $15''$ for all \bicep2 detectors.

When we compare the beam centers as fit from CMB maps at different boresight rotation angles,
we detect an offset in the elevation direction of an average of $1'$.  We interpret this offset as an
internal flexure of the focal plane assembly relative to the cryostat shell and the telescope mount.

\section{Observing strategy}
\label{sec:observing}
The \bicep2 observing strategy was based on deep integration in the region of
the sky least contaminated by polarized foregrounds.  The telescope spent 90\%
of its observing time on this CMB field, and the other 10\% on a secondary
Galactic field.  These observations were grouped in schedules of three sidereal
days, including a six-hour cryogenic service period.   Within one three-day
schedule the telescope scanned in azimuth at a fixed boresight angle---the
orientation of the telescope about its own axis.  The details of the observing
schedule were chosen to allow for control of possible systematics such
as drift in detector gain and ground-fixed signals.

\subsection{Observing fields}
\label{sec:fields}
\begin{figure} 
   \begin{center}
   \begin{tabular}{c}
   \def\svgwidth{7.7cm}
   \input{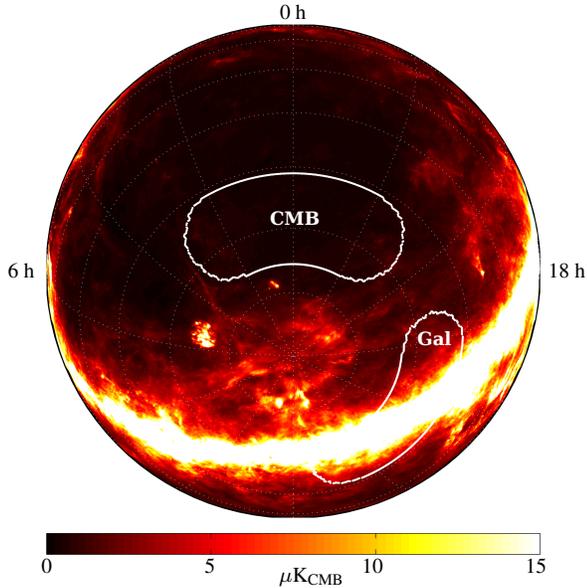}
   \end{tabular}
   \end{center}
  \caption{\bicep2 observing fields relative to the polarization amplitude predicted from FDS~\citep{fds99} model 8, assuming a 5\% polarization fraction.
  }
   \label{fig:syncrdust}
\end{figure}

\bicep2 spent most of its time observing the primary CMB field centered at 
($\mathrm{RA}=0~\mathrm{hr}$, $\mathrm{dec}=-57.5^\circ$).  This 1000-degree$^2$ field (2\% of the sky) lies well away from
the Galactic plane, within a larger region known as the ``Southern Hole'' where polarized foregrounds are expected
to be especially low.  The \bicep2 field is the same one observed by \bicep1.  It was selected for its very low
level of expected Galactic dust emission, less than $1\%$ of the sky median~\citep{fds99}
as shown in Fig.~\ref{fig:syncrdust}.  If the dust signal is polarized at 5\%, the resulting
contamination of the $B$-mode signal at $150~\mathrm{GHz}$ will be below $r=0.02$.
The faint synchrotron signal within the Southern Hole has not been well measured, but a scaling of \wmap\ data
at $23~\mathrm{GHz}$ implies that the $B$-mode contamination at $150~\mathrm{GHz}$ is at a level similar
to or below that from dust~\citep{nguyen08}.

The secondary \bicep2 field covered a part of the Galactic plane centered at
($\mathrm{RA}=\mathrm{15:42}~\mathrm{hr}$, $\mathrm{dec}=-55.0^\circ$).
Observations of this field can be used for Galactic science objectives~\citep{bierman11}
and as a bright, partially polarized source for use in instrument characterization.

These same two fields have also been observed by \bicep1\footnote{\bicep1 also observed
a third field in a different part of the Galactic plane.  This field has not been covered by
\bicep2 or the \keck.} and the \keck.  Coverage of the same fields
by the three experiments allows for consistency tests, cross-calibrations on the bright
Galactic signal, and the possibility of achieving greater map depth by stacking
CMB maps across multiple experiments.  The additional frequencies of \bicep1 and the
\keck\ (beginning in the 2014 season) also give spectral information needed to separate
any foreground signals from the CMB.

\subsection{Scan pattern}
\label{sec:scanpat}
The telescope scanned at $2.8^\circ/\mathrm{s}$ in azimuth, so that at an elevation of
57.5$^\circ$ a signal with frequency $f$ (in Hz) corresponds to a multipole $\ell=240f$.
This set the science band for the experiment: 0.05--1~Hz for $20\leq \ell \leq 200$ where the
inflationary $B$-mode signal is expected to peak, or $2.6~\mathrm{Hz}$ for $\ell=500$.

Each scan spanned $64.2^\circ$ in azimuth, of which the central $56.4^\circ$ (77.7\% of the
duration of the scan) was covered at uniform speed and is used for mapmaking.  The
region around each turn-around is excluded from CMB analysis.  The trajectory of each scan was optimized at the time of
\bicep2 deployment for a gain of 4\% in the usable, central part of the scan relative to \bicep1.
The elevation was kept fixed as the telescope executed 53 round-trip scans over a
period of 46 minutes.  During this single ``\scanset'' the telescope scanned back and forth within fixed limits
in azimuth, rather than continuously tracking the sky.  Each \scanset\ was preceded and followed
by bracketing calibrations as described in ~\S\ref{sec:scansetcals}, bringing the total duration of
each \scanset\ up to 50 minutes.

   \begin{figure*}
   \begin{center}
   \begin{tabular}{c}
   \def\svgwidth{17.65cm}
   \input{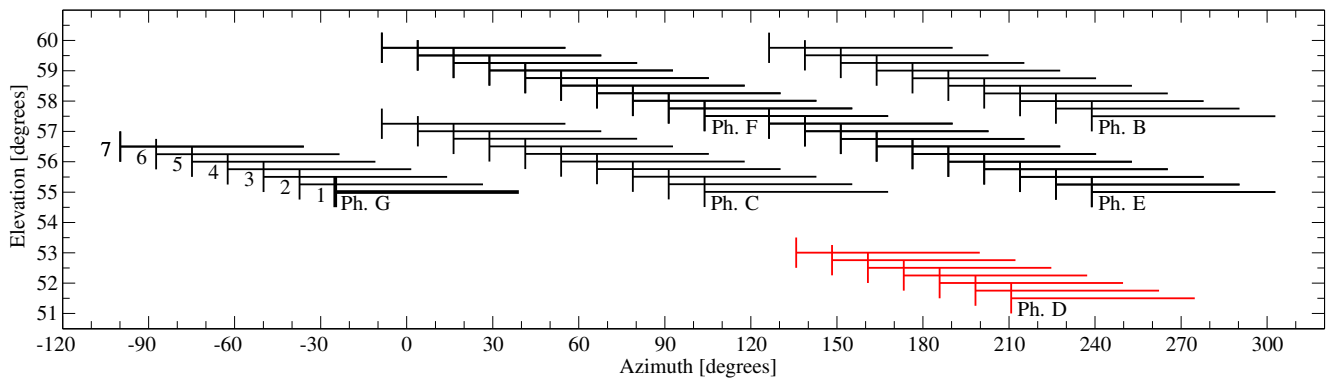}
   \end{tabular}
   \end{center}
   \caption[example]
   { \label{fig:schedpattern}
       Observing pattern of a typical three-day schedule.  Phase letters are as in Table~\ref{tab:phases}.
       The scansets of Phase G are numbered, with the
       first scanset at the lowest elevation.  The first scanset of Phase G is shown in bold, showing the
       throw of the field scans (horizontal line) and the bracketing elevation nods (vertical line).  The two six-hour phases
       can vary in elevation: the Galactic D phase is shown at the lowest of four elevation steps, and the CMB G phase
       is shown at the lowest of three elevation steps.  The H and I phases on the third LST day alternate between
       the B/C pattern and the E/F pattern.
   }
   \end{figure*}

At the end of each 50-minute \scanset, the telescope stepped up by $0.25^\circ$ in elevation and shifted the azimuth
of the scan center to follow the apparent motion of the field on the sky before beginning
the next \scanset.

This scan pattern deliberately covers a fixed range in azimuth within each 50-minute observing block,
rather than a fixed range in right ascension.  After 50 minutes the CMB has drifted by $12.5^\circ$ relative to
the ground.  Therefore, any pickup of ground-fixed optical power, the magnetic field of the Earth or nearby
structures, scan-fixed thermal fluctuations, or scan-fixed vibrational noise would all appear in the same locations
from scan to scan.  This allows us to remove these signals using a simple ground-subtraction algorithm~(\S\ref{sec:groundsub}).

\begin{table}[t]
\caption{Phases in a schedule}
\label{tab:phases}
\begin{center}
\begin{tabular}{ccl} 
\hline \hline
\rule[-1ex]{0pt}{3.5ex}  Phase & LST & Field \\
\hline
\rule[-1ex]{0pt}{3.5ex}  A & Day 0 23:00 & Cryo service \\
\rule[-1ex]{0pt}{3.5ex}  B & Day 1 05:30 & CMB (high el) \\
\rule[-1ex]{0pt}{3.5ex}  C & Day 1 14:30 & CMB (low el) \\
\rule[-1ex]{0pt}{3.5ex}  D & Day 1 23:00 & Galactic \\
\rule[-1ex]{0pt}{3.5ex}  E & Day 2  05:30 & CMB (low el) \\
\rule[-1ex]{0pt}{3.5ex}  F & Day 2 14:30 & CMB (high el) \\
\rule[-1ex]{0pt}{3.5ex}  G & Day 2 23:00 & CMB (variable el) \\
\rule[-1ex]{0pt}{3.5ex}  H & Day 3 05:30 & CMB (high / low el) \\
\rule[-1ex]{0pt}{3.5ex}  I & Day 3 14:30 & CMB (low / high el) \\
\hline
\end{tabular}
\end{center}
\end{table}

\subsection{Schedules and boresight angles}
\label{sec:schedules}
A three-day schedule was divided into groups of 50-minute \scanset s.
Each of these groups, called an observing phase, contained ten \scanset s (nine hours total) or
seven \scanset s (six hours total) along with the accompanying calibrations.  During one full
three-day schedule the telescope completed one six-hour cryogenic service phase,
six 9-hour phases and one 6-hour phase on the CMB field, and one 6-hour phase on the Galactic field, as
listed in Table~\ref{tab:phases}.
The azimuth/elevation pattern of a typical observing schedule is shown in Fig.~\ref{fig:schedpattern}.  
This represents only one of several
possible patterns: the Galactic phases alternated among four different elevation ranges; the
six-hour CMB phase on day two alternated among low, middle, and high elevation ranges; and
the day-three CMB phases alternated between following the day-one pattern (high then low) and the
day-two pattern (low then high).  These alternations ensured even coverage of each field and
uniform coverage of the CMB field at each azimuth range.  The low-elevation CMB phases had
boresight pointings from $55^\circ$ to $57.25^\circ$, and the high-elevation CMB phases had
boresight pointings from $57.5^\circ$ to $59.75^\circ$.

The three-day observing pattern was made possible by the long hold time of the \bicep2 helium bath and
$^3$He sorption fridge; in contrast, \bicep1 and the \keck\ have required cryogenic service every two days.
The \bicep2 cryogenic service period was used to refill the liquid helium bath, cycle the sorption
fridge, perform star observations to measure telescope pointing, and carry out other maintenance
tasks such as cleaning snow from the forebaffle and other exposed parts of the telescope.

Each schedule was taken with a fixed orientation around its axis or boresight angle.  We have selected
four boresight angles for standard observations:\footnote{A small amount of early \bicep2 data from
March 2010 used a different set of four boresight angles, with the same pattern but offset from the final
four.  These were $85^\circ$, $130^\circ$, $265^\circ$, and $310^\circ$.}  $68^\circ$, $113^\circ$, $248^\circ$, and $293^\circ$.
These include two pairs separated by $45^\circ$, so that it was possible to measure both Stokes $Q$ and $U$
using either $68^\circ$+$113^\circ$ or $248^\circ$+$293^\circ$.  Each of these angles also had a
counterpart that is $180^\circ$ away for characterization and control of instrument systematics such as
differential pointing.  The boresight angles are defined such that at $0^\circ$ the A detectors are sensitive
to vertical polarization and the B detectors to horizontal polarization, and vice versa at $90^\circ$.

\subsection{Integrated calibrations}
\label{sec:scansetcals}
Each 50-minute observing block began and ended with two types of integrated calibrations: an elevation nod (\elnod)
and a partial load curve.
The elevation nod measured detector response to a small change in atmospheric loading as
the telescope moved by a small amount in elevation.  The response to the \elnod\ gave a measure of the
detector response to a change in fractional air mass.  This responsivity measure
was used for three purposes: 1) to determine the relative gains of the two polarization
channels within a pixel, to allow pair differencing; 2) to correct for gain differences between
pixels before making maps; and 3) to reject data in which responsivity is anomalous or unstable.
The \elnod\ was the only source of gain calibration applied to individual detector time streams; there was no
calibration on an astrophysical source until the final absolute calibration, derived from the CMB
temperature anisotropy.  The details of the gain calibration are presented in ~\S\ref{sec:gaincal},
including an accounting for differences between the atmospheric and CMB
spectral energy densities.

In an ``up-down-up'' \elnod, as performed at the start of the observation, the telescope first stepped
upward by $0.6^\circ$, then downward by $1.2^\circ$, and upward again by $0.6^\circ$ to return to the starting
position.  This motion was performed slowly over about one minute.  The \elnod\ immediately following an observing
block was performed in the opposite order, ``down-up-down''.  

The second type of integrated calibration was a partial load curve.  The TESs were first driven
normal (heated above their superconducting transition) and the bias voltage as then stepped down to the standard operating point.  This gave an
$I$-$V$ characteristic for each sensor from which we can calculate the optical loading, resistance
in the operating and normal state, and Joule heating power.  The partial load curves were used to
reject data from periods of time when loading conditions placed the detectors outside the regime of
linear response, such as unusually cloudy days with very high atmospheric loading.

Extended versions of the \elnod\ and partial load curve were done at the beginning of each phase.
The elevation nod was replaced with a sky dip, in which the telescope slewed from an elevation of $50^\circ$ up
to the zenith and back down.  This provided a profile of atmospheric conditions.  The partial load curve was
replaced with a full load curve that covered the range from the high-bias normal state down to the
superconducting state at zero TES bias.  This gave a complete $I$-$V$ characteristic including the
entire transition region.  The full load curve was performed only once per phase because of the additional
time required to put the detectors back on transition from the superconducting state.

\subsection{Star pointing}
\label{sec:starpoint}
An optical camera was mounted at the top of the cryostat for star observations.
The camera has been described in~\citet{yoon06}.
The star pointing routine was performed as often as possible in order to measure and correct for the settling
of the DSL building on the moving ice sheet.  Winter star pointings observed 24 stars,
each at three boresight rotations.  They were performed every six days except when overcast
weather made it impossible to see the stars.  Star pointings were also performed in summer;
the optical camera was sensitive in the IR, so that it was possible to observe the brightest
12 stars on the list even when the Sun was above the horizon.

The star observations were used to fit a pointing model with seven parameters: the zero points of the
azimuth and elevation encoders; the tilt of the azimuth axis in two directions; the tilt of
the elevation axis; and two parameters for the collimation of the optical camera relative to the
telescope boresight.  This pointing model is applied to CMB observations to transform
the raw encoder coordinates into azimuth and elevation of the telescope boresight~(\S\ref{sec:pointing}).

\section{Data reduction}
\label{sec:pipeline}

Data reduction is performed using the analysis pipeline
developed for the \QUAD\ experiment~\citep{pryke09} and
subsequently adapted for \bicep1 and \bicep2.  In this section
we briefly review the data analysis process, focusing on
steps that relate closely to the performance of the instrument
and the production of the sensitivity figures that will be
presented in~\S\ref{sec:data}.  These are primarily the
low-level reduction and data quality cuts.  The mapmaking
procedure, $E$/$B$ separation, angular power spectrum
analysis, and simulation pipeline are described in detail
in the Results Paper  and the deprojection algorithm is presented in
the Systematics Paper.

\subsection{Transfer function correction}
\label{sec:deconv}

As the first stage of low-level data reduction the pipeline deconvolves the filters
that have been applied to the time streams~(\S\ref{sec:digifilt}).
The deconvolution kernel is an FIR filter designed from the known transfer functions of the MCE and \gcp\ filters.
We choose an FIR filter in order to ensure that any ringing from transients vanishes after a suitably short time.
This property allows the deglitching operation~(\S\ref{sec:deglitch}) to fully remove
the effects of glitches and flux jumps while excising only a small amount of data, $<2~\mathrm{s}$ per event.
The deconvolution kernel also includes a low-pass filter component to attenuate
signals $>3~\mathrm{Hz}$ that could be aliased during mapmaking.

\subsection{Deglitching}
\label{sec:deglitch}

There are several types of time stream glitches that must be flagged and removed in low-level data reduction.
It is necessary both to prevent the glitches themselves from being included in the maps and to ensure
that the time streams contain only well-behaved, stationary noise that can be represented by the noise model
used in simulations.  The majority of glitches in otherwise good channels were caused by cosmic ray hits
in the TES islands.  The energy deposited by ionization in the substrate thermalized and appeared as a brief
spike of power.  These transient events can be simply cut from the time streams.  We also observe occasional
step discontinuities in the DC levels when a cosmic ray event exceeded the bandwidth of the flux-feedback
loop and caused it to relock at a different point.  These events are very large steps equivalent to
$\sim 20~\mathrm{K}_\mathrm{CMB}$.     
The deglitching code for \bicep2 flags these events for additional correction.

In any given good detector, cosmic ray hits caused transient spikes large enough to be deglitched at
about one event per $3\times 10^4~\mathrm{s}$ and flux jumps at one per $7\times 10^5~\mathrm{s}$.
A small number of channels either had unstable detector bias or readout faults that caused more frequent flux
jumps~(\S\ref{sec:unstable}).

The time stream around any glitch is cut for $1~\mathrm{s}$ before and
after the glitch.  This time is calculated from the length of the FIR deconvolution kernel~(\S\ref{sec:deconv})
to ensure that the remaining time stream can safely be deconvolved without being affected by ringing.
For step-like glitches, the DC levels of the time stream before and after the excised
portion are level-matched in order to remove the low-frequency Fourier components of the step that could otherwise
contaminate noise simulations.  Finally, for large flux-jump steps, these measures are applied not only to the affected channel itself,
but to all nearby channels that could potentially be sensitive to crosstalk from the affected channel.

\subsection{Gain calibration}
\label{sec:gaincal}
The gain calibration is performed in two steps.  The first is a relative gain calibration derived from
elevation nods (\S\ref{sec:scansetcals}).  The second is an absolute
calibration derived from the CMB temperature itself and applied in mapmaking.
Given the fast detector time constants~(\S\ref{sec:timeconst}) there is no need for a frequency-dependent gain
calibration.

The relative gain calibration is based on the detector response to the \elnods\ performed immediately
before and after each observing block.  The calibration factor for a given detector is
its own \elnod\ response in raw digitizer units per air mass, divided by the median across
all detectors on the focal plane.  This procedure corrects for variation and drift in individual
detector response, while dividing out day-to-day changes in the overall brightness of the
atmosphere.  The relative gain correction is performed on the time streams as part of the low-level reduction,
before the polarization partner channels are combined to give the pair-sum and pair-difference
quantities.

The relative gain procedure allows the maps from all detector pairs to be coadded in the same
units.  This fully integrated map still must be converted into physical units of $\mathrm{\muK}_\mathrm{CMB}$.
We find the absolute gain factor for the entire data set by cross-calibrating the \bicep2
temperature maps against an external data set.  The overall intensity calibration thus comes from the CMB itself.
The absolute calibration factor $g$ is the ratio of two cross-spectra formed from the \bicep2
temperature map and two external data sets, the \textit{reference} and \textit{calibration} maps.
It is defined in the $i^\mathrm{th}$ multipole bin as
  \begin{equation}
  \label{eq:abscal}
    g_i = \frac{\sum_{\ell \in \left\{\ell_i\right\}} \left<a_{\ell m}^\mathrm{ref} ~ a_{\ell m}^\mathrm{B2}\right>}
          {\sum_{\ell \in \left\{\ell_i\right\}} \left<a_{\ell m}^\mathrm{ref} ~ a_{\ell m}^\mathrm{cal}\right>}
  \end{equation}
We choose to use separate reference and calibration maps, with uncorrelated noise, so that the cross-spectra
in the numerator and the denominator of Eq.~\ref{eq:abscal} are both free of noise bias.  The resulting
calibration is only very weakly sensitive to the reference map.
Before forming cross-spectra
we reobserve each with the \bicep2 observing pattern, pipeline, and averaged beam profile.
The consistency of the $g_i$ across all multipole bins therefore also serves as a confirmation of
the correctness of the applied beam profile.
For the final gain calibration number $g$ we take the mean of $g_i$ over five multipole bins in the range
$30\leq\ell\leq210$.

We consider the calibration of the \bicep2 three-year data set using external maps from both the
\wmap\ 9-year release~\citep{wmap9yr} and the \planck\ 2013 release~\citep{planckI,planckVI}.
For \wmap\ we use the V and W~band products as the reference and calibration maps, respectively.
For \planck\ we use the $100~\mathrm{GHz}$ and $143~\mathrm{GHz}$ products as the reference and calibration
maps, respectively.
In each case we deconvolve the \wmap\ or \planck\ beams before reobserving with
\bicep2 parameters.
The statistical error on the resulting estimates of $g$ is well below 1\%, with the overall
uncertainty dominated by the choice of calibration map.  The calculations from \planck\ and
\wmap9 differ slightly, with $g($\wmap9$)$ about $2\%$ higher than $g($\planck$)$.
This corresponds to the disagreement between the two experiments in the amplitude of the
CMB $TT$ spectrum~\citep{spergel2013}.  We adopt a central value of $g=3150~\mathrm{\muK}_\mathrm{CMB}$
per raw digitizer unit.  Given the tension between external calibration data sets this
result can be taken to have an uncertainty of $1\%$, which meets the requirements
for absolute temperature calibration of \bicep2.

\subsection{Pair difference}

   \begin{figure}
   \begin{center}
   \begin{tabular}{c}
   \def\svgwidth{7.7cm}
   \input{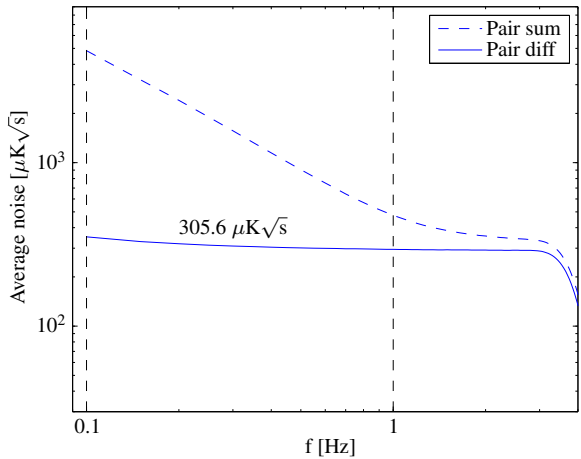}
   \end{tabular}
   \end{center}
   \caption[example]
   { \label{fig:sumdiffpsd}
      Average noise power spectra of all detector pairs with 2011--12 settings.  Cuts and inverse-variance weighting
      are applied as in standard mapmaking.  The pair-sum spectra show $1/f$ noise, which is removed by taking the
      pair difference.  The average NET per detector is taken from the average power spectral density in the pair-difference
      spectrum between 0.1 and $1~\mathrm{Hz}$.
   }
   \end{figure}

In addition to photon noise and the intrinsic noise of the
detectors and readout, there is an atmospheric signal.  This
appears as $1/f$ noise that dominates at low frequency in
the power spectrum from a single detector.  Because the
atmosphere is very unpolarized~\citep{hanany03},
the $1/f$ noise can be removed by differencing the A and B
detectors of each polarization pair.  We form pair-differenced
quantities for use in making polarization maps, and pair-sum
quantities for use in making temperature maps.  The pair-sum
time streams contain atmospheric $1/f$ noise while the
pair-differenced quantities do not, as shown in Fig.~\ref{fig:sumdiffpsd}.
The CMB $T$ maps accordingly have very different noise properties
and weaker sensitivity than the CMB $Q$ and $U$ maps, and the mapping speed 
in $T$ is much more strongly affected by weather.  Because
the primary science goal of \bicep2 is to make a polarization measurement, while the
CMB temperature in our field is already measured to high
signal-to-noise, the sensitivity and map depth calculations
in \S\ref{sec:data} will consider only the pair-differenced
or polarization quantities.

\subsection{Polynomial filtering}
\label{sec:polysub}
We remove remaining atmospheric $1/f$ noise at very low
frequencies and residual magnetic pickup by applying a polynomial
filter.  This is a third-order polynomial subtracted from the
time stream of each pair-sum and pair-difference channel
for each left-going or right-going scan.  It effectively
removes very low frequencies or very low multipole moments
in the scan direction (\ie~in right ascension).

\subsection{Ground subtraction}
\label{sec:groundsub}
We perform an additional filtering step to remove any signal
that is fixed relative to the ground rather than the sky.
We form a template in ground-fixed coordinates
for each pair-sum and each pair-difference channel over all
the scans in a scanset and subtract this template from the time streams.
Because the sky moves relative to the ground by $12.5^\circ$
of azimuth or right ascension during the 50~minutes of a scanset,
a sky-fixed signal is not strongly filtered.
This filtering is
accounted for in the power spectrum analysis as described in the
Results Paper.

Several types of signal enter into the ground-fixed template:
reflections from ground-fixed objects; any magnetic signal
not already removed by polynomial filtering; and the satellite
transmitter signal~(\S\ref{sec:satcom}).  There is some evidence
of a small atmospheric polarization dependence on the wind
direction.  Because the winds at the South Pole are steady over
long periods of time, this is also effectively removed by
ground subtraction.

\subsection{Data selection and cuts}
\label{sec:cuts}

\begin{table*}
\caption{Data Cuts}
\label{tab:cuts}
\begin{center}
\begin{tabular}{lccc}
  \hline \hline
  Cut parameter & Total time [10$^6~\mathrm{s}$] & Integration [10$^9~\mathrm{det}\cdot\mathrm{s}$] & Fraction cut [\%] \\
  \hline
  Before cuts & 36.5 & 14.8 &  -- \\
  Channel cuts & 36.5 & 13.2 & 10.9 \\
  Synchronization & 35.3 & 12.7 & 3.1 \\
  Deglitching & 33.6 & 10.7 & 13.8 \\
  Per-scan noise & 33.6 & 10.7 & $<0.01\%$ \\
  Passing channels & 33.3 & 10.7 & 0.22 \\
  Manual cut & 33.0 & 10.6 & 0.43 \\
  Elevation nod & 31.0 & 9.2 & 9.5 \\
  Fractional resistance & 31.0 & 9.2 & 0.16 \\
  Skewness & 31.0 & 9.1 & 0.41 \\
  Time stream variance & 30.9 & 9.0 & 0.52 \\
  Correlated noise & 30.9 & 9.0 & $<0.01\%$ \\
  Noise stationarity & 30.7 & 8.9 & 0.64 \\
  FPU temperature & 30.6 & 8.9 & 0.20 \\
  Passing data & 27.6 & 8.6 & 1.7 \\
  \hline
\end{tabular}
\end{center}
\end{table*}

The data set used for CMB mapmaking begins on 2010 February 15
and continues until 2012 Nov 6.  The period from 2011 January 1
through 2011 March 1 is excluded as it was used for special-purpose
calibrations and tuning of the 25~kHz data taking parameters.

We have developed a multistage data cut procedure which identifies and
removes data that suffers from pathologies related to detectors, multiplexing,
thermal control, noise properties, or data acquisition.  The goal is to ensure
that all data used in mapmaking is taken when the experiment is operating
properly and has only stationary, well-behaved noise.
We do not explicitly attempt to cut data which is noisier than average.
Instead, we apply inverse-variance weighting in mapmaking to form a nearly
optimal combination of partial maps made with detectors and time periods that
have differing noise levels.

A full list of the cut parameters is shown in Table~\ref{tab:cuts}.
The statistics listed are cumulative.
The first column gives the total time used for mapmaking after applying the named
cut and all cuts listed above it.
The second column is similar, but multiplies time by number of included detectors
to give a measure of the total map integration after applying each round of cuts.
The time and integration only give time actually accumulated into maps; they
exclude the time spent in elevation nods, scan turnarounds, etc.  The
integration column includes only detectors contributing to the polarization maps
for each observation.
The final column shows the fraction of the total
exposure that passes the previous cuts but fails the cut named on this line.
Because problematic data often fail several cuts, the fraction removed depends
on the order in which the cuts are listed.
(For reasons of convenience in implementation the pipeline applies some cuts
in a different order from that shown in the table.)

Before applying any cuts there are 384 good light channels in the early 2010
data set and 412 in late 2010 through 2012 (after TES bias optimization).  The first
cut is a channel cut that removes detectors with discrepant beam shape or
differences in pointing and beam shape between the A and B detectors.  It also
removes a small number of detectors in which there is a detectable amount of
leakage of unpolarized signal into the pair-differenced time streams because of
faults in the multiplexing and readout.  After applying channel cuts there are 334
remaining channels in early 2010 and 370 in late 2010 through 2012.   The second
line of Table~\ref{tab:cuts} shows that the channel cuts do not change the total
time, but they do remove $10.9\%$ of the exposure.

The synchronization cut excludes several discrete periods of time in 2010 and 2011
when the time stamps in the MCE data could not be matched to time stamps in the
telescope pointing because of failures in the hardware or control system~(\S\ref{sec:control}).

Cuts related to deglitching remove individual scans affected by flagged transient
events, both for the affected detector and (in the case of large glitches or flux jumps)
for neighbors potentially vulnerable to crosstalk.  Channels suffering an excessive
number of glitches within a scanset are removed
for the duration of the scanset.  This cut removes a substantial amount of data,
primarily because of the necessity of cutting the neighbors of unstable
detectors~(\S\ref{sec:unstable}).
The per-scan noise cut is set very loosely to exclude only severely high
noise also likely to be caused by a detector instability.

The passing channels cut requires that each individual scan have a sufficiently
large fraction ($\geq70\%$) of detectors within each multiplexing column passing
all cuts.  This is a further precaution against time stream irregularities that could
affect other channels through crosstalk.

The manual cut removes a small amount of data in 2010 that appear nominal
but may have nonstandard settings and is removed as a precaution.

The elevation nod cuts ensure that the leading and trailing elevation nods have
atmospheric response agreeing to within 30\% in each channel, or
10\% in the ratio of A to B detectors within a pair.  This ensures that the relative
gain is sufficiently stable for the pair differencing procedure to remove atmospheric
noise.  Detectors are also cut if their \elnods\ contain time stream glitches or
fail to match the expected atmospheric profile.  These cuts primarily remove
data in which the atmosphere changes too much over the course of a scanset,
for example when clouds are carried in or out by wind.

The fractional resistance cut removes detectors that have fractional resistance
outside the range $0.1<R_\mathrm{TES}/R_\mathrm{normal}<0.95$.  This excludes
detectors that have either latched into the superconducting state or have
been pushed out of transition by unusually high atmospheric loading.

The skewness, time stream variance, correlated noise, and noise stationarity
cuts remove detectors whose noise does not match the expected model during
a scanset.  This is usually caused by an irregularity in the readout system or by
unusually time-varying weather.

   \begin{figure*}
   \begin{center}
   \begin{tabular}{c}
   \def\svgwidth{17cm}
   \input{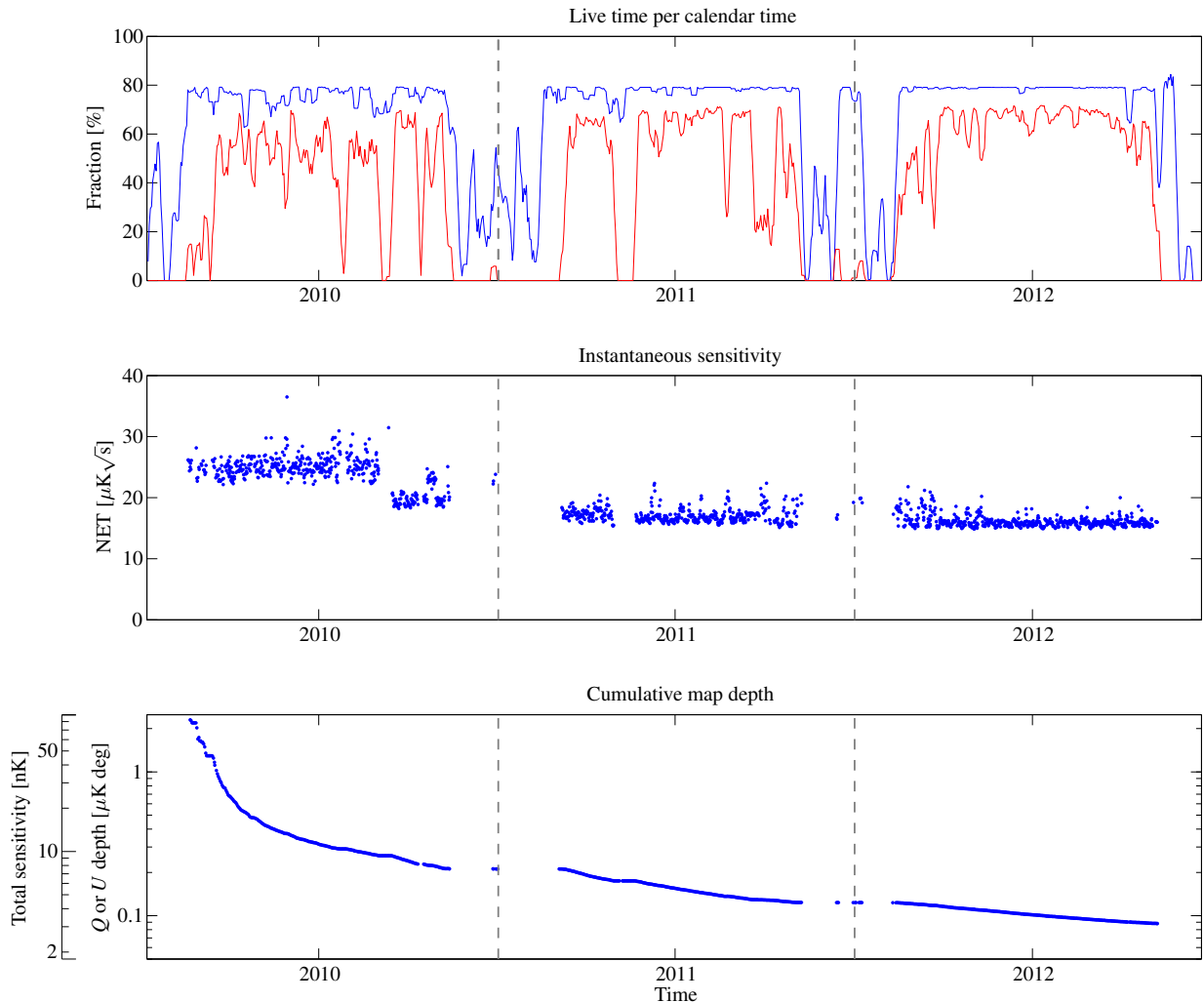}
   \end{tabular}
   \end{center}
   \caption[example]
   { \label{fig:integration}
Integration of the \bicep2 three-year data set.
\textit{Top panel:} Time per day spent in CMB scans, regular calibrations, and refrigerator cycling.
During austral summers (November--February), observing schedules have been interspersed with beam mapping
and other tests and calibrations.  During the austral winter, on-source efficiency (including Galactic observations) has been high,
never falling far below the ideal 79.2\% in the winter of 2012.
The lower, red curve includes data quality cuts.
\textit{Middle panel:} Mapping speed over time.  The improvement from early to late 2010 was
caused by the optimization of TES biases (\S\ref{sec:biasoptimization}), the improvement
from 2010 to 2011 was caused by the change to 25~kHz multiplexing (\S\ref{sec:muxoptimization}), and
the small improvement from 2011 to 2012 was caused by a small increase in the number of active channels
(\S\ref{sec:unstable}).  Each data point represents one observing phase of six or nine hours, as
described in Section~\ref{sec:schedules}.
\textit{Bottom panel:} Cumulative map depth over time as calculated in
\S\ref{sec:mapdepth}.
}
   \end{figure*}

The FPU temperature cut requires that the focal plane Cu plate be stably in the
range $200$--$300~\mathrm{mK}$ with a standard deviation no more than
$50~\mu\mathrm{K}$.  This removes a small number of scansets in which
the temperature had not settled sufficiently after the refrigerator recycling.

Finally, we demand a reasonable fraction of data to pass all of the above cuts.
This passing data cut ensures that the experiment is generally performing as
expected during a particular \scanset. If fewer than 50\% of channels pass all
the cuts, then the all of the data for that scanset is removed.

The overall pass fraction of the cuts is 75.6\% in terms of integration
time, or 58.1\% in terms of exposure when accounting for the number of
detectors passing cuts in each scanset.  The bottom-line figures in
Table~\ref{tab:cuts} give an average detector count of $311.6$ individual
detectors contributing to the map at any given time.

\subsection{Mapmaking}

The time streams are binned into equirectangular maps on a 
grid in right ascension and declination.  The map pixels are $0.25^\circ$ square
at the center of the field, declination $-57.5^\circ$.  The pair-sum time streams
are accumulated into the $T$ map, while the pair-difference time streams are accumulated
into the $Q$ and $U$ maps according to the polarization angles of the detectors
as transformed into celestial coordinates.  As the time stream for a single detector
pair in a single scanset is coadded into the map, it is weighted according to the
inverse of its time-stream variance across the entire scanset.

We ordinarily apply deprojection of one or more templates during mapmaking in
order to remove modes that have been contaminated by mismatches between the A
and B beams or relative gains.  The deprojection algorithm and its effectiveness
are described in the Systematics Paper.  For the remainder of the current paper
we apply deprojection of relative gain mismatch, differential pointing, and
differential ellipticity.

   \begin{figure*}[t]
   \begin{center}
   \begin{tabular}{c}
   \def\svgwidth{17cm}
   \input{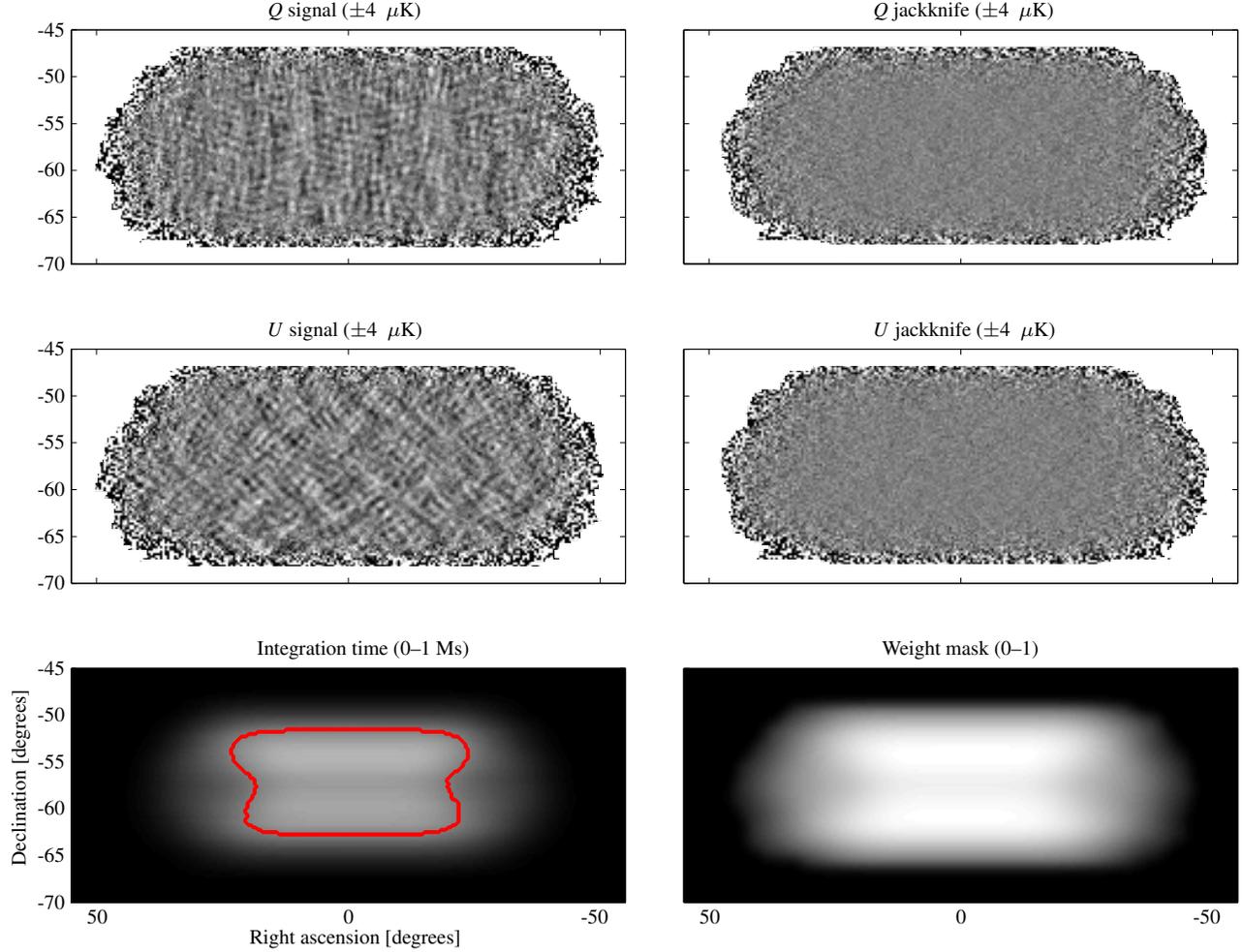}
   \end{tabular}
   \end{center}
   \caption[example]
   { \label{fig:qumaps}
Polarization maps and coverage maps used to calculate map depth (color scales in parentheses).
The maps are Stokes $Q$ and $U$
in the three-year data set, with full coadds on the left and temporal split jackknife
maps on the right.
The $Q$ maps show a horizontal and vertical pattern, while the $U$ maps show a diagonal pattern,
together revealing the dominant $E$-mode polarization pattern of the CMB.  The jackknife
maps contain no signal but only noise.  They are used to calculate the depth in our
polarization maps.
The lower left panel shows the integration time per 0.25$^\circ \times$0.25$^\circ$
pixel and the 70\% contour used
in older definition of the map depth, while the lower right panel shows the variance-weight
map used in the definition adopted here.
}
   \end{figure*}

\section{Three-year data set}
\label{sec:data}

\biceptwo\ was installed in the mount in DSL on 2009 December 22, and cooled to base
temperature.  It then ran for three years without warming up
or breaking vacuum.  During each austral summer, CMB observation was suspended for special-purpose
calibration data taking as described in \S\ref{sec:performance}.  Many of the calibrations required
the use of additional equipment such as the Fourier-transform spectrometer (\S\ref{sec:spectra}),
the far-field flat mirror (\S\ref{sec:beams}), the near-field beam mapping stage, 
and the dielectric sheet calibrator (\S\ref{sec:pol_response}).
Most of these were designed
to attach to the telescope without removing it from the mount (although some required temporary
removal of the absorbing forebaffle).  The telescope was finally removed from the mount on 2012 December 13.
The uninterrupted run allowed \biceptwo\ to maintain excellent stability and uniformity over the course
of its three-year CMB observations.

Although the telescope hardware remained unmodified during the entire run, some of the software-defined
operating parameters were optimized for greater sensitivity.  The detector biases were modified partway through the 2010 observing season,
and the multiplexing rate was increased at the end of 2010.  These optimizations increased the total number
of good detectors, improved the instrument noise, and also reduced detector crosstalk.  The three-year data
set can be divided into three epochs: early 2010, with conservatively chosen detector biases and multiplexing
rate; late 2010, with optimized biases; and 2011--2012, with optimized biases and multiplexing parameters.  Each
of the modifications resulted in a significant improvement in instantaneous sensitivity, with uniformly high
mapping speed for the entire 2011--2012 epoch.

The achieved efficiency in integrating on the CMB, the improvements in mapping speed, and the
progression of map depth are shown in Fig.~\ref{fig:integration}.

\subsection{Instantaneous sensitivity}
\label{sec:mapspeed}

The instantaneous sensitivity of the experiment can be expressed as a noise-equivalent temperature
(NET), or as a mapping speed ($\mathrm{NET}^{-2}$).  We have already calculated per-detector
NETs in~\S\ref{sec:noise}, using a limited set of good-weather data without applying cuts.
The achieved array NET for each phase in the three-year data set has been calculated
in the same way, following~\cite{brevik11}.  The results are shown in the middle panel of
Fig.~\ref{fig:integration}.  The overall trend shows the improvement in
mapping speed from early 2010 to late 2010 and then to 2011--12, the period with the
lowest noise.
Within each period the NET is generally stable, since the polarization
measurement is largely insensitive to variations in weather.  Although the instantaneous
sensitivity was almost unchanged between the 2011 and 2012 seasons, averaging $15.8~\mu\mathrm{K}\sqrt{\mathrm{s}}$,
the 2012 season makes the strongest contribution to the integrated sensitivity because of its
near-ideal live time fraction.

\begin{table*}[t]
\caption{Sensitivity by season}
\label{tab:seasons}
\begin{center}
\begin{tabular}{lcccc} 
\hline \hline
\rule[-1ex]{0pt}{3.5ex}  Season & Time [$10^6$~s] & Integration [$10^9$~det$\cdot$s] & Map depth [nK$\cdot$deg] & Total sensitivity [nK] \\
\hline
\rule[-1ex]{0pt}{3.5ex}  2010  &  8.1 & 2.3 & 213.2 & 7.70 \\
\rule[-1ex]{0pt}{3.5ex}  2011  &  8.5 & 2.6 & 148.6 & 5.37 \\
\rule[-1ex]{0pt}{3.5ex}  2012  & 11.0 & 3.7 & 124.0 & 4.47 \\
\rule[-1ex]{0pt}{3.5ex}  Total & 27.6 & 8.6 & 87.8 & 3.15 \\
\hline
\end{tabular}
\end{center}
\end{table*}

\subsection{Map depth}
\label{sec:mapdepth}

The map depth is a measure of the noise level in the polarization maps.  Together
with the area of the maps (solid angle on the sky), the map depth sets the final
sensitivity of the experiment under the assumption that it is statistically limited.
The coverage of the map is not uniform: the integration time is much higher in the
central region than near the edges.  (This is partly because of the large instantaneous
field of view of \bicep2.)  We choose to give the map depth $D$ in
the deepest, central part of the map, and calculate an effective area $A_\mathrm{eff}$
that accounts for the higher variance and lower weight in other parts of the map.

The effective area is calculated using the same apodization mask that is used in
the power spectrum analysis in the Results Paper.  This is constructed from
the maps of variance in $Q$ and $U$ as estimated from time stream noise.  The two
variance maps for the two Stokes parameters are averaged and then smoothed slightly
to ensure that they smoothly fall to zero at the edges.  The inverse of the
resulting smoothed variance map is the apodization or weight mask shown in the lower
right panel of Fig.~\ref{fig:qumaps}.  The effective area $A_\mathrm{eff}$ is the
integral of this weight mask over the entire field.  This calculation accounts for
the nonuniform coverage of the field, weighting each map pixel by its contribution
relative to the deepest, central part of the map.

The map depth is calculated from the variance in difference maps made from
two subsets of the data set.  
The differenced ``jackknife'' maps contain noise
at the same level as the fully coadded maps, but no signal.
We use a temporal split jackknife with equal noise levels in the two halves,
corresponding approximately to (2010 through late 2011) and (late 2011 through 2012).
The chronological jackknife $Q$ and $U$ maps are shown in the right-hand panels
of Fig.~\ref{fig:qumaps}.  (The chronological jackknife gives uncorrelated
noise in the two halves.  For this reason we prefer it to the scan-direction
jackknife we have used in some past map depth calculations, although
noise correlations are already very small in the polarization maps after
pair differencing.)

In past publications~\citep{brevik11,ogburn12,kernasovskiy12} we have reported map depths from only the best-covered,
central 70\% of the CMB field.  
We now choose instead to apodize the map with the same weight mask
used to define the effective area.  The single map depth number calculated in this way
also represents the depth in the deepest part of the field.  The combination
of $D$ and $A_\mathrm{eff}$ derived from the same apodization mask used
in the power spectrum analysis constitutes the most appropriate measure
of the achieved map depth as used in the final $B$-mode power spectrum
analysis.

Using the new definition, we calculate noise levels
in the Stokes $Q$ and $U$ maps of 86.7 and 87.7~nK in square-degree pixels
respectively (5.2 and 5.3~$\mu$K $\cdot$ arcmin), or an averaged map depth
of 87.2~nK$\cdot$degree.  The effective area of the map at this depth is
383.7 square degrees.

We also define a ``total sensitivity'' figure by combining the depths of
the Stokes $Q$ and $U$ maps and dividing by the square root of the effective
area, $T=D_\mathrm{min}/\sqrt{A_\mathrm{eff}}$.  This gives a single number in
temperature units that indicates the
level of $B$-mode signal that could be detected by the experiment within its
field and at the angular scales of interest.  For the \bicep2 three-year
data set we calculate total sensitivity of~$T=3.15~\mathrm{nK}$.  The progression
over the three-year data set is indicated by the second vertical axis in
the bottom panel of Fig.~\ref{fig:integration}, and the achieved map depth
and total sensitivity in each year and in the three-year data set are
listed in Table~\ref{tab:seasons}.

\section{Conclusions}
\label{sec:conclusion}
We have presented the design and performance characterization of \bicep2,
an experiment built to search for the inflationary gravitational wave
background through $B$-mode polarization on angular scales around $2^\circ$.
\bicep2 has completed three years of observation (2010--12) from the
South Pole.  The three-year data set has unprecedented map depth of
87.2~nK in square-degree pixels over an effective area of 383.7 square
degrees.  This corresponds to a total sensitivity level of
$T=3.15~\mathrm{nK}$.

The instrument has been extensively characterized, especially for possible
sources of systematic false polarization.  The Systematics Paper~\citep[in preparation]{b2syst14}
will show that these effects have been understood and controlled at a level
sufficient to remain dominated by integration time rather than systematics.
The subsequent Beams Paper~\citep[in preparation]{b2beams14} will provide additional performance
characterization for the main beams and sidelobes in \bicep2 and the
\keck.  The Results Paper~\citep{b2results14} presents the primary
scientific results of the three-year \bicep2 data set, making the first detection of
$B$-mode power at degree angular scales.

\acknowledgements

\bicep2 was supported by the US National Science Foundation under
grants ANT-0742818 and ANT-1044978 (Caltech/Harvard) and ANT-0742592
and ANT-1110087 (Chicago/Minnesota).  The development of antenna-coupled
detector technology was supported by the JPL Research and Technology
Development Fund and grants 06-ARPA206-0040 and 10-SAT10-0017
from the NASA APRA and SAT programs.  The development and testing of
focal planes were supported by the Gordon and Betty Moore Foundation
at Caltech.  Readout electronics were supported by a Canada Foundation
for Innovation grant to UBC.  The receiver development was supported
in part by a grant from the W. M. Keck Foundation.
The computations in this paper were run on the Odyssey cluster
supported by the FAS Science Division Research Computing Group at
Harvard University.  Tireless administrative support was provided by
Irene Coyle and Kathy Deniston.

We thank the staff of the US Antarctic Program and in particular
the South Pole Station without whose help this research would not have been possible.
We thank all those who have contributed past efforts to the \bicep /\keck\
series of experiments, including the \bicepone\ and \keck\ teams,
as well as our colleagues on the \spider\ team with whom
we coordinated receiver and detector development efforts at Caltech.
We dedicate this paper to the memory of Andrew Lange, whom we sorely miss.

\bibliographystyle{apj}
\bibliography{instr}

\end{document}